\newif\ifarxiv
\arxivtrue
\ifarxiv
\RequirePackage{snapshot}
\documentclass[
    reprint,
    superscriptaddress,
    amsmath,amssymb,
    aps,
    pra,
]{revtex4-2}
\renewcommand{\doi}[1]{\href{http://doi.org/#1}{doi:#1}}
\else
\documentclass[journal=jacsat,suppinfo,bookmarksdepth=subparagraph]{achemso}
\newcommand{\doi}[1]{\href{http://doi.org/#1}{doi:#1}}
\fi
\newif\ifshowlabels
\showlabelsfalse
\newif\iffigdraft
\figdraftfalse
\newif\ifgentoc
\newif\ifeditmode
\editmodefalse
\gentocfalse
\newif\ifshowparagraphinline
\showparagraphinlinefalse
\ifeditmode\else
\showparagraphinlinefalse
\fi
\ifarxiv
\usepackage[sectionbib]{bibunits}
\defaultbibliographystyle{new}
\defaultbibliography{references,msNotes}
\else
\fi
\usepackage{layouts}
\usepackage[acronyms]{glossaries}
\glsdisablehyper
\usepackage[procnames]{listings}
\usepackage{pygments} 
\usepackage{booktabs}
\providecommand{\tightlist}{%
  \setlength{\itemsep}{0pt}\setlength{\parskip}{0pt}}
\usepackage{amsfonts}
\usepackage{blindtext}
\let\oldblindtext\blindtext
\renewcommand{\blindtext}{\textcolor{gray}{\oldblindtext}}
\usepackage{cmap}
\usepackage{datetime}
\usepackage{environ}
\usepackage{etoolbox}
\usepackage[utf8]{inputenc}
\DeclareUnicodeCharacter{03BD}{\ensuremath{\nu}}
\DeclareUnicodeCharacter{2212}{--}
\DeclareUnicodeCharacter{03BC}{\textmu}
\DeclareUnicodeCharacter{039B}{\ensuremath{\Lambda}}
\DeclareUnicodeCharacter{03A9}{\ensuremath{\Omega}}
\DeclareUnicodeCharacter{03B2}{\ensuremath{\beta}}
\DeclareUnicodeCharacter{03B1}{\ensuremath{\alpha}}
\DeclareUnicodeCharacter{00B0}{\ensuremath{^\circ}}
\DeclareUnicodeCharacter{00B1}{\ensuremath{\pm}}
\DeclareUnicodeCharacter{2192}{\ensuremath{\rightarrow}}
\DeclareUnicodeCharacter{03C3}{\ensuremath{\sigma}}
\DeclareUnicodeCharacter{03C4}{\ensuremath{\tau}}
\DeclareUnicodeCharacter{2009}{ }
\DeclareUnicodeCharacter{2080}{\ensuremath{_0}}
\DeclareUnicodeCharacter{2082}{\ensuremath{_2}}
\DeclareUnicodeCharacter{2084}{\ensuremath{_4}}
\DeclareUnicodeCharacter{1F449}{\ensuremath{\rightarrow}} 
\DeclareUnicodeCharacter{1F448}{\ensuremath{\leftarrow}} 
\iffigdraft
    \usepackage[draft]{graphicx}
\else
    \usepackage{graphicx}
\fi
\usepackage{hyphenat}
\usepackage{ifxetex}
\usepackage{mathrsfs}
\usepackage{calc}

\newcommand{\ie}{\textit{i.e.\@}\xspace}
\newcommand{\vs}{\textit{vs.\@}\xspace}
\newcommand{\eg}{\textit{e.g.\@}\xspace}

\newcommand{\via}{\textit{via}\xspace}
\newcommand{\magn}[1]{\ensuremath{\times 10^{#1}}}

\ifeditmode
\usepackage{pdfcomment}

\else
\newcommand{\pdfcomment}[2][]{}

\fi
\usepackage{hyperref}
\hypersetup{colorlinks=true, pdfstartview=FitV, linkcolor=linkcolor, citecolor=dbluecolor, urlcolor=dgreencolor, bookmarksdepth=subparagraph}
\usepackage{bookmark}
\usepackage{sidecap}
\usepackage{soul}
\usepackage{textcomp}
\usepackage{url}
\usepackage{xcolor}
\usepackage{xspace}
\usepackage{multirow}
\usepackage{array}
\newcolumntype{R}{>{\vspace{2ex}\displaystyle}{r}}
\newcolumntype{L}{>{\displaystyle}{l}}
\definecolor{SUgrey}{HTML}{6F777D}
\definecolor{SUorange}{HTML}{D44500}
\definecolor{dbluecolor}{rgb}{.01,.02,0.29}
\definecolor{dgraycolor}{rgb}{0.50,0.50,0.50}
\definecolor{dgreencolor}{rgb}{0.0,0.4,0}
\definecolor{linkcolor}{cmyk}{0,0.7,0.5,0.5}
\usepackage{accsupp}    
\newcommand{\noncopynumber}[1]{%
    \BeginAccSupp{method=escape,ActualText={}}%
    #1%
    \EndAccSupp{}%
}

\usepackage[compact]{titlesec}
\ifshowparagraphinline
\titleformat{\paragraph}[runin]{\color{gray}\normalfont\bfseries\footnotesize}{}{3pt}{\hspace{0.75em}\ul{\footnotesize\thesubsection\theparagraph)\;}}[:]
\else
\renewcommand{\paragraph}[1]{\par\phantomsection\addcontentsline{toc}{paragraph}{#1}}
\fi
\makeatletter
\@ifundefined{linelabel}{%
\newcommand{\linelabel}[1]{}}{}
\makeatother
\makeglossaries
\renewcommand{\glossarysection}[2][]{}
\newacronym{odnp}{ODNP}{Overhauser Dynamic Nuclear Polarization}
\newacronym{adrosys}{ADROSYS}{Automated Deuterium Relaxation-Ordered SpectroscopY in Solution}
\newacronym{snr}{SNR}{signal-to-noise ratio}
\newacronym{nmr}{NMR}{Nuclear Magnetic Resonance}
\newacronym{esr}{ESR}{Electron Spin Resonance}
\newacronym{mps}{MPS}{Microwave Power Source}
\newacronym{ilt}{ILT}{Inverse Laplace Transform}
\newacronym{pre}{PRE}{Paramagnetic Relaxation Enhancement}
\newacronym{rosy}{ROSY}{Relaxation-Ordered SpectroscopY}
\newacronym{dss}{DSS}{Dynamic Stokes Shift}
\newacronym{md}{MD}{molecular dynamics}
\newacronym{ir}{IR}{infrared}
\newacronym{tms}{TMS}{tetramethylsilane}
\newacronym[plural=SPs,firstplural=electron spin probes (SPs)]{sp}{SP}{electron spin probe}
\newacronym[plural=RMs,firstplural=Reverse Micelles (RMs)]{rm}{RM}{reverse micelle}
\newacronym{aot}{AOT}{Aerosol-OT}
\newacronym{qens}{QENS}{Quasi-Elastic Neutron Scattering}

\renewcommand{\thesection}{\Roman{section}}
\renewcommand{\thesubsection}{\thesection.\arabic{subsection}}

\makeatletter
\renewcommand{\p@subsection}{}
\renewcommand{\p@subsubsection}{}
\makeatother
\newcounter{subfigure}[figure]
\newcounter{subfigurenonumber}
\newcounter{tempfigure}
\renewcommand\thesubfigure{
\arabic{tempfigure}\alph{subfigure}}
\renewcommand\thesubfigurenonumber{(\alph{subfigurenonumber})}
\newcommand{\subfig}[2]{%
    \setcounter{tempfigure}{\value{figure}}%
    \addtocounter{tempfigure}{1}%
    \refstepcounter{subfigure}%
    \setcounter{subfigurenonumber}{\value{subfigure}}%
    \expandafter\edef\csname ref#2\endcsname{\thesubfigurenonumber}
    \label{#1}%
    }
\newcommand{\titleblock}{
\newcommand\SUaffil{\affiliation{Department of Chemistry, Syracuse University, Syracuse, NY 13210, USA}}
\author{A. Guinness}
\author{Alec A. Beaton}
\author{John M. Franck}
\SUaffil
\email{jmfranck@syr.edu}
\title{Separate and Detailed Characterization of Signal and Noise Enables NMR Under Adverse Circumstances}
\date{\today}
}
\ifarxiv\else
    \titleblock
\fi
\newif\ifpoormancref
\poormancreffalse
\ifpoormancref
\usepackage[poorman]{cleveref}
\else
\usepackage{cleveref}
\fi
\crefname{equation}{Eq.}{Eqs.}
\crefname{table}{Table}{Tables}
\crefname{figure}{Fig.}{Figs.}
\crefname{section}{Sec.}{Sec.}
\crefname{subfigure}{Fig.}{Figs.}
\crefname{lstlisting}{listing}{listings}
\Crefname{lstlisting}{Listing}{Listings}

\usepackage{xr}
\ifshowlabels
\usepackage{refcheck}
\makeatletter
\newcommand{\refcheckize}[1]{%
  \expandafter\let\csname @@\string#1\endcsname#1%
  \expandafter\DeclareRobustCommand\csname relax\string#1\endcsname[1]{%
    \csname @@\string#1\endcsname{##1}\wrtusdrf{##1}}%
  \expandafter\let\expandafter#1\csname relax\string#1\endcsname
}
\def\@setmarginlbl{%
    \if@show@ref
        \if@labelled
            \set@fbox@par
            \if@unsdlbl
                \makebox[0pt][l]{\zero@height{$\,$\rotatebox{90}{\scalebox{0.6}{\mark@size
                {\bfseries\upshape?}\underline{\last@lbl}{k\bfseries\upshape?}}}}}%
            \else
                \makebox[0pt][l]{\zero@height{$\,$\rotatebox{90}{\scalebox{0.6}{\fbox{{\mark@size\last@lbl}}}}}}%
            \fi
        \else
            \if@show@unl@bld
                \makebox[0pt][l]{\zero@height{$\,$\rotatebox{90}{\scalebox{0.6}{\unl@bld@mark}}}}%
            \fi\fi
        \fi
        \global\@labelledfalse
    }
\def\@setnmmarginlbl{%
    \if@show@ref
        \set@fbox@par
        \if@unsdlbl
            \hbox to \textwidth{\makebox[0pt][r]{\rotatebox{90}{\scalebox{0.6}{\mark@size{\bfseries
                            \upshape?}$\langle$\last@lbl$\rangle${\bfseries
            \upshape?}}}$\,$}\hfill}%
        \else
            \hbox to \textwidth{\makebox[0pt][r]{\rotatebox{90}{\scalebox{0.6}{\mark@size$\langle$%
            \last@lbl$\rangle$}}$\,$}\hfill}%
        \fi
    \fi
    \global\@labelledfalse
}
\def\@bibitem@proceed@#1{%
    \@ifundefined{cit@#1}{\@warning@rc@{Unused bibitem `#1'}%
        \if@show@cite
            \gdef\@biblabel{\makebox[0pt][r]{\zero@height{\rotatebox{90}{\scalebox{0.6}{{\mark@size{\bfseries\upshape?}}%
                \underline{\@verbatim@{#1}}{\mark@size{\bfseries\upshape?}}}}$\,$}}%
            \@@@biblabel@@}%
        \fi
    }{%
        \if@show@cite
            \set@fbox@par
            \gdef\@biblabel{\makebox[0pt][r]{\zero@height{\rotatebox{90}{\scalebox{0.6}{\fbox{TESTTESTTEST\@verbatim@{#1}}}}$\,$}}\@@@biblabel@@}%
        \fi
}}%
\makeatother
\refcheckize{\cref}
\refcheckize{\Cref}
\fi
\begin{document}
\newlength\myfigwidth
\setlength{\myfigwidth}{3.5in}
\ifarxiv
\titleblock
\fi

\begin{abstract}
When deploying a spectrometer in an adverse
environment,
such as during a typical ODNP experiment or
other experiments that require low-volume low-field
measurements,
a clear and modern protocol for characterizing
and quantifying
the absolute signal and noise levels proves
essential.
This paper provides such a protocol.
It also highlights the clarity and insight
that come from
(1)
characterizing the spectral distribution of
the noise,
and (2) discussing NMR signal
intensities in (conserved) units of square root
power.
Specifically, the signal intensity is 
derived from a theory and notation
that combine elements of those developed for
NMR and ESR spectroscopy,
while the spectral noise density
``fingerprint spectrum'' can identify sources
of electromagnetic interference (EMI)
and definitively confirm which solutions do
and do not mitigate the EMI.

The protocol introduced here
should apply to a wide range of instruments, and
should prove especially useful in cases
subject to design constraints that require
integration with multiple other modules that are
not dedicated to NMR but that control other
forms of spectroscopy or other crucial aspects
of the measurement.
This protocol demonstrates firstly that
quantitative characterization of the noise
spectral density allows the identification of
successful noise mitigation techniques and
secondly that the absolute signal can be
accurately estimated from the theory,
allowing a systematic approach to instrument
design and optimization.
For the specific X-band ODNP design
demonstrated here
(and utilized in other laboratories),
the theory then
identifies the inefficient distribution of
fields in the hairpin loop probe as the main
remaining bottleneck for the improvement of
low-field, low-volume ODNP SNR.

\end{abstract}

\newcommand{\positionimage}[1]{\hspace*{-12pt}\raisebox{-\height}{#1}}
\newcommand{\figResponseZoomed}{%
    \begin{figure}
        \centering
        \subfig{fig:ResponseZoomedFour}{FourK}
        \subfig{fig:ResponseZoomedFourtyK}{FourtyK}
        \subfig{fig:ResponseZoomedTwoHK}{TwoHK}
        \subfig{fig:ResponseZoomedOneM}{OneM}
        \subfig{fig:ResponseZoomedFourM}{FourM}
        \subfig{fig:ResponseZoomedTenM}{TenM}
        \begin{tabular}{cccc}
            \refFourK & \hspace*{-12pt}\raisebox{-\height}{\includegraphics[width=0.45\linewidth]{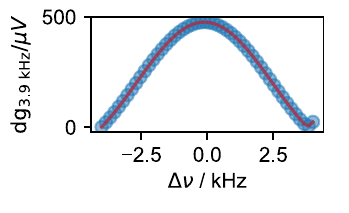}}
            &
            \refFourtyK & \hspace*{-12pt}\raisebox{-\height}{\includegraphics[width=0.45\linewidth]{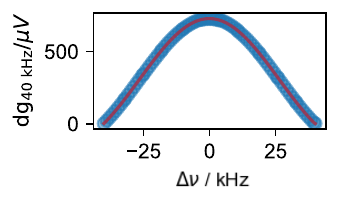}}
            \\
            \refTwoHK & \hspace*{-12pt}\raisebox{-\height}{\includegraphics[width=0.45\linewidth]{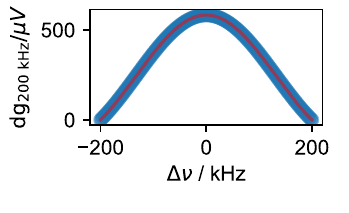}}
            &
            \refOneM & \hspace*{-12pt}\raisebox{-\height}{\includegraphics[width=0.45\linewidth]{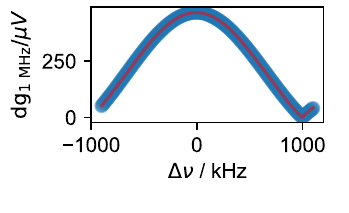}}
            \\
            \refFourM & \hspace*{-12pt}\raisebox{-\height}{\includegraphics[width=0.45\linewidth]{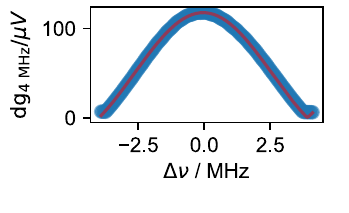}}
            &
            \refTenM & \positionimage{\includegraphics[width=0.45\linewidth]{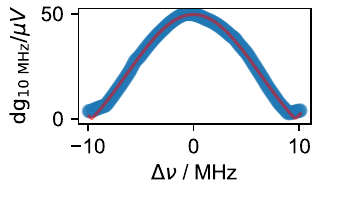}} 
        \end{tabular}
        \caption{An AFG
            outputs test signal at a range of frequencies ($\Delta \nu_{RF} = 2 \times SW$),
            which the SpinCore receiver board subsequently digitizes
            with a 14.9 MHz carrier frequency. 
            The maximum of the resulting PSD for each output
            frequency is plotted as a function of the offset
            and divided by the actual value of the test
            signal (which is approximately flat -- \cref{fig:AFG_output})
            and is plotted as a function of the offset from 14.9~MHz. 
            The transceiver response is plotted for \refFourK{}
            3.9 kHz, \refFourtyK{} 40 kHz, \refTwoHK{} 200 kHz, \refOneM{} 1 MHz,
            \refFourM{} 4 MHz, and \refTenM{} 10 MHz.}\label{fig:ResponseZoomed}
    \end{figure}}

\newcommand{\figPredictZoom}{%
    \begin{figure}
        \centering
        \subfig{fig:predictZoomOneM}{OneM}
        \subfig{fig:predictZoomTwoHK}{TwoHK}
        \subfig{fig:periodicPeaks}{PeriodicPeaks}
        \begin{tabular}{cc}
            \refOneM & \positionimage{\includegraphics[width=0.9\linewidth]{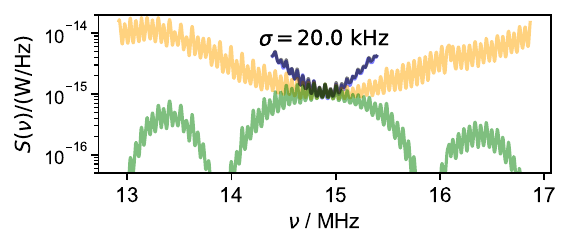}}
            \\
            \refTwoHK & \positionimage{\includegraphics[width=0.9\linewidth]{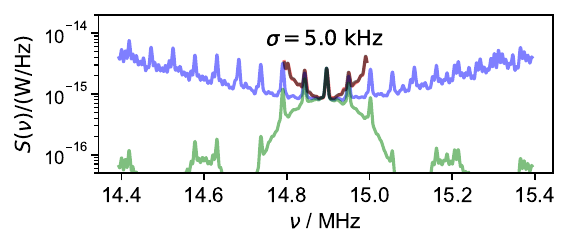}}
            \\
            \refPeriodicPeaks & \positionimage{\includegraphics[width=0.9\linewidth]{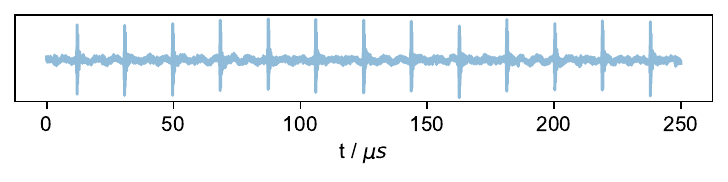}}
        \end{tabular}
        \caption{\refOneM{} The measured noise PSD for the 
            probe inserted between the magnets without chokes,
            acquired with
            SW = 4~MHz is divided by
            the square of the 4~MHz response function
            (see \cref{fig:ResponseZoomed})
            to yield the gold line.
            The product of the gold line
            with the square of the 1~MHz response function
            yields the green line. 
            Subsequent
            down-sampling in the time domain (see text)
            then results in a prediction
            (transparent grey/black line)
            that matches well with the measured 1~MHz PSD (blue). 
            Note that the transparent black prediction
            line overlaps the measured line
            almost exactly, making it appear
            a dark blue.
            \refTwoHK{}:
            The same
            1~MHz PSD data shown in \refOneM 
            (indicating the PSD has already been 
                divided by the square of the 1~MHz response
            function)
            convolved by a 5~kHz filter width
            (blue) is multiplied
            by the response function for a SW = 200~kHz (green).
            Downsampling of the green line results in 
            a PSD
            (transparent grey/black)
            that accurately predicts
            the actual 200 kHz PSD (red).
            \refPeriodicPeaks{}:
            Raw capture from the oscilloscope
            (acquired without
            noise mitigation)
            with spikes that repeat every
            $\sim 20\;\text{μs}=1/50\;\text{MHz}$.
        }\label{fig:predictZoom}
    \end{figure}}

\newcommand{\figPredictFourK}{%
\begin{figure}
    \centering
    \includegraphics[width=\linewidth]{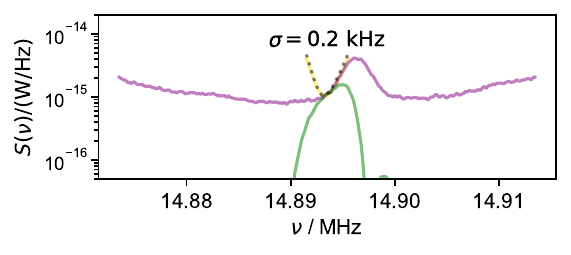}
    \caption{The PSD acquired on the transceiver board using
        SW = 40 kHz is divided by a sinc
        function appropriate for a spectral width of 40 kHz
        (purple) and subsequently multiplied by a narrower
        sinc function appropriate for a dataset with
        a spectral width of 3.9 kHz to yield a filtered
        noise PSD (green). 
        The downsampling of this product produces the 
        predicted noise PSD for a spectrum taken with a 
        spectral width of 3.9 kHz (dashed black line). 
        The actual acquired spectra (gold) having a 
        spectral width of 3.9 kHz shows the exact same PSD 
        as the predicted model. 
    }\label{fig:predict_4kHz}
\end{figure}
}

\newcommand{\figMagNoise}{%
    \begin{figure*}
        \centering
        \includegraphics[width=\linewidth]{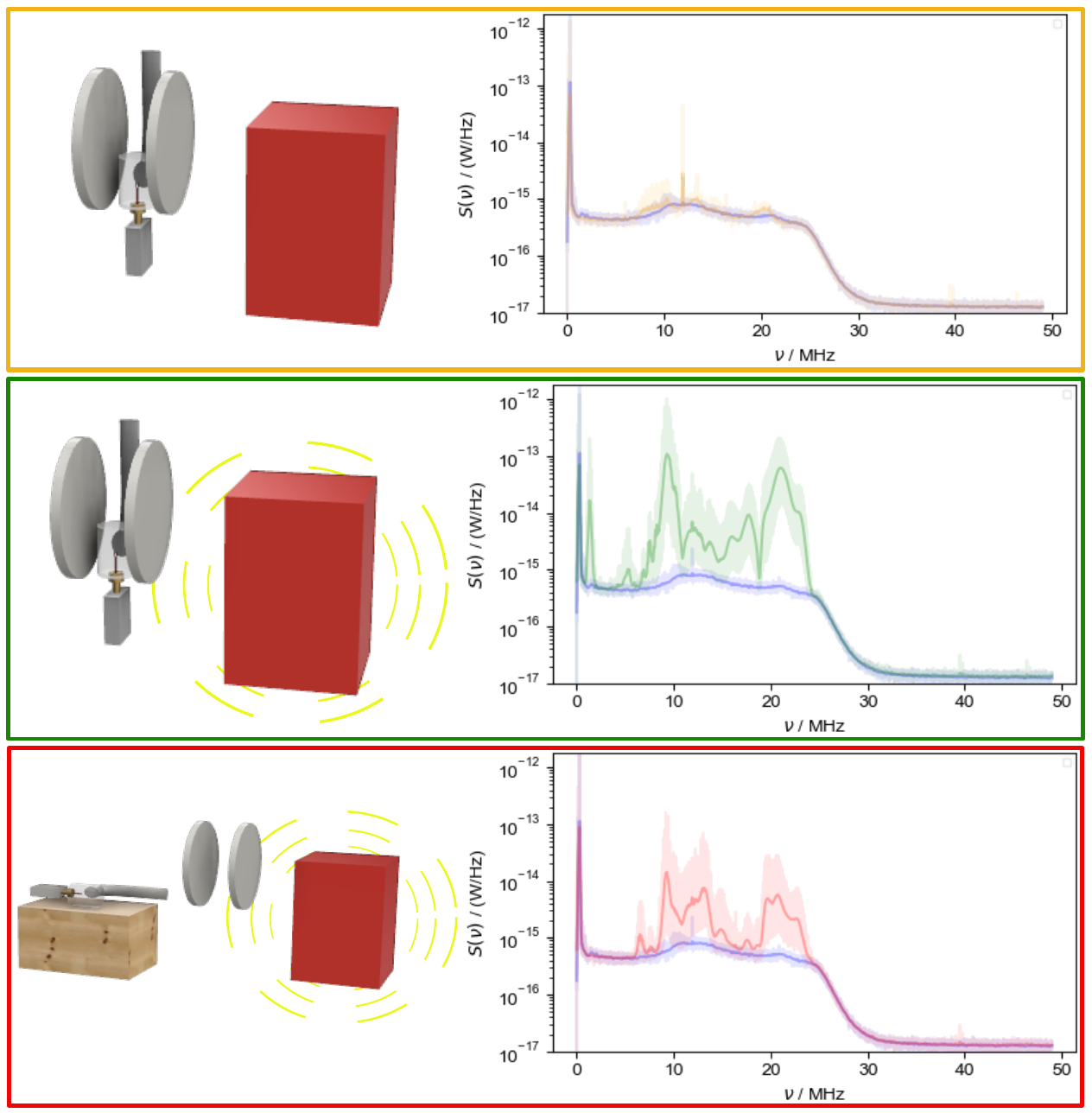}
        \caption{The noise PSD of the terminated 
            receiver chain (blue line) provides a baseline 
            reference for all plots.
            All other lines show the noise output by the
            receiver chain when connected to the NMR
            probe
            (grounded to the ESR cavity for gold and green).
            When the power
            supply is off (gold), the noise density closely
            approaches the noise levels of the
            terminated receiver chain (blue).
            Turning on the electromagnet power supply
            introduces dramatic levels of noise
            (green).
            Disconnecting the probe from the cavity,
            and placing it on a table
            5 ft from the
            power supply causes only subtle changes to
            the noise (red),
            as discussed in the text.
        }\label{fig:MagNoise}
    \end{figure*}
}

\newcommand{\figSincFilter}{%
    \begin{figure}
        \centering
        \subfig{fig:sincFilterTwoD}{TwoD}
        \subfig{fig:sincFilterOneD}{OneD}
        \begin{tabular}{cl}
            \refTwoD
            &
            \positionimage{\includegraphics[width=0.9\linewidth]{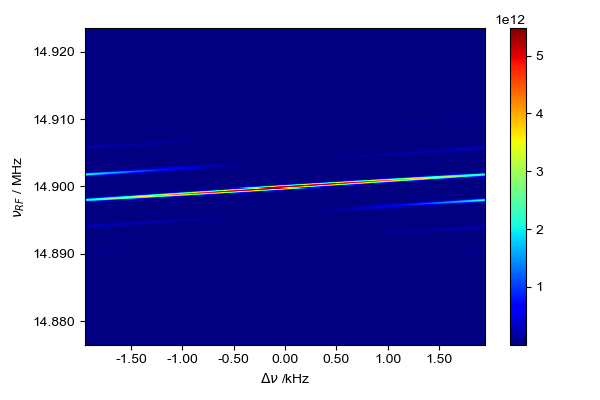}}
            \\
            \refOneD
            &
            \positionimage{\includegraphics[width=0.9\linewidth]{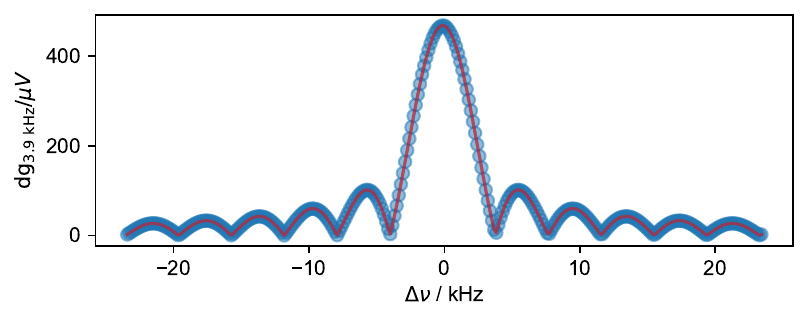}}
        \end{tabular}
            \caption{
             An AFG outputs varying frequencies of test signal (14.8766~MHz
             $\rightarrow$ 14.9234~MHz), which the SpinCore receiver board
             subsequently digitizes with a 14.9~MHz carrier frequency.
            \refTwoD{} A 2D plot shows how the intensity of the PSD 
            (a function of the offset, $\Delta \nu$, from the receiver
            carrier frequency)
            varies with the frequency of the test signal ($\nu_{RF}$).
            \refOneD{} The maximum of the PSD for each output
            frequency, acquired on the receiver, is divided by 
            the corresponding maximum of the PSD 
            acquired on the oscilloscope
            (which is approximately flat -- \cref{fig:AFG_output})
            and is plotted as a
            function of the offset from 14.9~MHz. 
            The red line shows the absolute value of a
            best-fit sinc function. 
        }\label{fig:sincFilter}
    \end{figure}}
\newcommand{\figRecipCirc}{%
    \begin{figure}
        \centering
        \subfig{fig:RecipCircLossless}{Lossless}
        \subfig{fig:RecipCircLossy}{Lossy}
        \begin{tabular}{cc}
            \refLossless & \positionimage{\includegraphics[width=0.7\linewidth]{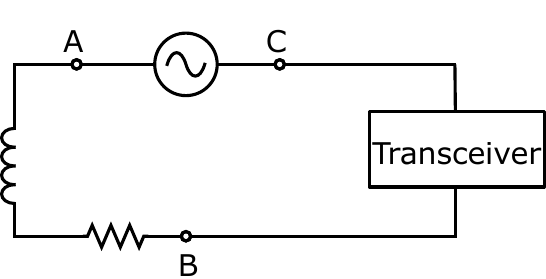}}
            \\
            \refLossy & \positionimage{\includegraphics[width=0.9\linewidth]{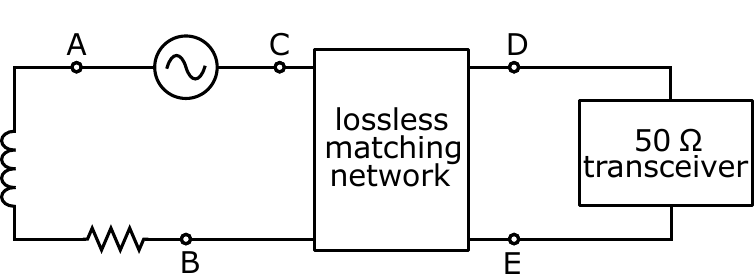}}
        \end{tabular}
        \caption{Example of a coil in 
            a(n): \refLossless{} ideal scenario where the coil is matched perfectly to a detector
            with no matching network required. The voltage loss between points A and B (across the coil 
            components) is equal to the voltage loss between points C and B on the side of the detector.
            \refLossy{} circuit that contains a lossless matching network between the coil and the
            detector, transforming the impedance of the coil to match the transceiver (50 Ω). In this case,
            the voltage drop from A to B is equal to the voltage loss from D to E.
        }\label{fig:RecipCirc}
    \end{figure}
}
\newcommand{\figGDSvSC}{%
\begin{figure}
    \centering
    \subfig{fig:GDSvsSCnofilt}{nofilt}
    \subfig{fig:GDSvsSChighpass}{highpass}
    \begin{tabular}{cc}
        \refnofilt &
        \positionimage{\includegraphics[width=\linewidth]{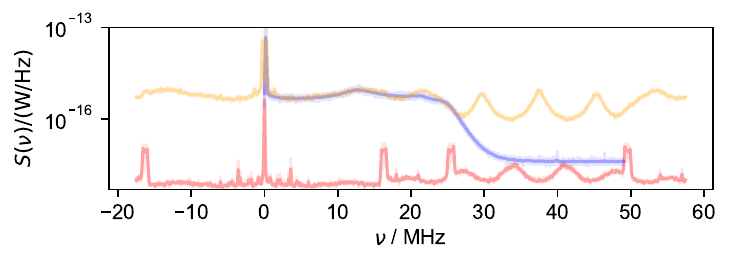}}
        \\
        \refhighpass &
        \positionimage{\includegraphics[width=\linewidth]{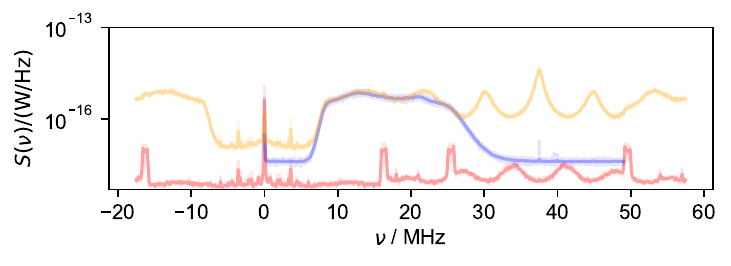}}
    \end{tabular}
    \caption{\refnofilt: The PSD of the terminated receiving chain is acquired on
        the oscilloscope (blue) as well as on the transceiver board
        with a carrier frequency of 20 MHz and a spectral width of
        75 MHz (gold).
        The noise PSD of the terminated SpinCore
        (red) shows the intrinsic noise of the transceiver board
        itself.
        \refhighpass: A high pass filter is inserted into the
        receiver chain after the
        low pass filter and the PSD is acquired again both on the 
        oscilloscope and the transceiver board (plotted with the same
        color scheme as \refnofilt{}).
    }\label{fig:GDSvSC}
\end{figure}
}

\newcommand{\figNutation}{%
\begin{figure}
    \centering
    \subfig{fig:NutationDCCT}{DCCT}
    \subfig{fig:NutationOneD}{OneD}
    \subfig{fig:NutationPulseShape}{PulseShape}
    \begin{tabular}{cc}
        \refDCCT &
        \positionimage{\includegraphics[width=0.9\linewidth]{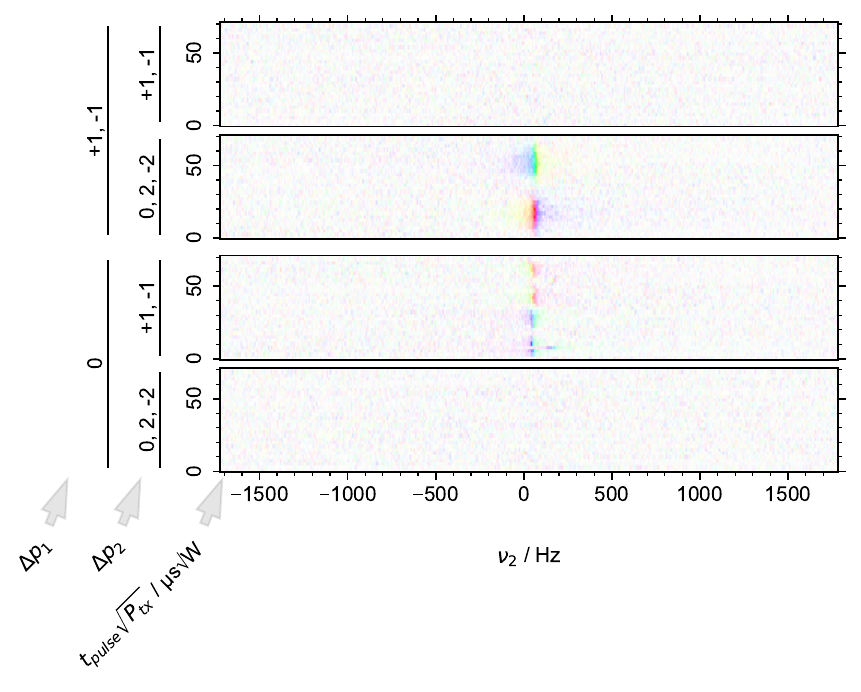}}
        \\
        \refOneD &
        \positionimage{\includegraphics[width=0.9\linewidth]{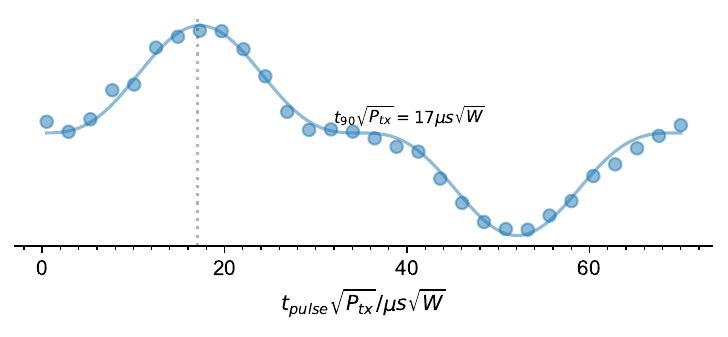}}
        \\
        \refPulseShape &
        \positionimage{\includegraphics[width=0.9\linewidth]{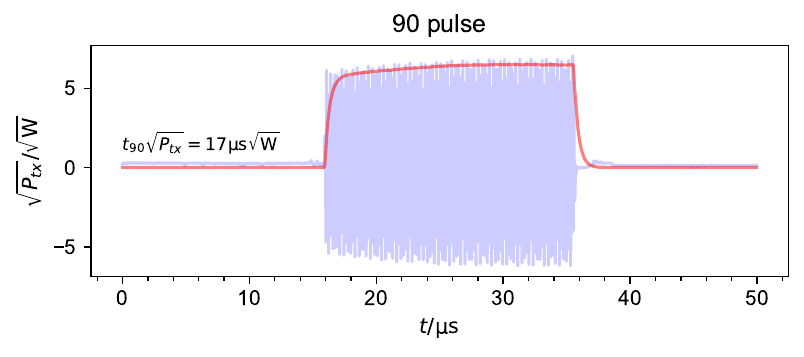}}
    \end{tabular}
    \caption{\refDCCT: The domain colored coherence transfer pathway (DCCT) of the 
        nutation experiment performed by varying the output
        $t_{90}\sqrt{P_{tx}}$ using a two step phase cycle on the $90^{\circ}$
        pulse and a two step phase cycle on the $180^{\circ}$ pulse. 
        The color change in \(\Delta p_1 = +1, \Delta p_2 = -2\) indicates a rise 
        in signal (yellow/red) followed by a null and finally an inversion to blue 
        at longer pulse times. A color wheel is depicted for reference of phase. 
        \refOneD: The data is integrated to extrapolate the first maximum, 
        indicative of a proper 90° tip at
        $17\;\text{μs}\sqrt{\text{W}}$. 
        \refPulseShape: Capture of the $90^{\circ}$ pulse on the oscilloscope.
        The analytic signal (blue) is frequency filtered using a lorentzian
        whose full width half-maximum is determined by measurement of $S_{1,1}$ on
        a network analyzer (red). Integration of
        the filtered pulse shows that the actual integral of the pulse
        ($t_{90}\sqrt{P_{tx}}$) as seen on the oscilloscope.
    }\label{fig:Nutation}
\end{figure}
}

\newcommand{\figPredictPSD}{%
\begin{figure}
    \centering
    \includegraphics[width=\linewidth]{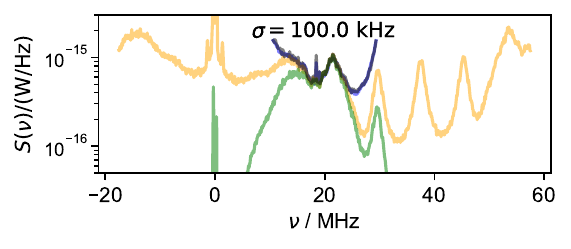}
    \caption{\cref{eq:convPSD} converts the digitized captures
        of the noise for the terminated receiver chain
        acquired on the receiver with the base sampling rate of 
        75 MHz, to a PSD (gold).
        Multiplication of the gold line
        by a sinc function appropriate for a dataset with
        a spectral width of 20 MHz yields a filtered
        noise PSD (green).
        Subsequent downsampling produces the 
        predicted noise PSD for a spectrum taken with a 
        spectral width of 20 MHz (transparent
        grey/black).
        This prediction
        matches the PSD measured at 20~MHz
        (blue);
        note that the transparent black prediction
        line overlaps the measured line
        almost exactly, making it appear
        a dark blue.
        }
        \label{fig:predictPSD}
\end{figure}
}

\newcommand{\figPSDwGain}{%
\begin{figure} \centering \includegraphics[width=\linewidth]{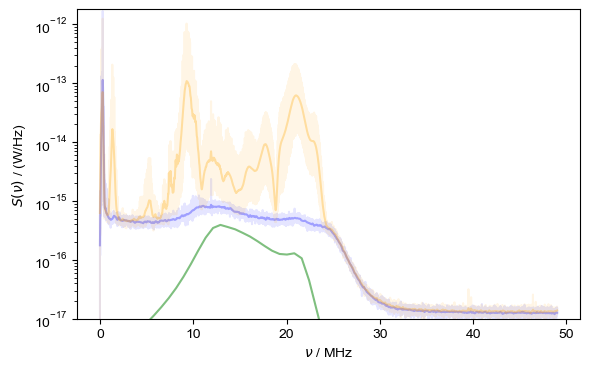} \caption{The noise PSD of the terminated receiver chain (blue)
rises slightly above the thermal noise limit (green)
over the sensitive range of the duplexer (14-15 MHz).
The shape of the thermal noise limit
is controlled by the gain of the receiver
chain (\cref{fig:Gain}), which is non-uniform here.
When the NMR probe is attached to the input of the receiver chain (gold),
very strong EMI is introduced.}\label{fig:PSDwGain}
\end{figure} 
}

\newcommand{\refHighResChokes}{%
\begin{figure}
    \centering
    \includegraphics[width=\linewidth]{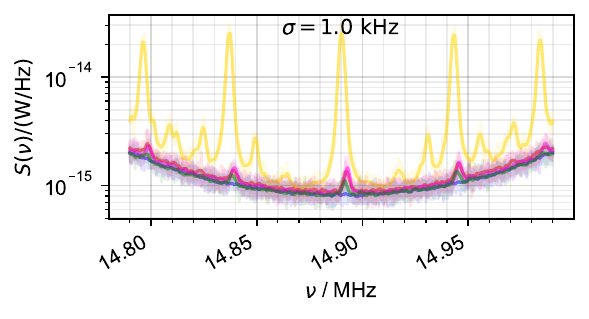}
    \caption{The noise PSD was acquired at a spectral width of 200 kHz
        and convolved using a convolution width of 1.0 kHz. The blue line
        illustrates the Johnson noise of the terminated receiver chain.
        Similar to fig.~\ref{fig:Chokes}, gold shows the PSD for the probe inserted
        into the cavity with the magnet on, and it is clear that magnitudes
        of noise are introduced. When 12 chokes are added (red) the noise
        is halved. The addition of 27 chokes (magenta) results
        in a very minimal decrease in noise compared to the 12 chokes.
        Replacing the chokes with a toroid (green) results in the largest
        reduction of noise. 
    }\label{fig:200kHz_chokes}
\end{figure}
}

\newcommand{\figChokesFID}{%
\begin{figure}
    \centering
    \subfig{fig:okayNoisePSD}{OKPSD}
    \subfig{fig:okayNoiseFID}{OKsignal}
    \begin{tabular}{cc}
        \refOKPSD &
        \positionimage{\includegraphics[width=0.9\linewidth]{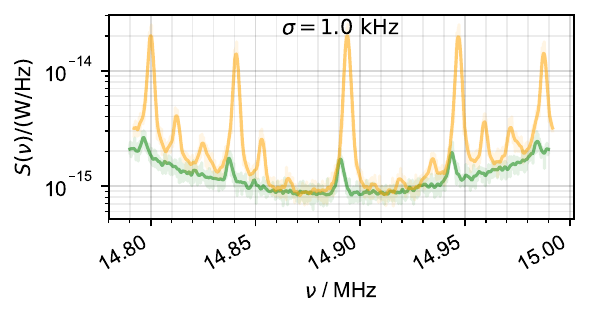}}
        \\
        \refOKsignal &
        \positionimage{\includegraphics[width=0.9\linewidth]{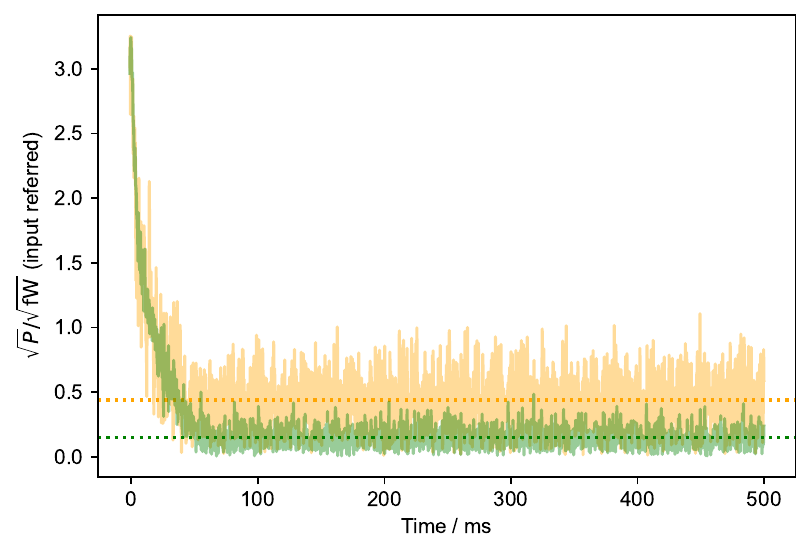}}
    \end{tabular}
    \caption{
        \refOKPSD: The noise PSD acquired with a 200kHz SW, without any form
        of mitigation (gold) rises magnitudes above the noise PSD acquired at 
        a SW of 200 kHz with the toroid in place and ideal positioning of coax.\@
        cabling (green).
        \refOKsignal: When an FID is acquired on resonance with the center spike in
        \refOKPSD, the signal level remains consistent for both 
        with and without mitigation techniques. However, the average
        noise power without the toroid (gold) is 0.44 $\mathrm{\sqrt{fW}}$ (input
        referred) illustrated by the dashed gold line while the average
        noise power with the toroid in place (green) is only 0.15 $\mathrm{\sqrt{fW}}$ 
        (input referred) shown by the dashed green line.
}
\label{fig:ChokesFID}
\end{figure}
}

\newcommand{\figBadFID}{%
\begin{figure}
    \centering
    \subfig{fig:BadNoisePSD}{BadPSD}
    \subfig{fig:BadNoiseFID}{Badsignal}
    \begin{tabular}{cc}
        \refBadPSD &
        \positionimage{\includegraphics[width=0.9\linewidth]{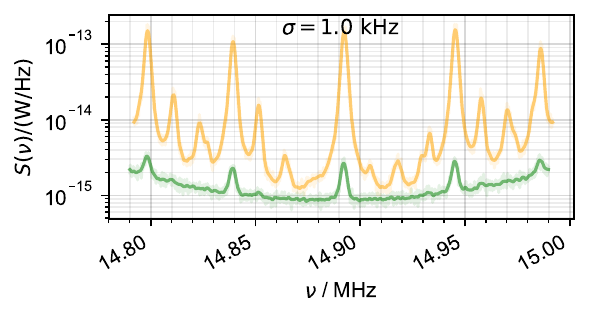}}
        \\
        \refBadsignal &
        \positionimage{\includegraphics[width=0.9\linewidth]{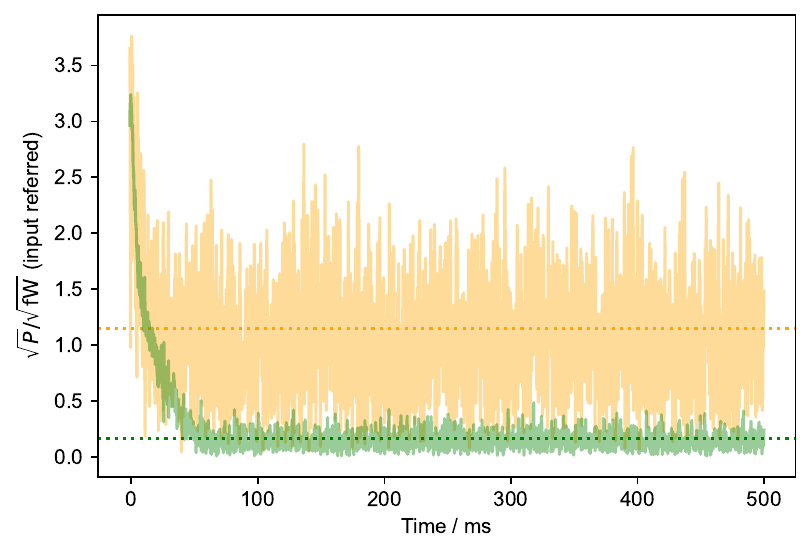}}
    \end{tabular}
    \caption{\refBadPSD: The noise PSD acquired with a SW of 200 kHz, without
        any form of mitigation (gold) rises magnitudes above the noise PSD
        acquired when optimal cable positioning and a ferrite toroid is in
        place (green). 
        \refBadsignal: When the signal is acquired on the central spike,
        with no form of mitigation techniques (gold) the average power of the
        noise is 1.15 $\mathrm{\sqrt{fW}}$ (input referred, gold dashed line)
        obfuscating the 3.3 $\mathrm{\sqrt{fW}}$ of signal. 
        When the toroid and optimal positioning of BNC
        cabling is used (green) and again the signal is acquired on the central
        spike, the noise level is reduced to 0.16 $\mathrm{\sqrt{fW}}$ (input
        referred, green dashed line) and the signal is clearly defined.
}
\label{fig:BadFID}
\end{figure}
}

\newcommand{\figVariedTau}{%
\begin{figure}
        \centering
        \includegraphics[width=\linewidth]{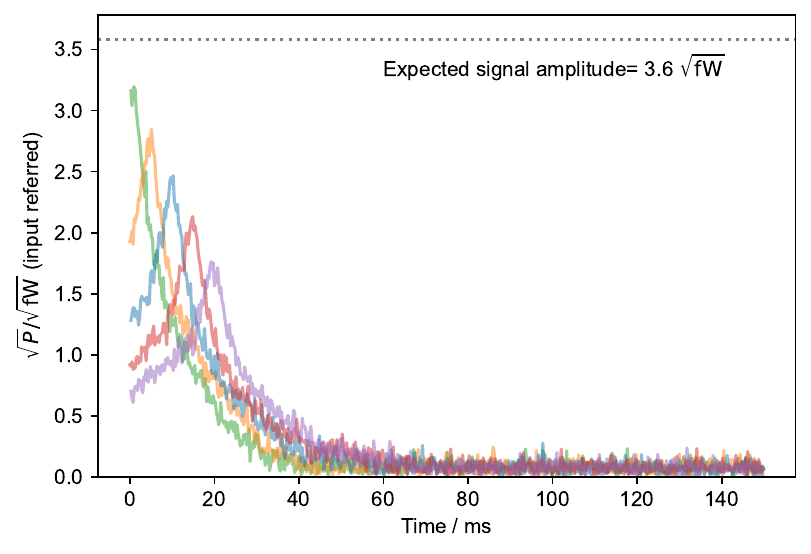}
        \caption{An experiment with varied
            echo length performed 
            on the same 27 mM TEMPOL sample.
            Signal comes from an echo
            sequence
            comprising a $\pi/2$ pulse,
            a delay $\tau$ that is varied for
            the differently colored signals,
            and a $\pi$ pulse.
            A dashed black horizontal line indicates the 
            predicted signal amplitude of 3.6 $\mathrm{\sqrt{fW}}$
            (input referred). }
\label{fig:VariedTau}
\end{figure}
}
\newcommand{\figBiot}{%
\begin{figure}
    \centering
    \subfig{fig:BiotActual}{Actual}
    \subfig{fig:BiotDiff}{Diff}
    \begin{tabular}{cc}
        \refActual &
        \positionimage{\includegraphics[width=0.9\linewidth]{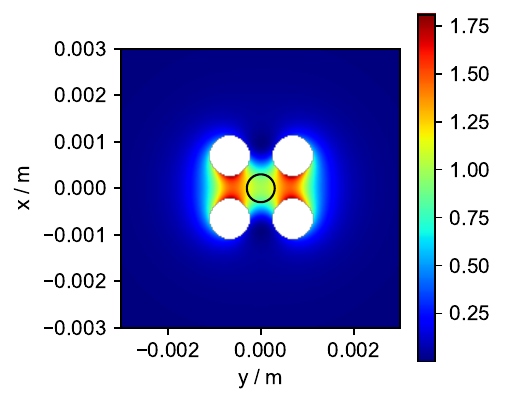}}
        \\
        \refDiff &
        \positionimage{\includegraphics[width=0.9\linewidth]{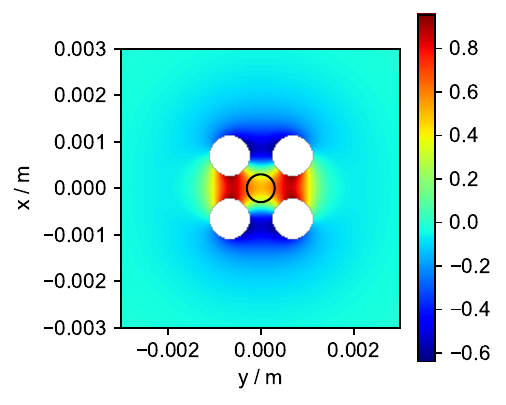}}
    \end{tabular}
    \caption{A very simple simulation assuming infinite wires perpendicular to
        the page that generate a field of magnitude $\mu_0 I/2 \pi r$
        (where $r$ is the distance to the wire)
        and direction determined by Biot-Savart cross-product.
        The white circles indicate the location of the wires.
        \refActual: The magnetic field energy density normalized against
        the magnetic energy density due to the desired field
        (\ie the plot shows $|B|^2_{1}/4 B^2_{1,avg}$,
            where $2 B_{1,avg}$ is the field in the lab frame).
        The black circle indicates the sample region
        (and the $y$-component of the field within the circle
        is averaged to determine $2 B_{1,avg}$).
        \refDiff: The same calculation, subtracted from the fields
        generated by the four wires, one a time.
    }
    \label{fig:Biot}
\end{figure}
}

\newcommand{\figOscDig}{%
\begin{figure}
    \centering
    \subfig{fig:osc_mV}{mVvariation}
    \subfig{fig:osc_dig}{oscDig}
    \begin{tabular}{cc}
        \refmVvariation &
        \positionimage{\includegraphics[width=0.9\linewidth]{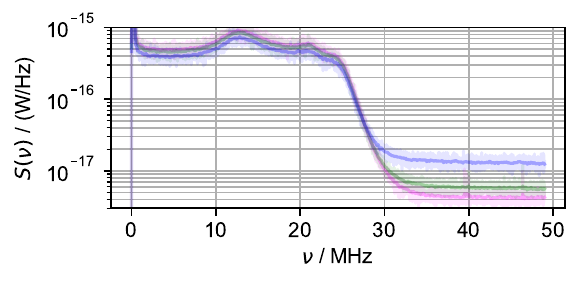}}
        \\
        \refoscDig &
        \positionimage{\includegraphics[width=0.9\linewidth]{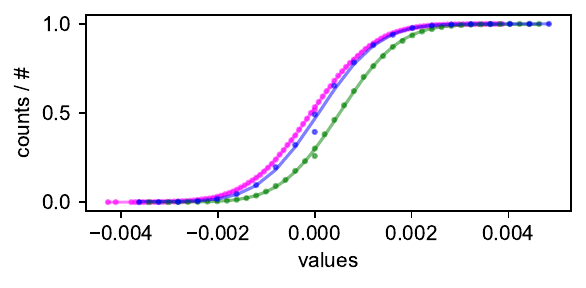}}
    \end{tabular}
    \caption{\refmVvariation: PSD of the terminated receiver chain acquired 
        on the oscilloscope with a voltscale of 2 (magenta), 
        5 (green) and 10 (blue) mV. Here the y axis of the plot 
        is zoomed to emphasize the difference in PSD. 
        \refoscDig: The normalized gaussian cumulant of each measurement fit
        with a spline, plotted using the same colors as A. 
}
\label{fig:oscDigitization}
\end{figure}
}

\newcommand{\figCavityProbe}{%
    \begin{figure}
        \centering
        \includegraphics[width=\linewidth]{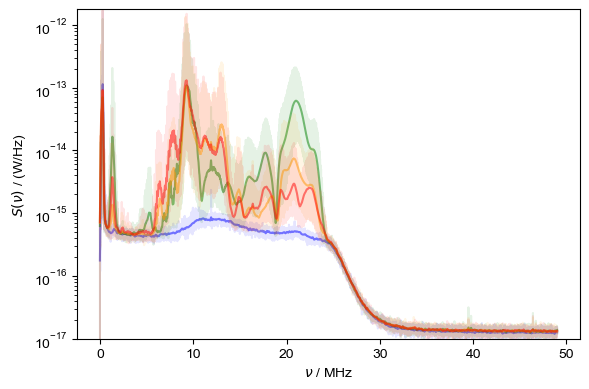}
        \caption{In the standard configuration
            (green -- same as the configuration for green in
            \cref{fig:MagNoise}),
            large and distinctive clusters of peaks
            appear.
            By
            detaching the waveguide/bridge connection (gold) the noise
            at frequencies higher than 13 MHz drop an order of
            magnitude, but the noise density at some lower frequencies
            slightly increases.
            Entirely detaching the NMR probe from the cavity (red) and
            holding it between the magnets does not yield
            dramatic changes in the noise density.
            (Terminated receiver chain noise shown in blue for
            reference.)
        }\label{fig:cavityProbe}
    \end{figure}
}
\newcommand{\figChokes}{%
\begin{figure}
    \centering
    \includegraphics[width=\linewidth]{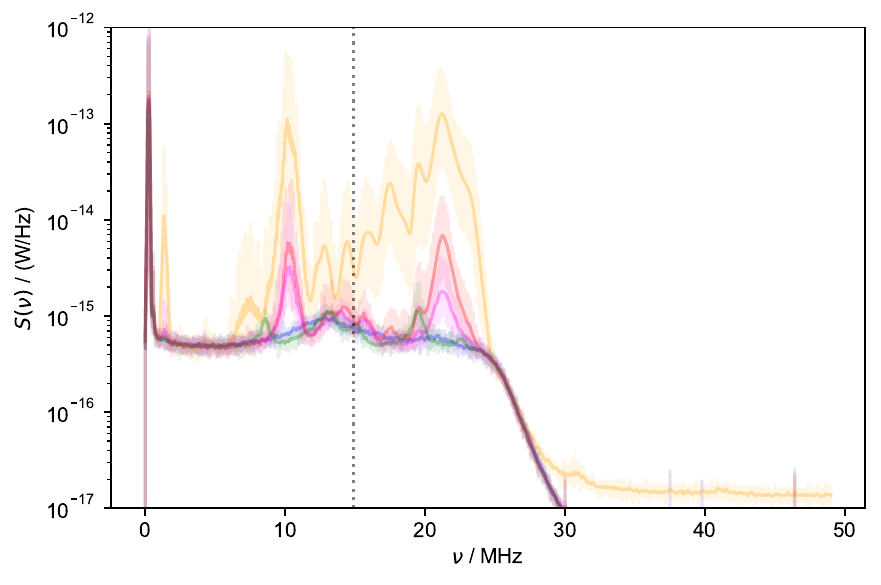}
    \caption{Common-mode chokes mitigate the EMI,
        as illustrated by the noise PSD from the
        standard configuration with 0 (gold -- equivalent to the setup shown
            in green
            of \cref{fig:MagNoise}),
        12
        (red) and 26 (magenta) ferrite snap-on chokes added
        to the coaxial cable connecting the probe to
        the receiver chain. Replacing the chokes with a simple
        ferrite toroid (green) results in further mitigation of 
        noise.
        (Baseline noise of terminated receiver chain shown in blue.)
    }\label{fig:Chokes}
\end{figure}
}
\newcommand{\figCalcGain}{%
\begin{figure}
    \centering 
    \subfig{fig:AFG_output}{AFGoutput}
    \subfig{fig:RX_output}{RXoutput}
    \subfig{fig:RX_gain}{RXgain}
    \begin{tabular}{cc}
        \refAFGoutput &
        \positionimage{\includegraphics[width=0.9\linewidth]{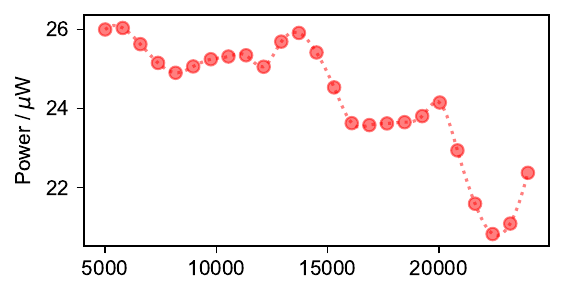}}
        \\
        \refRXoutput &
        \positionimage{\includegraphics[width=0.9\linewidth]{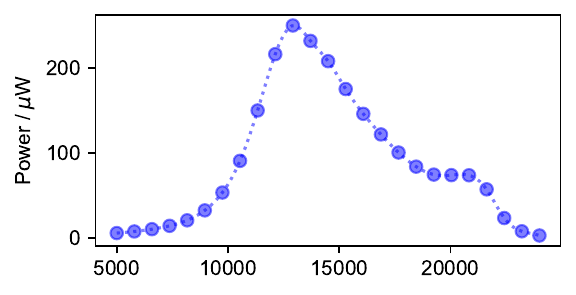}}
        \\
        \refRXgain &
        \positionimage{\includegraphics[width=0.9\linewidth]{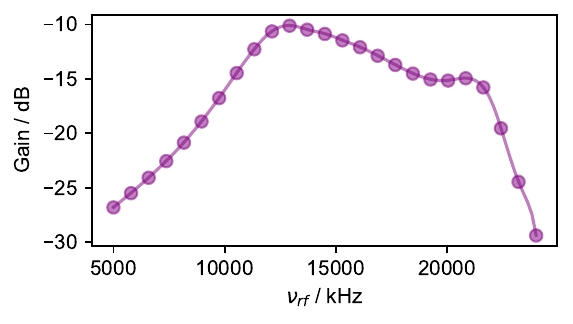}}
    \end{tabular}
    \caption{\refAFGoutput: The AFG outputs test signal output directly to the
        oscilloscope and the acquired power as a function of frequencies is 
        fit to a cubic spline. \refRXoutput: The test signal from \refAFGoutput
        is injected into a 40.021 dB attenuator assembly prior to the input of the 
        receiver chain. The oscilloscope captures the amplified test signal at the 
        output of the receiver chain and is plotted as a function of the frequencies. 
        \refRXgain: Gain in dB calculated with the attenuator taken into consideration, 
    as a function of the input frequency.}
    \label{fig:Gain}
\end{figure}
}

\ifarxiv\begin{bibunit}\fi
\maketitle
\ifarxiv\glsresetall\fi
\section{Introduction}
New physical techniques and modern technologies
are reinvigorating the magnetic resonance
community.
This is especially true of low-field
NMR.
For example, ODNP performs
detailed analyses of hydration dynamics on the
surfaces and interiors of samples with intricate
structure at the nanoscale: ranging from proteins
to porous
materials~\cite{Franck2019OveDynNuc,Berkow2023CoaFunAct,Ortony2013AsyColBio,Armstrong2009OveDynNuc,Beaton2022ModVieCoh,Franck2013QuaCwOve,Franck2013NonScaSur,Dunleavy2023ChaDisPat,Saun2020NanWatDyn,Doll2012LiqStaDNP}. 
ODNP extends to applications outside of hydration dynamics 
as well including resolving $^1$H chemical shift differences~\cite{Uberruck2020ComXBaODN,Keller2020HigOveDyn},
studying structural transitions~\cite{chaubeyAssessmentRole2Trifluoroethanol2020}, 
assisting in ``diagonal'' suppressed homonuclear 2D correlation~\cite{georgeChemicalShiftResolved19FNMR2014}, 
studying the performance of potential polarizing agents~\cite{Perras2022MetDriOveDyn}, 
and interrogating the time course
of nuclear spin polarization buildup~\cite{Dey2019UnuOveDyn}.
However, it imposes specific requirements on the
experimental setup.
In particular, while integration with a standard
cw ESR spectrometer offers clear advantages,
it is reasonably well recognized that
standard ESR instrumentation introduces
a sea of noise that renders the isolation and
optimization of signal increasingly difficult.
ODNP thus serves as a good example of
what can be termed
``NMR under adverse
circumstances.''
It serves as a case study for a field that includes
a diverse set of applications:
single-sided NMR for materials studies~\cite{Marble2007ComPerMag},
NMR in the presence of electromagnetic noise~\cite{Kremer2022RevAcqSiga},
NMR in oil bore wells~\cite{Kleinberg2001NMRWelLog,Jin2024NewInsDat},
NMR techniques in hydrogeological investigations~\cite{Lin2019RanNoiSup,Lin2018EnaSurNuc,Kremer2022RevAcqSiga}, 
and portable NMR~\cite{Blumich2021WheMOULea}.

While the realm of low-field NMR, generally,
and even X-band ODNP specifically,
has become quite crowded
with the development of an ever-increasing array
of designs to resolve this issue, 
the authors are aware of few clear and
robust protocols to aid in the development of
these designs.
This situation is exacerbated by the fact that
actions that solve noise issues on one
system
(\emph{e.g.} for ODNP), such as adding hardware to ground
transmission lines to the waveguide~\cite{Kaminker2015ChaSixOve,Franck2019OveDynNuc,Franck2013QuaCwOve},
changing the model of the commercial
electromagnet power supply~\cite{Teucher2023SpeOrtSpi},
deriving magnetic
field from permanent
magnets~\cite{Morin2024LowCerMag,Marble2007ComPerMag,Uberruck2020ComXBaODN,Keller2020HigOveDyn},
or developing customized
hardware~\cite{Uberruck2020ComXBaODN,Doll2019PulConMag,Keller2020HigOveDyn}
do not yield the same success in different
laboratory environments.
Practically, even the best new hardware designs in
the literature might not be compatible with the
requirements, capabilities, or environment of
a different laboratory.

Similar problems have been seen in other techniques such
as atomic force microscopy.
One previously presented solution~\cite{Hoepker2011DieFluPol}
that is treated as a prime method for noise
analysis~\cite{Fitzpatrick2018Cha14Noi},
utilizes
the power spectral density to isolate environmental noise
from the frequency noise induced by molecular motions. 
This technique
facilitated drastic decreases
to the sample-related force 
noise and frequency noise in magnetic resonance force
microscopy~\cite{boucherNonPerturbativeLowNoiseSurface2023}.

Meanwhile, the optimization of signal to noise in the
NMR community can often be treated as an art.
A variety of strategies are available,
including utilization of pre-polarization pulses~\cite{Lin2018EnaSurNuc,Hiller2020UtiPreEnh},
incorporation of better
shielding~\cite{Chung2020MatEleInt,boucherNonPerturbativeLowNoiseSurface2023,Radeka1998ShiGroLar,Sankaran2018RecAdvEle,Andris2016NoiIntMea,Morin2024LowCerMag},
active shielding of low frequency noise~\cite{Yang2022ActEMISup},
improved grounding~\cite{Radeka1998ShiGroLar},
additional shim coils~\cite{Keller2020HigOveDyn},
and improved amplifier noise figures~\cite{Andris2019AnaNMRSpe,Andris2014NoiMeaPre,Ardenkjaer-Larsen2015FacOveSen},
while the post-processing stage of data analysis
is also implemented to improve the SNR~\cite{Lin2019RanNoiSup,Goryawala2020EffApoSmo,Ebel2006EffZerApo,Traficante2000OptWinFun,Ernst2004AppFouTra,Cavanagh2007CHAEXPASP}.
The fundamental science governing the
SNR is well established,
but is frequently presented with design-specific details
that distract from the essential clarity that
can be synthesized from a couple seminal
papers
(Hoult~\cite{Hoult1976SigRatNuc} and
Mims~\cite{Mims1972EleSpiEch}).
As a result, the NMR community requires not just new
instrumental schematics, but also (as provided
here) a simple, flexible, and fundamentals-driven protocol for
both systematically diagnosing and mitigating the
sources of noise and for identifying the elements
of the design that limit the absolute signal
intensity.

This paper begins with a theory section (\cref{sec:theory})
then the results are arranged in the order of the
recommended protocol:
First, common off-the-shelf equipment, especially when
coupled with open-source software, can easily
characterize the noise over a broad bandwidth
(\cref{sec:FirstPSD}).
Digitization of the noise power spectral density
identifies the relative contribution of various
noise sources in the laboratory according to their
respective spectral fingerprints (\cref{sec:EMInoise}).
Noise mitigation efforts can then be reproducibly
tested until the noise approaches the thermal
noise limit (\cref{sec:chokes}).
Characterization of the receiver response
then leads to quantitative agreement between the
noise measured on the receiver \vs
with standard off-the-shelf equipment (\cref{sec:wideSC}).
In particular, this allows identification and
correction for the idiosyncrasies of the receiver,
and avoidance of extra noise introduced by the
receiver (\cref{sec:SCresponse}).
Finally, the adapted theory can
make a very accurate prediction of the signal that
will be observed in the customized system,
and can also identify the most promising next
steps (\cref{sec:predictSignal}).

\section{Theory}\label{sec:theory}
This section first presents equations for NMR signal intensity
based on relatively few assumptions.
In particular, it emphasizes that a great deal
of conclusions can be derived exactly if two principles are combined:
(1) the concept of reciprocity (from Hoult~\cite{Hoult1976SigRatNuc},
\cref{sec:recip}) 
and (2) the relationship between the stored energy and the definition of the
$Q$-factor (from Mims~\cite{Mims1972EleSpiEch}, \cref{sec:etap}).
These concepts integrate most naturally
with the aid of a definition from the ESR literature: the
conversion factor ($\Lambda$, \cref{sec:theconversionfactor}).
The theory also reviews the concept
of Johnson-Nyquist noise (\cref{sec:JohnsonNoise}).
Finally, because part of the results emphasize the
benefit of characterizing the power spectral
density of the noise, the theory section provides equations to
generate such plots from data acquired on
a standard digital capture oscilloscope
(\cref{sec:NoiseProc}).

\subsection{The Conversion Factor}\label{sec:theconversionfactor}
The term the ESR literature denotes as the
``conversion factor'' (i.e., ``efficiency
parameter'')~\cite{Mett2008DieMicRes,Rinard1994RelBenOve,Neudert2016ComXBaRes,Annino2009HigDouStr},
$\Lambda$, with units $[\text{T}/\sqrt{\text{W}}]$, gives the ratio of 
the rf magnetic field to the
pulse power:
\begin{equation}
    \Lambda = \frac{B_{1,avg}}{\sqrt{P_{tx}}}
    \label{eq:LambdaB}
\end{equation}
where $P_{tx}$ ($[\text{W}]$) is the average
transmitted power required to achieve a rotating frame
magnetic field of $B_{1,avg}$
(where $B_{1,avg}$
$[\text{T}]$ is averaged over the sample)~\cite{Neudert2016ComXBaRes,Rinard1994RelBenOve,Annino2009HigDouStr}. 
By applying the simple relationship, $B_{1,avg} =
\frac{\theta}{\gamma t_p}$ where $\theta$ is the tip angle
$[\text{rad}]$, $\gamma$ is the gyromagnetic ratio 
$[\text{rad} \cdot \text{s}^{-1} \cdot \text{T}^{-1}]$, 
and $t_p$ $[\text{s}]$ is the pulse length required to achieve $\theta$,
one sees that the value of $\Lambda$ can be determined
from a standard calibration of the $90^{\circ}$ pulse time,
namely:
\begin{equation}
    \Lambda = \frac{\pi}{2\gamma t_{90}\sqrt{P_{tx}}}
    \label{eq:Vsig90t}
\end{equation}
where $t_{90}\;[\text{s}]$ gives the time
it takes a pulse of average power
$P_{tx}$ to tip the spins from the $z$-axis
into the transverse plane ($\theta$ = $\frac{\pi}{2}$).

\subsection{Reciprocity in terms of Λ}\label{sec:recip}

With the assumption that the spins are uniformly excited,
the peak voltage of the signal ($\xi_{sig}$) generated 
during an echo with minimal relaxation can be determined
from
\begin{equation}
    \xi_{sig} = \left( \frac{B_{1,lab}}{I_{coil,p}} \right) \omega M_{0} V_{s},
    \label{eq:HoultV}
\end{equation}
where $\omega$ is the Larmor frequency
{[}$\text{rad}/\text{s}${]},
$V_s$ is the volume 
of the sample $[\text{m}^{3}]$,
and $B_{1,lab}$ is the
average magnetic field $[\text{T}]$ that is generated
perpendicular to the static field
when a current of $I_{coil,p}$ $[\text{A}]$ flows through
the coil;
note that the ratio $B_{1,lab}/I_{coil,p}$ appears in
\cref{eq:HoultV} as a consequence of the
principle of reciprocity
(derived in \cref{sec:derive_recip},
    consistent with
literature~\cite{Hoult1976SigRatNuc}).
\Cref{eq:HoultV} is equivalent to Eq.~4 of
Hoult's seminal paper on
reciprocity~\cite{Hoult1976SigRatNuc},
assuming $K=1$.
$M_{0}$ is the nuclear 
magnetization which,
in the absence of hyperpolarization,
is given by
\begin{equation}
    M_{0} = \frac{N\omega \gamma \hbar^{2}I(I+1)}{{3k_B T}},
    \label{eq:M0}
\end{equation}
where $N$ is the number of spins per unit
volume,
$I$ is the quantum
number of the spin (\emph{e.g.}, I = 1/2 for $^{1}$H, 
1 for $^{2}$H), and $T$ is the temperature $[\text{K}]$
of the sample.

Of course,
a lossless matching network attached to
the coil will alter this voltage $\xi_{sig}$,
but will conserve
the square root of the peak signal power
\begin{equation}
    \sqrt{P_{sig}} = \frac{\xi_{sig}}{\sqrt{R_{coil}}} = \left(
    \frac{B_{1,lab}}{I_{coil,p}} \right) \frac{\omega M_{0}
    V_{s}}{2\sqrt{R_{coil}}},
    \label{eq:HoultP}
\end{equation}
where $R_{coil}$ is the resistance of the
coil.
Because power is conserved,
the same $\sqrt{P_{sig}}$
will be measured at the terminals of any
lossless matching network (\cref{fig:RecipCircLossless}).
Note there is an equal voltage drop (of $\xi_{total}/2$)
across the lossy elements of the probe and the
transceiver~\cite{Conradi2022NMRInsPri} in \cref{fig:RecipCirc}
resulting in the factor of two in the denominator
of \cref{eq:HoultP}.
Furthermore, a collection of terms in \cref{eq:HoultP}
can be substituted with the conversion factor
($\Lambda$) by noting that
the only source of resistive impedance
in a circuit such as the one illustrated
in \cref{fig:RecipCircLossless},
is $R_{coil}$, and thus
\begin{equation}
    \frac{B_{1,lab}}{I_{coil,p}\sqrt{R_{coil}}}
    =
    \frac{\left( 2 B_{1,avg} \right)}{(2\sqrt{2} I_{coil,rms})\sqrt{R_{coil}}}
    =
    \frac{1}{\sqrt{2}} \Lambda,
    \label{eq:sqrtTwoLambda}
\end{equation}
where $B_{1,lab}=2 B_{1,avg}$ accounts for
the two counter-rotating components of the
magnetic field,
while $I_{coil,rms} = I_{coil,p}/\sqrt{2}$
converts from current amplitude to rms
current.
\Cref{eq:HoultP}
then becomes
\begin{equation}
    \sqrt{P_{sig}} = \frac{1}{\sqrt{2}} \omega M_{0}
    V_s\Lambda
    \label{eq:Psig_Eff}.
\end{equation} 
This expression also holds for any circuit
equivalent to \cref{fig:RecipCircLossy},
since the attenuation will decrease $\Lambda$ equally in the
context of the pulse generated and of the signal detected.
\Cref{eq:Psig_Eff} quantifies in a more convenient
form what is already stated by
Hoult~\cite{Hoult1976SigRatNuc}
and well-known in ESR spectroscopy:

$\Lambda$ is not merely an expression of
how efficiently the probe converts
pulse power into magnetic field,
but also determines the amplitude of the signal
that the probe will generate.

\figRecipCirc

\subsection{\texorpdfstring{Factors affecting $\Lambda$}{Factors affecting \textbackslash Lambda}}\label{sec:etap}
Having established the centrality of the
conversion factor ($\Lambda$)
as a figure of merit for probe design,
a careful analysis of existing literature (both
NMR and ESR) concludes that only three
quantities control the value of $\Lambda$:
the volume of the sample, the $Q$-factor,
and the
``field distribution factor'' $\eta'$
(a precisely defined quantity that takes the
place of previous approximations for the filling factor).

On resonance,
the probe oscillates between storing
the same amount of energy as magnetic and electric 
field energy.
As noted by Mims~\cite{Mims1972EleSpiEch}, the energy stored in an
impedance-matched resonant rf circuit can thus be
\emph{exactly} quantified with the relation:
\begin{equation}
    \frac{1}{2\mu}
    \iiint\left|\textbf{B}_{1}(\textbf{r})
    \right|^{2}d\textbf{r} =
    \frac{Q P_{tx}}{\omega}
    \label{eq:FieldQ}
\end{equation}
where the three dimensional $\mathbf{r}$ integral
spans all space
(excluding fields inside the matched transmission line),
$\omega$ is the resonance 
frequency, and $P_{tx}$ is the average power that
generates the $\mathbf{B}_1$ field.
The left-hand-side
of \cref{eq:FieldQ},
is a standard expression for
magnetic field energy,
while $Q P_{tx}/\omega$
gives the energy stored inside the resonator,
based on the definition of $Q$
(ratio of energy stored to energy dissipated
per radian)
and the fact that a matched resonator must be
dissipating $P_{tx}/\omega$ Joules of
energy per radian~\cite{Mims1972EleSpiEch}.
\Cref{eq:FieldQ} defines
$\mathbf{B}_{1}(\mathbf{r})$ distinctly 
from $B_{1,avg}$ in \cref{eq:LambdaB} and integrates 
over all magnetic fields generated by all 
circuit elements within the impedance matched 
probe. 
$B_{1,avg}$ represents the average 
only over the portion of the rf field capable 
of generating nutation: 
\begin{equation}2 B_{1,avg} = \frac{1}{V_s}\iiint_{V_s}\frac{\left|\textbf{B}_{0}(\mathbf{r})\times\textbf{B}_{1}(\textbf{r})\right|}{\left| \mathbf{B}_{0}(\mathbf{r}) \right|}d\textbf{r}\label{eq:SampleB1}\end{equation}
where the vector $\mathbf{B}_{0}$ is the static 
magnetic field \footnote{\cref{eq:SampleB1} only assumes that
the spins do not diffuse over the timescale of
the experiment between regions with
significantly different orientations of $B_{1}$,
so that any rf field perpendicular to $B_{0}$
excites spins in an equivalent fashion. Also
note, the fields are only integrated over the
sample volume.}.

Mims' analysis~\cite{Mims1972EleSpiEch} was developed
for resonators that even lacked well-defined
lumped circuit elements.
In various contexts,
such an analysis can offer crucial insight;
for example, many X-band ODNP
setups employ a hairpin loop coil with a very small inductance that
one can suspect suffers competition from stray
inductances within the rest of the tuning
circuitry.
A traditional definition of the ``filling
factor''
($\eta$) frequently refers to the extent to which
a sample fills a sample coil
(\cite{Hill1968LimMeaMag}),
although the very earliest NMR literature
acknowledges that this term is actually
defined from first principles in terms of the
distribution of the magnetic field~\cite{Bloembergen1948RelEffNuc,
BloembergenNucMagRel,Ladjouze1973FilFacTwo}.
There are many variations on this definition
in an attempt to more accurately estimate probe
sensitivity~\cite{Doty2015ProDesCon,1975CriFacDes,Rinard1999AbsEPRSpi,Bloembergen1954RadDamMag,McDowell2007OpeNanSca,McDowell2007OpeNanSca}.
However, by integrating elements from Mims'
analysis~\cite{Mims1972EleSpiEch},
one can see that the definition
\begin{equation}\eta' = \frac{4B_{1,avg}^{2}V_s}{\iiint\left| \textbf{B}_{1}(\textbf{r}) \right|^{2}d\textbf{r}}\label{eq:FillFactorField}\end{equation}
proves more general and/or accurate than other
options.
The value of $\eta'$ is unitless:
both the numerator and
denominator have units of $[\text{T}^{2}\cdot
\text{m}^{3}]$
(with field in the lab frame).
In the idealized
case where the only rf magnetic
field that exists outside the transmission line
is a uniform field amplitude of $2 B_{1,avg}$ that exists
only inside the coil volume,
then
$\eta'$ would
correspond to the extent to which the sample
fills the inside of the coil: \emph{i.e.} under such
idealized (and non-physical) conditions, one
would find $\eta' \rightarrow
V_{sample}/V_{coil}$. 
However, as the results will indicate, very 
constrained designs such as the hairpin loop utilized 
in many ODNP experiments fall very far 
from this condition.

With regards to the definition of $\eta'$,
some historical comparison and contrast may prove fruitful.
Note that the definition of $\eta'$ is very similar
to that of the ``magnetic filling factor''~\cite{Doty2015ProDesCon},
but
to distinguish from many other,
subtly different definitions of
filling factor,
and to choose a more accurate terminology,
$\eta'$ is referred to here as the ``field 
distribution factor.''
Note that
it corresponds exactly to the quantity $V_{s}/V_{c}$
in Mims' (and therefore much ESR) nomenclature~\cite{Mims1972EleSpiEch}.
Assuming that the definition of ``effective resonator volume''
in ESR literature matches that of Mims,
several definitions of ``filling factor'' in ESR
literature~\cite{Borbat1997MulTwoFou,aliDesignPlanarMicrocoilbased2017,Bloembergen1954RadDamMag}
in fact match
the definition of the field distribution factor
almost exactly.
Crucially, however, note that the original
``$V_c$'' of Mims 
(exactly equal to $V_s/\eta'$)
is an ``effective cavity volume''
that is defined as the volume integral of
$|B_1(\mathbf{r})|^2/(2 B_{1,avg})^2$.
Unlike in Abragam and others,
it is distinct from the coil volume
(pg. 74 of~\cite{Abragam1961PriNucMag}).
Rather, it gives the effective volume
over which the field is distributed,
and it is never equal to the volume of
a coil, cavity, or any other
physical object.

Some simple substitutions offer further
insight into the nature and importance of $\eta'$.
First, substitution of \cref{eq:FieldQ} into
\cref{eq:FillFactorField},
yields
\begin{equation}
    \eta' = \frac{2 \omega B_{1,avg}^2
    V_s}{\mu Q P_{tx}}.
\end{equation}
Note this expression relies on few
assumptions,
being simply a ratio between
the (Joules of) energy that would be required to
generate a rotating-frame $B_{1,avg}$ under
idealized circumstances
($2 B_{1,avg}^2 V_s/\mu$),
and
the total energy stored in the probe,
$Q P_{tx}/\omega$.
Subtitution of \cref{eq:LambdaB}
simplifies and
relates $\eta'$ to $\Lambda$:
\begin{equation}
    \eta'
    =
    \frac{2\omega V_s \Lambda^2}{\mu Q}.
    \label{eq:etaFundamental}
\end{equation}
Rearranging \cref{eq:etaFundamental}
yields an equation that decomposes
the conversion factor ($\Lambda$) into contributions
from the field distribution factor ($\eta'$),
the $Q$-factor, and the volume of the
sample: 
\begin{equation}
    \Lambda = \sqrt{\frac{\mu \eta' Q}{2\omega V_s}}
    \label{eq:LambdaEta}
\end{equation}
This expression explains how to optimize the
sensitivity of the probe,
and it therefore
naturally resembles many equations that one
finds in the probe design
literature~\cite{Doty2015ProDesCon,Rinard1999AbsEPRSpi}.
However, as a result of following the analysis of Mims~\cite{Mims1972EleSpiEch}
and the resulting definitions,
\cref{eq:etaFundamental} is exact and can be
generalized to unusual
probe designs -- in fact,
to any impedance-matched rf resonator.
Substituting \cref{eq:LambdaEta}
into \cref{eq:Psig_Eff} to yield
\begin{equation}
    \sqrt{P_{sig}} = \frac{1}{2} M_{0}
    \sqrt{\mu \omega \eta' Q V_s}
\end{equation} 
particularly emphasizes that
more efficiently distributing the
magnetic field (increasing $\eta'$),
decreasing the loss of the circuit (increasing $Q$),
or increasing the sample volume (increasing $V_s$),
will all have equivalent
effects on the signal amplitude.
Furthermore,
improvement in any one of these factors
will allow for the same SNR with proportionately less
signal averaged transients
(as $\text{SNR}\propto\sqrt{\text{\# transients}}$).
Because the needs of the application
typically dictate the resonance frequency,
the number of scans that are possible,
and the volume of the sample,
the probe designer concerns themselves with
optimizing the field distribution $\eta'$ and
the $Q$-factor.

Finally note that a measurement of the
$Q$-factor, and the $t_{90}$ at a specific
power ($P_{tx}$) provides sufficient
information
that one can deduce the value of the field
distribution factor ($\eta'$)
without relying on any approximations or
assumptions.
Specifically, substitution of \cref{eq:Vsig90t} into 
\cref{eq:etaFundamental}
shows that the field distribution factor:
\begin{equation}
    \eta'
    =
    \displaystyle\left(\frac{\pi^2}{2 \mu \gamma^2}\right)
    \left(\frac{ \omega V_s}{Q t_{90}^{2} P_{tx}}\right)
    \label{eq:FillFactorMeas}
\end{equation}
is a function of physical constants ($\mu$ and $\gamma$) and
quantities that one can measure ($\omega$, $t_{90}$, $Q$,
$P_{tx}$ and $V_s$).
\subsection{Johnson-Nyquist Noise}\label{sec:JohnsonNoise}

Of course, this article concerns itself, in part,
with the origins of noise that can obscure
low-field NMR signal. 
The Johnson-Nyquist noise~\cite{Nyquist1928TheAgiEle} that
a receiver chain detects when connected
to an impedance-matched probe
is given by~\cite{Rinard1999AbsEPRSpi}:
\begin{equation}\frac{P_{noise}}{\Delta\nu} =
k_{B}T\label{eq:NoisePower}\end{equation} where $k_{B}$
is Boltzmann's constant $[\text{J/K}]$, $T$ 
is the temperature {[}K{]}, and $\Delta \nu$ is the 
bandwidth {[}Hz{]} of the detector. 
In contrast to 1/f noise from active components or 
EMI from other
instruments in the environment, the Johnson-Nyquist 
noise is white noise -- \emph{i.e}, 
does not vary with frequency~\cite{Fitzpatrick2018Cha14Noi,Andris2019AnaNMRSpe}.

Of course, low-noise amplifiers will also add
a noise level consistent with their noise
temperature, as governed by the Friis equation,
\begin{equation}F_{total} = F_{1} + \frac{F_{2}-1}{G_1} + \frac{F_{3}-1}{G_{1}G_{2}} + ... + \frac{F_{n}-1}{G_{1}G_{2}...G_{n-1}}\label{eq:Friis}\end{equation}
where $F_{n}$ is the noise factor of the $n^{th}$ stage of a multistage
amplifier, and $G_{n}$ is the
gain of the nth stage of the multistage amplifier~\cite{Friis1944NoiFigRad}. 
As a large portion of the results focus on
identifying and reducing EMI;
we refer to the noise that results from
\cref{eq:NoisePower} and the LNA noise (\cref{eq:Friis})
as the ``thermal noise limit.''

\subsection{Standard Equations for Signal Processing}\label{sec:NoiseProc}

Several equations assist in relating real-valued oscilloscope data to
quadrature data and/or noise power spectral densities.

The standard analytical signal transformation effectively
converts real-valued signal $s(t)$ into
analytic signal $s_a(t)$.
Specifically,
\begin{equation}
    \tilde{s}(\nu)= \sum_{j=0}^{N-1}
    e^{-i2\pi\nu j \Delta t} s(j\Delta t) \Delta t
    \label{eq:FourierTransform}
\end{equation}
\begin{equation}
    s_a(t)
    =  \sum_{j=0}^{N-1} (2-\delta_{\nu,0}) e^{+i2\pi j \Delta\nu t} \tilde{s}(j \Delta\nu) \Delta\nu
    \label{eq:RealToAnalytic}
\end{equation}
where
$\Delta t$ is the spacing between the time domain
points,
$N$ is the number of data points
(and $t_{acq}=(N-1)\Delta t$ is the acquisition
length),
$\delta_{\nu,0}$ is zero except at $\nu=0$
(where it is 1),
and the spacing between the frequency domain
points is given by $\Delta \nu = 1/N\Delta t=1/(t_{acq}+\Delta t)$.
(\Cref{eq:RealToAnalytic} corresponds to multiplication in
the frequency domain by a Heaviside function and
downsampling.)
The analytic signal provides instantaneous amplitude
($|s_a(t)|$), power ($|s_a(t)|^2$), and phase
($\mbox{angle}(s_a(t))$) information,
while 
\begin{equation}
    \Re[s_a(t)]=s(t)
    \label{eq:analVsReal}
\end{equation}
(with $s_a(t)$ sinc interpolated to twice the sampling rate)
corresponds to the signal captured on the
oscilloscope.

Note specifically that the average power of
an arbitrary analytic signal is then
\begin{equation}
    P_{avg} = \frac{1}{2 Z_c t_{total}}
    \int_{0}^{t_{total}}
    \left| s_a(t) \right|^2 dt,
    \label{eq:analyticPavg}
\end{equation}
where $Z_c$ is the characteristic impedance $[\text{Ω}]$,
and $t_{total}$ is the duration of the analytic signal.
When considering an rf signal with constant
    amplitude,
    from \cref{eq:analVsReal},
    the complex magnitude of the analytic signal
    $|s_a|$ will be equal to the rf
    amplitude.
Thus \cref{eq:analyticPavg} becomes 
$P_{avg} = \frac{\left| s_a \right|^2}{2Z_c}$.
Pointing out
    that $|s_a|/\sqrt{2}$ is equivalent to
    the rms voltage may allow the reader to see the
    equivalence between
    \cref{eq:analyticPavg} and more
    well-known expressions.

Furthermore, the analytic signal derived from the
oscilloscope data corresponds to signal digitized
on a quadrature digital receiver ($s_{rx}(t)$),
\begin{equation}
    s_{rx}(t) = (s_a(t)\exp(-i 2 \pi \nu_{c} t))\otimes f_{DF}(t)
    \label{eq:transceiverTimeDomain}
\end{equation}
where $\nu_{c}$ is the carrier
frequency of the NMR signal
(the digital equivalent of a local oscillator
(LO)),
$f_{DF}(t)$ is the time-domain convolution filter
that is applied by the digital receiver,
and $\otimes$ represents the convolution
operation.

Even though noise is, of course, phase incoherent,
the power of the noise
($P_{noise}(\nu) \propto |\tilde{s}(\nu)|^2$)
still
adds linearly, allowing for the signal averaging
and/or convolution that are important to clarifying
the shape of $P_{noise}$ and deriving meaning from
it.
Specifically, the complex
magnitude of the Fourier transform
of time-domain noise acquired by an oscilloscope
or digital receiver
is divided by the characteristic impedance and the
length of the acquisition to convert to
a PSD ($P_{noise}(\nu)$) with units
$[\text{W}/\text{Hz}]$.
A convolution
(normalized Gaussian of width $\sigma$)
in the frequency domain also smooths
the data:
\begin{equation}
\begin{array}{rl}
    \displaystyle P_{noise}(\nu) &=
    \displaystyle \frac{1}{t_{acq}Z_{c}\sigma N \sqrt{2\pi}}\sum_{i=1}^{N}\sum_{j=0}^{N-1} e^{-(\nu-j\Delta\nu)^2/2\sigma^2}
    \\
    \displaystyle &\quad
    \displaystyle \times \left|\sum_{t=0}^{t_{acq}}e^{-i2\pi\nu' t}n_{i}(t)\Delta t \right|^2\Delta \nu
    \\
    \displaystyle &=
    \displaystyle \frac{1}{t_{acq}Z_{c}\sigma N \sqrt{2\pi}}\sum_{i=1}^{N} e^{-\nu^2/2\sigma^2}
    \\
    \displaystyle &\quad
    \displaystyle \otimes\left|\sum_{t=0}^{t_{acq}}e^{-i2\pi\nu t}n_{i}(t)\Delta t \right|^2
\end{array}\label{eq:convPSD}
\end{equation}
where $n_{i}(t)$ {[}$\text{V}${]} is the time-domain noise voltage,
$Z_{c}$ {[}$\Omega${]} the 
characteristic impedance of the transmission 
line (typically 50 $\Omega$),
$t_{acq}\;[\text{s}]$ the length of the acquisition,
$\sigma\;[\text{Hz}]$ the convolution width,
and $N$ is the number of averaged scans. 
Here, scripts that store data to disk always
convert the (real valued) oscilloscope data
to analytic signal \via \cref{eq:RealToAnalytic}.
\section{Methods}
\subsection{Samples}
A 4.38~μL sample of 27~mM
4-hydroxy-2,2,6,6-tetramethylpiperidine-1-oxyl
(TEMPOL, Toronto Research Chemical), prepared in
deionized water, was loaded into a capillary tube
with an inner diameter of 0.6 mm and outer diameter
of 0.8 mm, and flame
sealed for use as the standard sample,
with a typical relaxation time of about 100 ms.

\subsection{Hardware}\label{sec:Hardware}

Roughly following the design shown in~\cite{Kaminker2015ChaSixOve}, the tuning box
containing adjustable capacitors for the probe's
circuit attaches to the bottom of the ESR
cavity (which has a threaded connector) by way of
a threaded collar piece that presses against
a conductive neck protruding from the tuning box.
The threaded collar thus brings the tuning box
into electrical contact with the shield
surrounding the ESR cavity. 
A multimeter was used to perform
a continuity check between the ground
of the ESR cavity and the ground of the probe.

As \cref{fig:hardwareSetup} shows, the ESR cavity sits
between the plates of the electromagnet, attached
to a waveguide. 
The bottom of the probe's tuning box connects 
\emph{via} BNC connectors and 50~Ω coaxial cable 
(RG58/U 20~AWG) to the input of an aluminum 
enclosure containing the receiver chain (yellow box
in \cref{fig:hardwareSetup}).
The receiver chain comprises a 
home-built passive duplexer, two standard LNAs
(MiniCircuits ZFL-500LN+), a low-pass
filter (MiniCircuits SLP-21.4+) and (for select
measurements) a high-pass filter (Crystek
Corporation CHPFL-0010-BNC). 
Note, that all coaxial cabling presented
in \cref{fig:hardwareSetup} is 50~Ω with BNC
connectors.
When employed, common-mode chokes of the snap-on
Ferrite style
(broadband operation up to 300~MHz, 6.35~mm i.d.,
12.2-19~mm o.d.)
attach to the coaxial cable
connecting the tuning box to the receiver chain
enclosure.
Alternately,
a 9 foot coaxial cable
connecting the
tuning box to the receiver chain enclosure
threads through
a MnZn PC40 toroidal ferrite core 
(Hondark HK Limited).

\begin{figure}
\centering
\includegraphics[width=\linewidth]{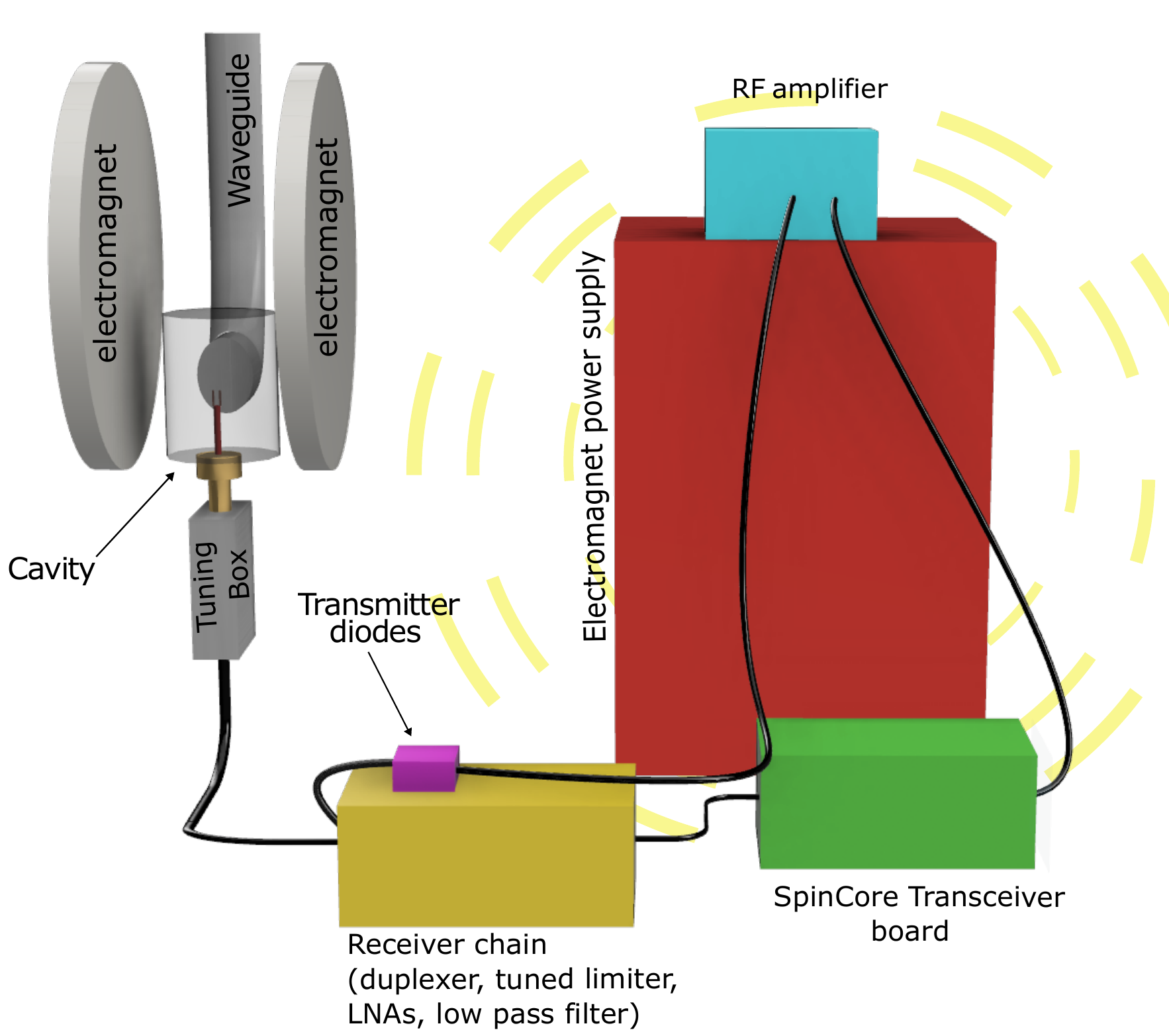}
\caption{Instrumental setup:
The power supply of the electromagnet (red) sits
adjacent to the magnet itself, connected by
heavy-gauge DC supply lines (not shown). 
The RF amplifier (blue box)
    outputs pulses to the transmitter diodes
    (pink box),
    which in turn connect via
    cable to a tee connector at the input of the receiver
    chain (yellow).
The silver box labeled ``tuning box'' contains
the capacitors for the probe circuit, with the
hairpin loop rf coil extending upwards through a hole
into the ESR cavity.
A cable connects the NMR probe to the tee connector
at the front of the receiver chain before the
transceiver (combined transmitter/pulse-programmer and receiver)
board (green) digitizes it. 
The electromagnet power supply emits interference
noise as indicated by the yellow dashed circles.
This drawing omits the microwave amplifier
and parts of the ESR instrumentation
in favor of emphasizing the elements crucial to
the protocol.}\label{fig:hardwareSetup}
\end{figure}

The Bruker E500 hardware~\cite{JiangEle500Use}, in conjunction with
the XEPR software and a custom script employing the Bruker Python API,
controls the magnetic field. 
It employs a Hall sensor for magnetic 
field detection. 

During typical NMR experiments, the SpinCore
RadioProcessor-G serves as the transceiver board
(green box in \cref{fig:hardwareSetup}),
offering both pulse programming capabilities and a real-time
oversampling capability during detection. 
All communication with the SpinCore
transceiver occurs \emph{via} a Python API
extension developed by the authors from the C language API
and examples supplied by the vendor~\cite{InstNotes}.
Importantly, note that the protocol here
applies to a variety of possible receivers
that offer oversampling,
and is not restricted to the model of the NMR receiver.

The SpinCore transceiver board outputs the desired
pulses that are subsequently amplified by a SpinCore RF
amplifier (PA75W-M, 75 W RMS, blue box in \cref{fig:hardwareSetup}).
The amplified pulses are then sent via BNC
cabling to an aluminum box containing transmitter diodes
(pink box in \cref{fig:hardwareSetup}),
which are connected with a BNC tee both
to the front of the receiver chain
and
(\via a 1~m cable or \via a 3~m cable
threaded through a ferrite toroid)
to the NMR probe.

Data that requires
independent off-the-shelf test and
measurement equipment
relies on a minimal set of instrumentation:
an oscilloscope (GW-Instek
GDS-3254, with bandwidths up to 250 MHz and sampling
rates up to 5 GSPS),
an Arbitrary Function 
Generator (GW-Instek AFG-2225),
and a handheld
Vector Network Analyzer (NanoVNA-H, version 3.5).

It is worth noting that every laboratory setup
differs in a way that may affect noise generation
and transmission,
and that in the authors' lab, by necessity,
all the ESR hardware connects to
a different circuit (different voltage supply)
than the NMR hardware, and a transformer
isolates it from the electrical mains supply. 
Also, while the flooring surface is
insulating, it rests over continuous metal
sheeting.

\subsection{Details of Oscilloscope-Based NoiseMeasurement}\label{sec:oscDetails}

For the widest applicability, this section covers
detailed considerations of performing noise PSD
measurements with a standard oscilloscope.
All data is acquired using the library
available at
\url{https://github.com/jmfrancklab/FLinst}
that specifically adopts a strategy of enabling
python context blocks for the connection to the
oscilloscope, and providing the captured data as
an object containing all axis coordinate and unit
information
(in \href{https://github.com/jmfrancklab/pyspecdata}{pySpecData} format).
The strategy should be extensible to a wide range
of digital capture oscilloscopes, as well as to
spectrum analyzers capable of digital capture
(in addition to the direct USB RS232 communication
employed here, the libraries also offer context
blocks for handling GPIB communications over
Prologix connectors).
The authors have previously provided~\cite{Beaton2022ModVieCoh}
a demonstration that the same
software and oscilloscope employed here are capable of
capturing a phase-cycled spin echo.

For an oscilloscope-based measurement of noise,
the receiver chain needs to include a low-pass filter (serving
both in reducing high frequency noise and as a point of
reference for the noise PSD) after the low-noise
amplifiers within the receiver chain.
The sampling rate of the oscilloscope must then
exceed twice the high-frequency edge of the
low-pass filter (\emph{e.g.} in \cref{fig:voltscale} the
sampling rate exceeds 70 MHz).
A plot of the noise PSD (\cref{eq:convPSD})
then clearly displays the edge of
the low-pass filter, as shown by the red, green,
and gold curves in \cref{fig:voltscale}, which was
acquired with a 25 MHz low-pass filter in place.

While standard oscilloscope measurements offer an
easy comparison to, and smooth transition to,
measurements on the NMR receiver, they do suffer
the drawback of a limited dynamic range.
Therefore, the vertical scale of the
oscilloscope requires optimization
(as shown in \cref{fig:voltscale}).
A small voltage scale (vertically zoomed in)
maximizes the difference between the amplified
noise coming from the receiver chain and the
intrinsic noise of the oscilloscope that is seen
for frequencies above the threshold of the low-pass
filter ($\ge 30\;\text{MHz}$ in \cref{fig:voltscale}).
However, a voltage scale that is too small results
in a time-domain waveform that is (vertically) clipped,
leading to a clear distortion of the PSD (shown in
blue).
\Cref{fig:voltscale} clarifies this process
of dynamic range optimization with data
that was acquired in a relatively high-noise
configuration.
Specifically, a 10 mV voltage
scale (gold)
maximizes the difference between the
intrinsic noise of the oscilloscope and the noise of
the receiver chain while still avoiding the
clipping-induced distortion seen at 5 mV (blue).
\Cref{sec:optGDS} includes further details
about this dynamic range optimization.

\begin{figure}
\centering \includegraphics[width=\linewidth]{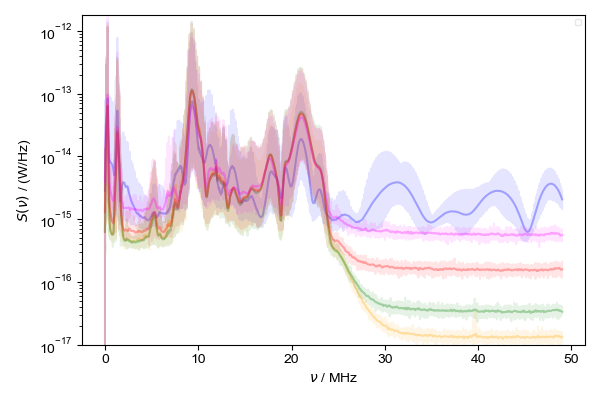} \caption{An oscilloscope set to a sampling rate of 100
MSPS and a voltage scale of 5 (blue), 10 (gold),
20 (green), 50 (red) and 100 mV (magenta),
digitized 100 traces of noise
acquired from the output end of the receiver
chain when hooked up to the
tuning box (\cref{fig:hardwareSetup}).
In each case, a 100~kHz Gaussian convolution (\cref{eq:convPSD}) yields
a smoothed version of the line direct average,
shown as the
more transparent line of the same color.}\label{fig:voltscale}
\end{figure}

\Cref{fig:voltscale} additionally demonstrates that even after
averaging the PSD from 100 captures
(acquired in under three minutes)
to yield the
faint, more transparent lines, these lines still
appear broad/thick due to rapid
frequency-dependent oscillations in the noise.
The Gaussian convolution chosen for broadband measurements
(here 100~kHz)
smooths out these features
to identify general trends in the data while also
requiring minimal or no signal averaging
in order to produce a smooth and reproducible PSD.

\subsection{Generation of Test Signal}\label{sec:genTestSignal}

A variety of sources can generate a test signal at
a fixed voltage and frequency (the results here
employ the previously mentioned programmable AFG)
which serves as a convenient control case.
However, testing with the duplexer in place
requires signal at the level of NMR signal -- at the
μV to nV level, which is much smaller than the 
minimal output of the average source.
Therefore, it becomes necessary to connect
precisely calibrated attenuators to the
output of the frequency source.
This is achieved by designating
``attenuator
assemblies'':
the combination of an
attenuator with attached cabling
and/or adapters that begins with a BNC jack (F) and ends with a
BNC plug (M).
With the frequency source set to a high
voltage,
an oscilloscope measures the output of
each such assembly to allow calculation of
the ratio between the unattenuated voltage 
and the voltage with the attenuator assembly in place.
(A more typical attenuation $[\text{dB}]$ value
can be used if one takes care to remember the unusual rules for
significant figures of logarithmic quantities.)
The measurement is repeated several times,
to check for variations due to cable
positioning.

To generate the test signal,
the frequency source is then set to a mV-level voltage
(and measured on the oscilloscope),
and
several attenuator assemblies,
which (by design) do not involve
insertion losses from any
intervening cabling,
are then
connected as needed.
In order to speed this calibration,
a simple script captures the oscilloscope
data and automatically converts to analytic
signal, filters, and measures the rf voltage
amplitude (\cref{single_GDS_acq}).
\section{Results and Discussion}
The central result presented here is a
protocol that enables detailed
characterization and optimization of the
absolute signal and noise levels
of a custom NMR system.
The sections below, in order,
outline an example of such a protocol.
Over the course of this protocol,
the validity and utility of the reorganized
theory (\cref{sec:theory}) are also explored,
as well as various specific observations
pertaining to the low-field ODNP system
investigated here.

\subsection{Noise PSDs from Test and Measurement Equipment}
A single unprocessed capture of the noise on the 
NMR receiver or an oscilloscope tends to yield a relatively uninformative 
picture, allowing calculation of the standard 
deviation of the noise voltage and little
else~\cite{Andris2019AnaNMRSpe, Yang2022ActEMISup}.
Spectrum analyzers can quantify noise with a high dynamic 
range,
and prove particularly useful for LNA noise 
factor quantification,
while VNAs can identify
cross-talk between components ($S_{12}$)
\cite{Rytting2001NetAnaAcc,Nordmeyer-Massner2011NoiFigCha,Micheli2017ChaEleCha,Sankaran2018RecAdvEle}.
However, both for limiting the necessary
instrumentation,
and for easier comparison to the NMR receiver,
the first steps of this protocol rely primarily on
a standard digital
oscilloscope to capture the noise
that is
output by the receiver chain.
\Cref{eq:convPSD} then converts the captured
time-domain traces to smooth PSDs.

\subsubsection{Intrinsic Noise of the Receiver Chain}\label{sec:FirstPSD}
The first step of the protocol involves checking that the metal
box enclosing the receiver chain provides adequate shielding.
The green line in \cref{fig:PSDwGain} shows
the gain of the
receiver chain (measured in \cref{sec:gainQuant}, \cref{fig:Gain})
multiplied by the theoretical Johnson-Nyquist noise
(\cref{eq:NoisePower}).
With a 50 Ω terminator attached to its input,
the (shielded) receiver chain emits the
noise density shown in blue.
The noise cutoff at 25 MHz arises from the presence of
the low pass filter at the end of the
receiver chain,
while the steep rise at frequencies
below 1 MHz
comes from 1/f noise of the LNAs.
At intervening frequencies,
the noise density appears approximately flat.
The fact that the measured noise \emph{vs.}
the thermal noise limit are separated by a factor of $\sim 2$,
and only for a limited range around the intended
resonance frequency of 14-15 MHz, is expected
here;
the duplexer contains various
frequency-sensitive components that filter
signal outside the 14-15 MHz range to varying extents,
while the LNAs are broadband and will
continue to transmit a noise level
appropriate for their noise figure
(specified as 2.9~dB).
\figPSDwGain
\subsubsection{Noise with Attached Probe}
\Cref{fig:PSDwGain} also highlights a major unexpected issue:
attaching the NMR probe to the receiver chain
introduces electromagnetic interference (EMI)
that exceeds the thermal noise limit by two orders of magnitude.
Such high levels of EMI prevent the
isolation of
small amounts of signal -- especially when
considering, \emph{e.g.}, the un-enhanced reference scan for an ODNP experiment
that comes from a 4.38~μL sample with only
thermal (Boltzmann) levels of spin polarization.
However, as will be demonstrated in the next
step (\cref{sec:EMInoise}),
the distinctive shape and frequency
specificity of this measurement allows the
spectroscopist to systematically identify and
mitigate the source of the EMI,
identifying it similarly to how one might identify
a chemical compound from the fingerprint region of a spectrum.

Note that frequency-domain
convolution (\cref{eq:convPSD}) of the noise PSD
proves crucial to developing clear results.
As mentioned in \cref{sec:oscDetails},
the apparent thickness of the
lines from the
un-convolved data (the transparent lines in
\cref{fig:PSDwGain})
arises from
high-resolution oscillations.
These oscillations make it difficult to extract
identifying details like the $1/f$ noise or
distinctive frequency variations arising from
particular sources of EMI.
The origins of these oscillations will be
interpreted later;
at this stage in the protocol,
however,
an empirical observation of the line
``thickness'' drives
the choice of convolution width in
\cref{eq:convPSD} (in \cref{fig:PSDwGain},
$\sigma=100\;\text{kHz}$).
Notably, over longer periods of time
or when equipment is moved slightly,
it is not uncommon for the distribution of the noise
density to decrease slightly at some frequencies while
increasing slightly at others (not shown).
This underscores another important
element of the noise measurement protocol introduced here:
It is important to start by recording the noise PSD over a
broad frequency range.
Measures that truly mitigate noise will tend
to do so over a broad frequency range.
In contrast,
exclusively observing noise densities
zoomed in to more specific frequency ranges
could make it appear that the noise was
increasing or decreasing when various changes
are made when, on average, it is not.
\subsubsection{Example: Investigating Source of Noise}\label{sec:EMInoise}
\figMagNoise
Many magnetic resonance
systems that integrate complex components
will have to contend with the resulting added noise
contributions such as those seen in \cref{fig:PSDwGain}.
For example,
ODNP spectroscopists are relatively familiar
with the fact that a commercial electromagnet
power supply
(here the Bruker E500 system)
can emit large amounts of
EMI.
The protocol here enables a laboratory
with almost any level of experience and equipment,
access to measure and to confirm the nature
of this noise.
\Cref{fig:MagNoise} expressly
shows that the electromagnet power supply
introduces EMI
that travels through free space
and that this EMI
exceeds the thermal noise limit by
two orders of magnitude.
Specifically, the act of turning on
the electromagnet power supply
(gold → green of \cref{fig:MagNoise})
introduces the
overwhelming majority of the noise
that was seen in \cref{fig:PSDwGain}.
Removing the probe from the cavity and
placing it on a nearby table breaks any
potential direct current path between the
probe and the ESR power supply.
One would expect that this action would
remove any noise brought on by spurious
electrical connections,
ground loops, \emph{etc.}
However, as expected only for EMI traveling through
the air,
moving the probe out of the cavity yields only a slight decrease 
of the noise and a redistribution in
frequency space.

The following three subsections demonstrate 
how a reasonably processed broadband noise PSD
enables one to systematically choose between a selection of
rather simple noise mitigation options,
to identify a solution that works
well for a particular lab's experimental needs.

\subsubsection{Test 1: Analysis of Shielding}
\figCavityProbe
An implied advantage of the design adapted
from Kaminker~\cite{Kaminker2015ChaSixOve} is that the
grounded elements of the probe and cavity
would together form a Faraday cage that would
shield the probe from EMI.
The fact that removing the probe from the
cavity and exposing it to the environment
(green → red in \cref{fig:MagNoise}) results in
reduced noise is thus surprising.

Therefore, a series of measurements tested the
integrity of the shielding scheme.
Three noise PSDs were
acquired (\cref{fig:cavityProbe}):
one with the NMR probe
secured and properly grounded to the cavity,
a second where the probe-cavity-waveguide assembly is
detached from the ESR microwave bridge and simply held between the
magnets, and a third where the probe is completely
disconnected from the cavity and held between the
magnet plates.
These PSDs show a 
lack of a dramatic change in the overall noise
density,
thus indicating that the cavity likely provides
insufficient shielding for the NMR probe.

Thus, \cref{fig:cavityProbe} identified that
improved shielding might prove
an effective strategy for noise mitigation.
However, optimization of connections between 
NMR probe and cavity -- including improving the metal-to-metal
contact between various parts and adding
conducting foil over joints -- yielded
negligible improvement (not shown).
Additionally, insertion of a copper plate into the waveguide connection in an
attempt to improve the grounding between
the NMR probe and the cavity/waveguide assembly
(a strategy that
was shown to work well in the Han lab~\cite{Franck2013QuaCwOve}),
demonstrated no improvement (not shown).
While a secondary shielding
enclosure may still prove successful
for mitigating the EMI noise,
this was beyond the scope of basic
improvements,
and, as shown in subsequent subsections,
proved unnecessary.

\subsubsection{Test 2: Balanced Probe Design}

A balanced probe design (green circuit in \cref{fig:BalVSone})
aims to concentrate the current within the
circuit near the coil and to mitigate the antenna
effect by approximately equalizing the impedance to ground
on either end of the coil~\cite{Mispelter2006NMRProBio,Stringer54BALMOD}. 
Here, the probe containing a
standard tank circuit (gold circuit in \cref{fig:BalVSone})
is referred to as the ``single-sided tank probe'' for contrast.
Both coils have an internal volume of 8.6~μL,
utilize a double hairpin loop,
and integrate
specifically into a Bruker Super High Sensitivity
Probehead X-Band resonator (ER 4122 SHQE).
In addition to the circuitry, the two probes differ in shielding:
a thicker 6.5 mm aluminum box encases the single-sided tank circuit
while a 1.5 mm thick aluminum box encases the balanced circuit.
When the single-sided tank probe is connected in place of the
balanced probe (\cref{fig:BalVSone}),
the receiver picks up slightly less interference noise.
Therefore all following
measurements utilize the single-sided probe.

\Cref{fig:BalVSone} emphasizes that the
EMI transmitted by the power
supply here does not interact
solely with the coil;
rather, it interacts with the
probe and ESR cavity assembly as a whole.
The slight differences between the noise PSD of the two
probes (\cref{fig:BalVSone}) may arise either from the thicker
shielding of the single-sided tank probe,
or from incidental changes to
the rf cross-section of the probe.

\begin{figure} \centering \includegraphics[width=\linewidth]{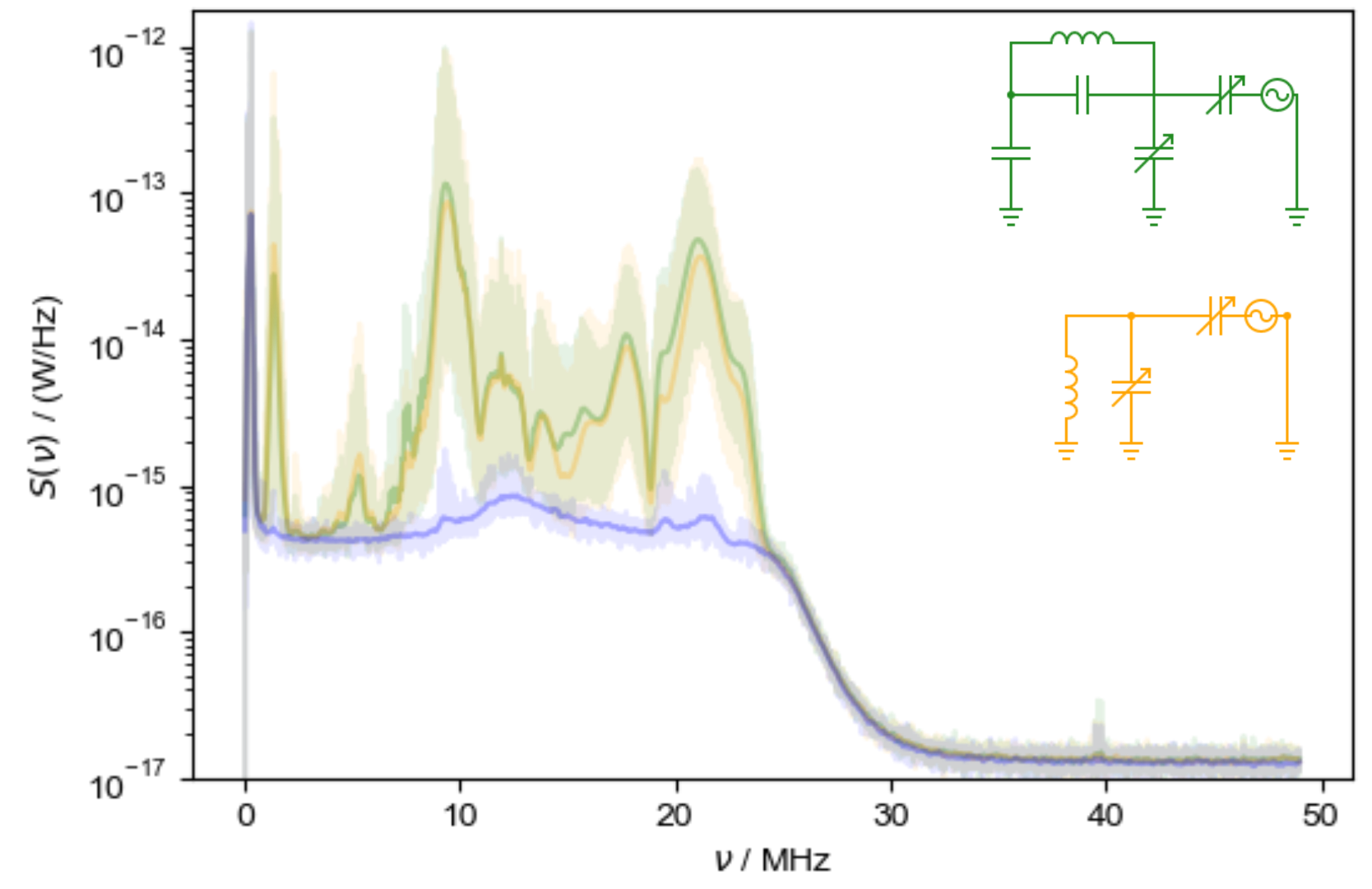} \caption{Different probe circuitry does not
aid in preventing EMI pickup.
When the DC supply of the Bruker E500 is
powered on,
the resulting interference noise transmits
slightly more efficiently into the
balanced NMR probe (green) 
\emph{vs} the single-sided tank probe (gold).
(Terminated receiver chain noise shown in
blue for reference.)}\label{fig:BalVSone}
\end{figure}
\subsubsection{Test 3: Ferrite Chokes Mitigate Electromagnetic Interference}\label{sec:chokes}
Toroidal ferrite chokes or snap on chokes are a common
solution for mitigating high-frequency EMI by
selectively increasing the impedance of common mode
transmission
(associated with noise)
while leaving the impedance of the
desired transverse (TEM) mode unaltered.
\Cref{fig:Chokes} shows how successive addition of
snap-on chokes to the coaxial cable
connecting the probe to the receiver chain
progressively mitigates the EMI in the example system here.
However,
a toroidal ferrite core can
accommodate multiple loops of cable
and thus offer a very large common-mode
impedance.
Simply threading the coaxial cabling around a 
single toroid produces the same
benefit as XX snap on chokes,
but with less weight pulling on the
cables.

As the noise of the system is decreased,
the voltscale
should be adjusted to further maximize the difference
between the noise of the receiver chain and the
intrinsic noise of the oscilloscope
(\cref{sec:oscDetails}). 
Consequently, the noise PSD in \cref{fig:Chokes} acquired
with no form of mitigation (gold) and with a voltage scale of 
10 mV, has a seemingly higher
level of noise at frequencies greater than the low pass filter
threshold compared to the PSDs acquired with 
chokes applied (which were acquired with voltage scales of 5 or 2
mV). 
Additionally, while subtle, the 
noise power of the probe with the chokes present dips
below the noise PSD of the terminated receiver chain at
frequencies surrounding 14.9 MHz (the frequency which
the probe was tuned to with the VNA). 
This is explained by noting that the reactive
capacitance only matches the reactive inductance at the
resonant frequency (14.9 MHz). 
However, at other frequencies, the reactive capacitance
does not match the reactive inductance, causing the
noise PSD of the probe to dip below the PSD of the
terminated receiver chain.
\figChokes
\subsection{Noise PSDs Acquired by NMR Transceiver}

There is little quantitative value to the
broadband noise PSD measurements unless they
correspond to the noise actually digitized on
the NMR receiver (transceiver).
Therefore, the protocol advocated here next
quantifies the noise density on the receiver
and compares it to the broadband
PSD measurements.
\subsubsection{Receiver calibration}
Receivers frequently collect and present acquired data
with arbitrary units intrinsic to the board
(denoted here as ``dg'' for ``digitizer units'')
that must be converted
to standard voltage units ($\text{V}$).
Initial calibration of the receiver
was performed by injecting
an rf test signal (\cref{sec:genTestSignal})
close to the transceiver carrier frequency
with an amplitude of
$\approx 15\;\text{mV}$
(exact value verified by oscilloscope using \cref{single_GDS_acq})
directly into the input of
the SpinCore transceiver board.
A nonlinear fit of the resulting waveform
to a complex exponential
(\emph{cf.} \cref{eq:transceiverTimeDomain}),
yields the
amplitude in {[}dg{]};
this is compared to the
amplitude in {[}V{]} of the same signal
measured on an oscilloscope (\cref{fig:calibFit}).
For the widest spectral width
(75~MHz, the base digitizer frequency),
a calibration factor of $6.054\;\text{dg}/\text{μV}$
was measured.

\subsubsection{\texorpdfstring{Comparison of Oscilloscope \emph{vs} Transceiver Noise Measurements}{Comparison of Oscilloscope vs Transceiver Noise Measurements}}\label{sec:wideSC}

The next step of the protocol here
compares the noise PSD from the
terminated receiver chain acquired on the NMR receiver
to the PSD acquired on the oscilloscope (\cref{fig:GDSvSC}).
Specifically, the SpinCore API Python extension~\cite{InstNotes}
captures a time-domain trace using the same
functions that are employed for NMR signal
acquisition,
with a very wide spectral width (75 MHz,
the base digitization frequency).
Division by the calibration factor
$[\text{dg}/\text{V}]$
appropriately converts the units to volts.
Subsequently, just as for oscilloscope captures,
\cref{eq:convPSD} converts the time-domain data
to a PSD.

\figGDSvSC

When digitizing the noise from the terminated
receiver chain, 
the receiver observes approximately the same noise
density as the oscilloscope measurements over 
frequencies less than $\approx 18\;\text{MHz}$.
However,
at frequencies higher than 20
MHz,
the noise PSD of the terminated receiver chain,
as acquired on the transceiver board,
differs drastically from the PSD
acquired on the oscilloscope.
Specifically,
peaks with a noise density similar to or
greater than the noise coming from the
receiver chain appear at frequencies $\ge
26\;\text{MHz}$
(\ie, above the low pass filter threshold),
even though the oscilloscope measurement
shows that the receiver chain does not
transmit noise in that frequency range.
Notably, these contributions are only developed
by the transceiver in response to a significant
power density of noise.
As mentioned in the SI (\cref{fig:attenPSD}),
these high-frequency noise peaks are found to scale
approximately linearly with the input.
In the absence of amplified input
(\textit{i.e.} with only a terminator
attached to the receiver input),
the noise density recorded by the receiver
drops several orders of
magnitude and changes in frequency (\cref{fig:GDSvSC}, red).
Therefore, the high-frequency noise contributions
observed on the receiver board might arise
from accidental mixing of the input with internal
clock signals, \emph{etc.}
Interestingly, as shown in \cref{fig:GDSvsSChighpass},
addition of a high pass filter
does mitigate
low-frequency noise,
but actually exacerbates the contribution of
the internally-generated high-frequency receiver
noise peaks.
Fortunately, for the X-band ODNP measurements
targeted here,
only 12-16 MHz carrier frequencies are required;
therefore, if the high-frequency peaks can be
digitally filtered,
they are not of concern.

\subsubsection{Testing Oversampling Performance}\label{sec:SCresponse}
\figSincFilter
\figResponseZoomed
Smaller spectral widths yield higher
resolution,
as longer time-domain traces can be
captured and stored using the same on-board
memory.
With proper oversampling
they also exclude noise
significantly outside the receiver bandwidth.
In the exact context employed here,
the default oversampling scheme
(specifically $f_{DF}(t)$ in \cref{eq:transceiverTimeDomain})
was actually
unknown.
With all receivers,
measuring and validating the oversampling
performance proves an essential step to
accurately quantifying and optimizing
the exact signal and
noise power.

To measure the receiver response,
a single script (\cref{sec:CodeRecResp}) varies the frequency
of a continuous sine wave test signal
(\cref{sec:genTestSignal}) with
$\approx 5\;\text{mV}$ amplitude
and captures the response on the receiver.
The 2D signal captured on the NMR receiver
(\cref{fig:sincFilterTwoD})
demonstrates how the receiver does record an alias
of signals that exceed the spectral width,
and, as expected,
the digital filter does decrease the amplitude
of these aliases to ultimately negligible levels.
The receiver response function quantifying this behavior
(\cref{fig:sincFilterOneD})
is determined from
the square root of the ratio between
the peak of the PSD recorded on
the receiver
\textit{vs.}
the peak of the PSD
for the input signal
(measured on an oscilloscope).
The resulting 1D response plot,
shows a good fit to the absolute value of a sinc
function,
implying that the default oversampling scheme for
this NMR receiver is to
box-car average the time-domain datapoints.
Here, the response specifically fits to
a sinc function
with a width of twice the spectral width;
this characterization holds for all spectral
widths (\cref{fig:ResponseZoomed}),
with only slight deviations for the widest
spectral widths.
The overall amplitude of the response
(here the amplitude of the sinc function)
can be determined with greater precision
through multiple measurements
(\cref{sec:Calibration30mV}),
resulting in the values given in
\cref{tbl:calibValues}.
\subsubsection{Noise PSDs at Different SW}\label{sec:predictPSD}
\figPredictPSD
\figPredictZoom
Full characterization of the response of the NMR receiver
(as in \cref{fig:ResponseZoomed})
exactly explains the relationship between
the noise PSD observed at different
spectral widths.
Specifically,
\cref{fig:predictPSD,fig:predictZoom}
demonstrate that
the following procedure can predict
the noise density at a smaller SW
from the noise density acquired with a larger SW:
\begin{enumerate}
        \def\labelenumi{\arabic{enumi}.}
        \tightlist
    \item Convert the time domain trace for
        the broader SW to a PSD via
        \cref{eq:convPSD}.
    \item Divide by the square of the
        receiver response [$\text{dg}/\text{V}$]
        specific to the larger SW,
        yielding the true noise PSD of the
        broader SW -- \eg
        the gold line in
        \cref{fig:predictZoom}.
        (In \cref{fig:predictPSD},
            the response of the gold line
            is assumed to be approximately flat,
            since the SW corresponds to the
            base sampling rate.)
    \item Multiply by the square of the
        response for the smaller,
        desired spectral
        width -- yielding, \textit{e.g.}, the green lines in
        \cref{fig:predictPSD,fig:predictZoom}.
    \item Move to the time domain (inverse
        Fourier Transform)
        and interpolate the data with a
        cubic spline function
        (because the faster sampling rate is
        not always an integer multiple
        of the slower sampling rate).
    \item Sample the interpolation function at
        the slower sampling rate
        corresponding to the narrower SW,
        and move back into the
        frequency domain -- yielding, \textit{e.g.}, the
        dark grey lines in
        \cref{fig:predictPSD,fig:predictZoom}
        that compare well with the noise density 
        measured at the narrower SW.
\end{enumerate} 
This procedure can explain unusual features
in the noise density.
For example, the small noise spike both
observed and predicted near 19~MHz
in \cref{fig:predictPSD} arises
from aliasing of the $1/f$ noise,
as evidenced by the fact that the
filtered (green) spike near 0~MHz still has
sufficient amplitude to contribute to the
PSD after it is filtered and aliased during
downsampling.

This process of predicting and observing
PSDs of subsequently narrower SWs
reveals the impact and importance of the choice
of digital filters employed by the receiver board.
From the measurements of
\cref{fig:predictZoom,fig:predictPSD},
one can observe first that
the noise density of the 
smaller SW always
matches the wider SW measurement at the carrier
and, second, that the PSDs exhibit a strong
upward curvature off-resonance.
The first observation -- consistency of the PSD at
the carrier frequency -- depends on the digital filter.
Here, the designers of the receiver board
have chosen
a sinc of width $2\times SW$
as the digital filter;
it has the property
of falling to zero
(having a node)
at all offsets equal
to an integer multiple of $SW$
(\ie, $n\times SW$).
Thus,
at zero offset from the carrier frequency,
the contributions from all
aliases of the filtered signal
(\ie, all aliases of the green line in 
\cref{fig:predictPSD,fig:predictZoom})
fall to zero.
For another filter to have no aliased noise
at zero offset from the carrier frequency,
it must also equal zero
for all offsets $n\times SW$.
The second observation --
the growing increase of the noise PSDs
for the more narrow SW
upon moving off-resonance
-- occurs because the sinc filter
(and therefore the aliased noise contributions)
rises smoothly away from the nodes.
With the important caveat that,
for many receiver boards, the sinc filter
is likely to be the only filter available
at a SW close to the base
sampling rate
(because boxcar averaging in the time domain can be
implemented with very few samples),
it is worth noting that the upward curvature 
of the narrow SW PSD would be
mitigated by employing an optimized FIR filter
(not shown here)
that would cause the response to
both fall abruptly to zero and remain near zero
at the edge of the SW.
Note also that, importantly to ODNP
and other quantitative methodologies,
precise quantification
of signals whose offset is a significant
fraction of the SW also requires
correction by the receiver response function,
regardless of whether a boxcar (sinc) or more optimal FIR
filter is employed.
\subsubsection{Mitigation of Noise (Observed by the Receiver)}\label{sec:SCmitigation}
\refHighResChokes
Because oversampling allows the digitization of more
noise energy over a 
smaller bandwidth,
it allows \cref{eq:convPSD}
to employ a smaller convolution filter
while still
measuring a smooth noise PSD,
as illustrated in
\cref{fig:predictZoom,fig:200kHz_chokes}.
\Cref{fig:predictZoom} shows the PSD acquired with a
1~MHz SW (blue)
using two different convolution widths: 20~kHz, which is more
appropriate for the 4~MHz PSD (\cref{fig:predictZoomOneM}),
and 5~kHz, which is more appropriate for the higher 
resolution 1~MHz (\cref{fig:predictZoomTwoHK}).
Notably,
the decrease in convolution width
reveals the presence of periodic spikes
with a spacing of 50~kHz.
As previously alluded to, 
these periodic spikes contribute to the
apparent (vertical) breadth 
of the un-convolved (transparent) lines in
\cref{fig:voltscale,fig:PSDwGain,fig:GDSvSC,fig:MagNoise,fig:cavityProbe,fig:BalVSone,fig:Chokes}.
As expected, for cases with high EMI,
this feature can also be seen as periodic spikes
in the noise signal captured by the oscilloscope
-- e.g., \cref{fig:periodicPeaks}.
However, the larger convolution widths employed at larger SW
(\eg, for the larger SW shown in
\cref{fig:GDSvSC,fig:MagNoise,fig:PSDwGain,fig:cavityProbe,fig:Chokes,fig:BalVSone,fig:voltscale})
spread these spikes into each other,
contributing to a higher PSD when they are present,
even though the individual spikes are not resolved.
These spikes, when resolved at smaller SW,
rise above a baseline that is not too
far above the amplified Johnson noise
($9\magn{-16}\;\text{W}/\text{Hz}$).

As noted previously,
the frequency distribution of the EMI does
change over the course of a day,
and the frequencies of the noise peaks in,
\eg \cref{fig:200kHz_chokes}
do change in frequency
even over the course of minutes or hours,
though their intensity tends to remain
approximately constant.
Therefore, PSD measurements
that encompass several such peaks,
as in \cref{fig:200kHz_chokes},
offer more reproducible comparisons than
PSDs acquired at smaller SWs approaching those used for
signal acquisition (as in \cref{fig:predict_4kHz}).

In keeping with the improvement shown 
in \cref{fig:Chokes},
the addition of 12 chokes
(gold $\rightarrow$ red in both
\cref{fig:Chokes,fig:200kHz_chokes})
does reduce the large noise spikes
from more than an order of magnitude above the baseline,
down to a factor of about 2 or 3 above the
baseline.
The addition of chokes even reduces the baseline
level of the PSD
(between the intense peaks).
\Cref{fig:200kHz_chokes} furthermore clarifies that
in the range of interest for the current example
(14-15 MHz),
an increase from 12 to 27 snap-on chokes does not
significantly decrease the noise;
whereas the broadband measurement in \cref{fig:Chokes}
clarifies that this does decrease the intense
noise spikes at other frequencies.

Similar to \cref{fig:Chokes},
the toroid choke removes the same
(or even slightly more) amount of EMI noise
\vs the 27 snap-on
chokes in \cref{fig:200kHz_chokes}.
Acquisition of signal points to another
benefit of the toroid:
the snap-on chokes occasionally decrease
the $T_{2}^*$ (not shown),
while the toroid
(located outside the main static magnetic field)
does not.
A final note is that cable placement and length prove
to be an essential component in mitigating EMI noise.
For example,
further PSD measurements not shown here
guided the 
shortening of the cables between the output of the 
transmitter diodes and the receiver chain,
as well as to test and secure the exact
pathway of the cable running from the
receiver chain enclosure to the probe.

When performing ODNP measurements,
the microwave resonance frequency of the cavity
effectively fixes the NMR resonance frequency.
(The cavity resonance,
    which typically cannot be adjusted,
    determines the $B_0$ field
    at which ESR resonance will be obtained,
    and therefore also determines the NMR frequency.)
Therefore,
it is important to achieve reasonable
SNR even when
the NMR resonance frequency happens
to line up with one
of the EMI spikes
(\cref{fig:ChokesFID}). 
\footnote{The signal has been averaged across the
transients of the four step phase cycle
in \cref{fig:ChokesFID},
so that the diplayed signal amplitude equals
that of a single transient,
while the noise amplitude is $1/\sqrt{4}$ the
noise amplitude of a single transient.}

\Cref{fig:ChokesFID} illustrates that
the addition of rf chokes does not affect the
overall amplitude of signal acquired on the NMR
receiver and strictly acts to reduce noise,
thus improving the overall SNR and sensitivity.
Even though the SNR of the FID without chokes (gold in
\cref{fig:ChokesFID}) seems reasonable, there is still
a clear benefit of the inclusion of rf chokes 
by the further reduction of noise (green in \cref{fig:ChokesFID}).
Unfortunately, there are 
situations in which the noise levels without mitigation
strategies in place are much higher
than those shown in \cref{fig:ChokesFID}, in which case
the signal is barely recognizable (\cref{NoisyCase}).
These instances further emphasize the practical use of
the presented protocol in finding the optimal
mitigation techniques that are suitable for the given
setup of the laboratory.
\figChokesFID
\subsection{Absolute signal intensity matches prediction}\label{sec:predictSignal}
At this point,
the noise has been
characterized with a level of quantitative
detail that allows one to test and select between 
different mitigation strategies.
While attempts to improve the absolute ODNP signal
amplitude will require the construction of
new and involved hardware beyond the scope of
this study,
this section demonstrates the aforementioned
adapted theory for predicting the 
absolute signal and breaking down the contributing
factors to further improve the signal intensity.

Only the conversion factor, $\Lambda$, must
be known to predict the absolute signal 
intensity (\cref{eq:Psig_Eff}).
This involves a simple measurement
of the power and pulse length required
to yield a 90° pulse
(\cref{eq:Vsig90t}).

In the authors' lab, the rf amplifier outputs
imperfect waveforms resulting in a pulse that
is not a perfect rectangle and thus obfuscates this
seemingly simple measurement.
Therefore,
it is necessary to calibrate 
the integral of $\sqrt{P_{tx}(t)}$
(see \cref{eq:integratedTPProduct}),
and not possible (as is typical) to simply rely
on an assumption of a rectangular pulse shape
coupled with a single measurement of
$P_{tx}$.
Here,
the $t_{90}\sqrt{P_{tx}}$ that generates
optimal excitation is $17\;\text{μs}\sqrt{\text{W}}$
(\cref{sec:t90_Ptx}),
corresponding to
$\Lambda=3.5\times 10^{2}\;\text{μT}/\sqrt{\text{W}}$
(\cref{eq:Vsig90t}) .

\figVariedTau

Based on this $\Lambda$ value and a sample
volume of 4.38~μL,
the predicted value for the square root of the signal power
(input-referred) is $3.6\;\sqrt{\text{fW}}$
(\cref{eq:Psig_Eff}).
Experiments (\cref{fig:VariedTau})
agree with this prediction fairly well.
Specifically, \cref{fig:VariedTau} shows
signal from
a series of echoes with varying
inter-pulse delays ($\tau$) with the same averaging
across the transients of the phase cycle that is
applied to \cref{fig:ChokesFID}.
The signal appears to extrapolate to the
expected value ($3.6\sqrt{\text{fW}}$)
at $\tau=0$.
Thus, a simple measurement of the $\pi/2$ pulse (in units of
$\mathrm{s\sqrt{W}}$),
can determine the signal levels fairly accurately,
as expected by the principle of reciprocity.

Having confirmed the ability of the theory to predict the absolute
signal level,
one can proceed to calculate the field distribution
factor, $\eta'$, from \cref{eq:etaFundamental}.
After a (short-open-load) calibration,
a nanoVNA (inexpensive pocket-sized VNA) measures the $Q$-factor of
the probe to be 10.1.
From this value, and the measured
$\Lambda$, \cref{eq:etaFundamental}
determines a field distribution factor of
$\eta'=7.4\times 10^{-3}$.
This number, at first,
seems unreasonably low.
However, this result is not unreasonable if
one considers the standard
hairpin loop design employed in
ODNP~\cite{Armstrong2009OveDynNuc,Franck2013QuaCwOve,Franck2019OveDynNuc,Kaminker2015ChaSixOve}.
The design consists of two loops of wire that are threaded parallel to the capillary tube,
through two teflon spacers at opposite ends of the capillary tube.
Measurements of the inductance of the hairpin loop coil
(84~nH employing a resonant RLC circuit)
reveal that the hairpin loop inductance is close to
the theoretical inductance of a
straight wire before winding into the hairpin loop structure.
Therefore, the stored magnetic field energy of stray
inductances compete with that of the coil.
Specifically, the wires and circuit board within the tuning circuit
and especially in the thin lead wires that lead from the tuning circuitry into the center of the cavity
likely generate competing inductances.
Furthermore,
because the overwhelming majority of the stored magnetic field
energy of a straight wire is stored very near to the wire,
even with the wires relatively close to the sample,
it is likely that a greater amount of magnetic field energy
is stored near to the wires rather than in the volume containing
the sample.

\figBiot

\Cref{fig:Biot} presents a very basic calculation of the fields
in the typical hairpin loop structure.
Note that extended regions closer to the wires store $1.5-1.7\times$ as much 
energy as the field flowing through the sample
(\cref{fig:BiotActual}).
The field distribution factor calculated from \cref{fig:BiotActual} is 7\%
(according to \cref{eq:FillFactorField}).
\Cref{fig:BiotDiff} demonstrates that the mutual inductance
of the wires does increase the field at the sample,
but not as much as in other regions outside the sample.
Also, because cancellation of the fields at the top and bottom
leads to less energy storage,
the finite inductance per unit length in this calculation 
(calculated from the 2D magnetic energy density per unit current integrated over a finite grid)
is actually less than the sum of inductances achieved with one of the
four wires active at a time.
Therefore, this basic calculation supports
at least one order of magnitude loss of the field distribution factor
and, more importantly,
supports the idea that the inductance of the hairpin loop is so low that it competes with single-wire inductances elsewhere in the tuning circuit.

As, by improving this surprising low value of
$\eta'$,
one could improve the signal amplitude by
$40\times$,
moving forward,
one is motivated to pursue new ODNP probe designs
that employ higher inductance coils to more
effectively concentrate field into the sample.
Admittedly, this strategy does present several challenges.
In the past,
the need to avoid perturbation of the
microwave field inside the cavity
was the primary driver for ODNP probe
designs at X-band.
However, by pointing to the fact that
$\eta'$ is orders of magnitude less than the
$Q$-factor,
the current study opens up new possibilities,
such as resorting to very fine-gauge wire
that might allow multiple loops without
perturbing the microwave field.

\section{Conclusion}
This article quantitatively
explored the status quo of the absolute signal
and absolute noise intensities in an
ODNP instrument with a standard
home-built NMR probe.
A reproducible protocol shows how to reduce
the noise density to its theoretical lower
limit (the thermal noise limit),
while a reorganization of the theory not
only demonstrates a predictive understanding
of the absolute signal level,
but -- for this case study of ODNP -- identifies
clear areas for future improvement of the SNR
by up to $40\times$.

Open-source software, paired with
very standard instrumentation and some
limited data-processing,
successfully enabled a detailed
characterization of the absolute noise
density,
as well as a measurement of the response of
the receiver.
The protocol introduced here thus identified
which new components were introduced to the
noise density by the receiver,
whether internally generated (high frequency)
components,
or aliasing arising from imperfections in the
oversampling filters.

The general strategy outlined here could be
adapted to any low-field system.
The literature abounds with interesting and novel designs,
with many new and exciting opportunities for advanced and portable magnet
systems,
for customized spectrometer circuitry,
and even for adapting open-source software and software-defined radio
hardware to develop customized NMR
transceivers~\cite{Michal2018LowMulSof,Doll2019PulConMag,Griffin1993LowNMRSpe}.
The current paper enables better cross-pollination between such
developments in different laboratories.
With specific respect to ODNP development,
it should facilitate more systematic deployment of ODNP at different fields and resonance frequencies, 
enabling a more comprehensive exploration of dynammics.





\appendix
\ifarxiv
\putbib
\end{bibunit}
\pagebreak
\onecolumngrid
\begin{center}
\makeatletter
\textbf{\large Supplemental Materials for:\\
\@title}
\makeatother
\end{center}
\twocolumngrid
\begin{bibunit}
\setcounter{equation}{0}
\setcounter{figure}{0}
\setcounter{table}{0}
\setcounter{page}{1}
\renewcommand{\theequation}{S\arabic{equation}}
\renewcommand{\thefigure}{S\arabic{figure}}
\renewcommand\thesubfigure{S\arabic{tempfigure}\alph{subfigure}}
\renewcommand{\thesection}{S\arabic{section}}
\renewcommand{\thepage}{S\arabic{page}}
\renewcommand{\bibnumfmt}[1]{[S#1]}
\renewcommand{\citenumfont}[1]{S#1}
\glsresetall
\newcommand{\maincref}[1]{\cref{#1} (main text)}
\newcommand\Csl{\texorpdfstring{\ensuremath{C_{SL}}\xspace}{[SL] }}
\newlength{\mylength}
\newlength{\mysecondlength}
\newcommand{\ubpair}[2]{%
\newcommand*{\ubpairtext}{\mbox{\begin{footnotesize} #2\end{footnotesize}}}
\settowidth{\mylength}{\ubpairtext}%
\settowidth{\mysecondlength}{\ensuremath{\displaystyle #1}}%
\setlength{\mylength}{\maxof{\mylength-\mysecondlength}{0pt}/2}
\hspace{-\mylength}%
\ensuremath{{\color{dbluecolor}\underbrace{\normalcolor #1}_{\ubpairtext}}}%
\hspace{-\mylength}%
}
\newcommand{\obpair}[2]{%
\newcommand*{\obpairtext}{\mbox{\begin{footnotesize} #2\end{footnotesize}}}
\settowidth{\mylength}{\obpairtext}%
\settowidth{\mysecondlength}{\ensuremath{\displaystyle #1}}%
\setlength{\mylength}{\maxof{\mylength-\mysecondlength}{0pt}/2}
\hspace{-\mylength}%
\ensuremath{{\color{dbluecolor}\overbrace{\normalcolor #1}^{\obpairtext}}}%
\hspace{-\mylength}%
}
\newcommand{\Cmacro}{\ensuremath{C_{macro}}\xspace}%
\newcommand{\kHH}{\ensuremath{k_{HH}}\xspace}%
\newcommand{\krhop}{\ensuremath{k_{\rho}(p)}\xspace}%
\newcommand{\Tw}{\ensuremath{T_{1,w}}\xspace}%
\newcommand{\DeltaTw}{\ensuremath{\Delta T_{1,w}}\xspace}%
\section{Supporting Information}\label{sec:SI}
\subsection{Broadband Noise PSD Summary}
The problem of DC power supplies acting as noise sources for
ODNP is not a new observation, and various individual
solutions have been implemented, frequently at the expense
of great experimental time, but the same solutions do not
apply to every unique setup in each
laboratory~\footnote{For example, the authors attempted to implement the solution that had been implemented in \cite{Franck2013QuaCwOve}, by passing the coaxial line through a pass-through BNC grounded to the waveguide, but this actually made the noise worse.}. The overarching observation
here is that performing a noise PSD measurement of
reasonable quality allows one to: (1) identify primary
sources of noise (2) systematically identify which solutions
help to mitigate those noise sources and which do not, and
(3) identify how closely a particular noise density comes to
the idealized thermal (Johnson-Nyquist) noise limit (\emph{vs.}
how much interference noise or poor noise figures dominate
the noise spectrum).

Importantly,
the measured noise PSDs
are expected to be idiosyncratic for
the particular laboratory setup where these
measurements were acquired;
it is the measurement protocol itself rather than the
particular data that is expected to prove
transferable to other laboratories and types of
experiments.

\subsection{Digitization Rates of Oscilloscope}\label{sec:optGDS}
\figOscDig
The voltscale setting of the oscilloscope dictates the vertical
scale of acquisition. Depending on the setting, the resulting 
PSD for the noise before the low pass filter threshold
may appear lower in power than actuality, while the PSD for the
noise after the threshold may appear higher (e.g. the blue line
in \cref{fig:osc_mV}). This was investigated by analyzing the 
Gaussian cummulant of unique datasets with varying voltage 
settings (\cref{fig:osc_dig}).
Notably, the Gaussian cummulant of the larger voltscales (the 
blue and green lines) are much wider and include a jump
for values about 0. Whereas the Gaussian cumulant for the
2 mV (smallest voltscale setting on the GDS) demonstrates
a typical Gaussian noise distribution and is relatively smooth 
transitioning from lower values to higher values.
When comparing the broadband noise PSDs to the narrow
noise PSDs acquired on the receiver board, this is an
important note to take into consideration as the two
PSDs will inadvertently show a mismatch in power if the
broadband measurements are acquired with a larger
voltscale.

\subsection{Receiver Calibration Factor}\label{sec:Calibration30mV}

\begin{table}
    \begin{tabular}{c|c}
        \hline
        Spectral width & Receiver Calibration\\
        (kHz) & (dg / μV) \\ \hline\hline
        4 & 476.4 \\
        40 & 727.8 \\
        200 & 583.0 \\
        1000 & 466.0 \\
        4000 & 117.7 \\
        10000 & 49.73 \\
        75000 & 6.054 \\ \hline
    \end{tabular}
    \caption{\label{tbl:calibValues}%
        The appropriate receiver calibration factor in units of
        $\text{dg}/\text{μV}$, acquired by measuring the ratio between the
        raw digitizer units and the voltage amplitude of the
        $\approx 15\;\text{MHz}$ sine wave
        ($\approx 15\;\text{mV}$) injected into receiver.
        Amplitudes determined from the fit of the
        captured sine wave, and significant figures
        reflect variance of five measurements.
        These values were later validated by plotting
        the response function of the digital filter
        of the SpinCore divided by the output voltage
        as acquired on the oscilloscope (\cref{sec:SCresponse}).
    }
\end{table}

\begin{figure}
        \centering
        \includegraphics[width=\linewidth]{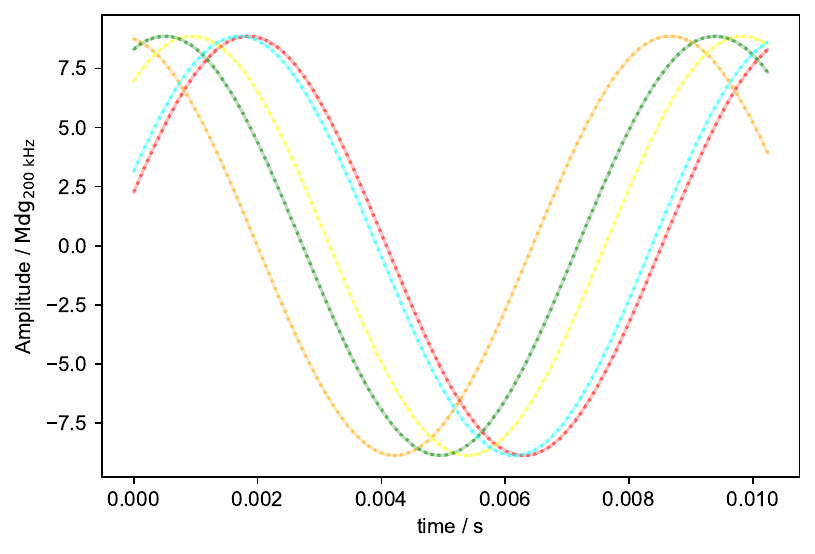}
        \caption{Test signal with a voltage amplitude of 15.21 mV
is injected into the SpinCore transciever board and 4
transients are acquired using a spectral width 
of 200 kHz. 
Each complex transient is fit to Euler's
formula to extract an amplitude of
$8.876\times10^{6}\;\text{dg}$ 
in arbitrary digitizer-specific units [dg] assigned by the transceiver
board.}
\label{fig:calibFit}
\end{figure}
The AFG injected 1 $V_{amp}$ into 2 calibrated 
attenuation assemblies followed by the receiver chain
to generate a final test signal with a voltage of 15.21
mV and a frequency of 15 MHz, which was verified first
by acquiring on the oscilloscope.
The SpinCore transceiver board then acquired the test signal 
at the set spectral widths (in the case of
\cref{fig:calibFit} - 200kHz)
with four transients. The amplitude in digunits, is
obtained by fitting each complex transient to Euler's
formula and taking the average of the fit amplitudes. 
The final receiver calibration factor is determined by
taking the ratio of the averaged amplitudes
($[\text{digunits}]$) to the real voltage value of the
injected test signal ($[\text{V}]$, measured by the
oscilloscope).

The values of \cref{tbl:calibValues}
were later validated by tracing the response
function of the SpinCore's digital filter
using the AFG source which output a sine wave
with an amplitude of 5 mV at a range of
frequencies that spanned 2X the spectral width
of the SpinCore data (\cref{afg_receiver_response}).
Each frequency acquired was averaged 25 times
and the max of the noise PSD was plotted as
a function of offset from the carrier frequency.
The resulting curve was then divided by the
PSD for each respective output frequency as
acquired on the oscilloscope.
The max of the square root of the curve (corresponding
to 0 offset from the carrier) produces the final
transceiver calibration factor for the respective
spectral width ($\mathrm{dg / \mu V}$).
These values match well with the factors
found when fitting the time domain captures of the
15.21 mV test signal to a complex wave function.

\subsection{High-frequency Noise of Transceiver}\label{sec:AttenHighF}

After observing the higher frequency oscillations in \cref{fig:GDSvSC},
the PSD was further investigated by varying the attenuation between 
the output of the receiver chain and the input of the transceiver 
board (\cref{fig:attenPSD}) and acquiring at a spectral width of 75 MHz.
This verifies that the noise does scale with the input and 
supports the idea that these oscillating peaks arise from the receiver
board itself.
\begin{figure} \centering \includegraphics[width=\linewidth]{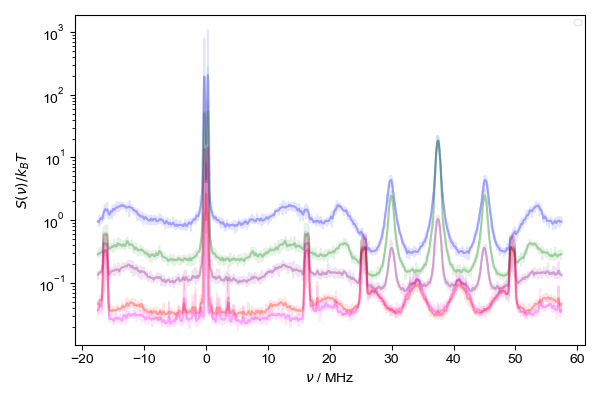} \caption{A variable attenuator 
        placed between the output of the terminated receiver chain and
        the input of the SpinCore generated attenuations of 3 dB (blue), 10
        dB (green), 13 dB (purple), and 23 dB (red).  The terminated
        SpinCore (magenta) served as a reference.  Both the noise and high
        frequency harmonics scale with attenuation up to 13 dB. At 23 dB
        the oscillations shift frequency and resemble the 
terminated SpinCore PSD in \cref{fig:GDSvSC}.}\label{fig:attenPSD}  \end{figure}

\subsection{Predicting wide spectral width PSD}
\figPredictFourK
The spectral widths with predictions are acquired 
and decreased step-wise until a spectral width of
3.9 kHz is eventually reached (figure~\ref{fig:predict_4kHz}).
As stated previously this noise peak appears to
be a standalone spike at this spectral width.
Had the wider spectral width PSDs not been taken
one might not realize that this peak arises from 
through space EMI or how to mitigate the effect
to isolate the desired signal at this zoomed
in spectral width.

\subsection{Quantification of Gain}\label{sec:gainQuant}

 The gain of the receiver pathway was calculated by
outputting a measured test signal over a range of
rf frequencies into the receiver pathway and the
amplified noise was collected on the oscilloscope (\cref{fig:Gain},
acquired with 
a variation of \cref{sec:CodeRecResp}).

\figCalcGain

The plot of measured test signal indicates a 
fairly consistent amplitude output of the AFG (\cref{fig:AFG_output}),
while the output power of the receiver chain (\cref{fig:RX_output}) shows
a clear rise in power at a frequency of 12.5 MHz.
The ratio of the output power of the receiver chain to the output
power of the AFG produces the final calculation of the gain as a
function of frequencies (\cref{fig:RX_gain}).
This trend in gain variation is most likely due to the
analog components of the receiver chain (e.g., the 
included pi circuits and tuned limiter) acting as 
additional filters. This measurement is key to properly
quantifying the signal and noise in later steps to 
calculate the input-referred signal.

\subsection{Cases of Extreme Noise}\label{NoisyCase}
\figBadFID
In a given ODNP experiment, the user often does not have
control over which NMR frequency to tune to (since
    it is largely determined by the cavity resonance
which typically cannot be adjusted). 
This puts the user in an unfortunate situation when
that frequency lays on an intense noise spike. In
extreme cases of noise the signal is buried in the
noise and may not even be noticable to the untrained
eye (\cref{fig:BadNoiseFID}).
Use of a toroidal ferrite choke and noise PSDs
(\cref{fig:BadNoisePSD}) to find the ideal positioning of
cables yields a ten-fold decrease in the noise so that
the signal clearly rises above the noise levels in the
FID.

\subsection{Deriving the Principle of Reciprocity}\label{sec:derive_recip}
\textbf{AG: this portion will need revision and relates to this slack convo:
\href{https://jmfrancklab.slack.com/archives/C06LNFNP2HM/p1724362168598509?thread_ts=1724359787.551599&cid=C06LNFNP2HM}{here}}
Faraday's law forms the basis for the calculation of the produced emf in the 
receiving coil via:
\begin{equation}
    \xi_{NMR} = \frac{d\Phi_{rcvg\ coil}}{dt}
    \label{Faraday1}
\end{equation}
where $\Phi_{rcvg\ coil}$ is the magnetic flux of the receiving coil.  Note this is
the induced emf from the magnetic moment of the spins within the sample. For
the purposes of this derivation, we replace the spins by a current loop
(\cref{fig:SI_circuit}) and the total sample magnetic moment is:
\begin{equation}
    m_{spins} = I_{spins}A_{spins}
    \label{spin_mag}
\end{equation}
where $I_{spins}$ is the current in the current loop representing the
spins, and $A_{spins}$ can be considered the area of the current loop.

\begin{figure} \centering \includegraphics[width=\linewidth]{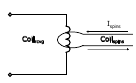} \caption{Simple
diagram of a NMR receiving coil (left) interacting with a current loop representing the 
spins of the sample that have magnetic moment $m_{spins}$ (right).}\label{fig:SI_circuit}  
\end{figure}

\Cref{Faraday1} however, introduces a problem: the $\Phi_{rcvg\ coil}$ is an unknown.
Fortunately, we can solve for it by inciting the law of mutual inductances:
\begin{equation}
\mathscr{M}_{rcvg\ coil,spins} = \mathscr{M}_{spins, rcvg\ coil}
\label{mutual_L}
\end{equation}
where the magnetic flux of the coil induced by the current flowing through 
the current loop, $\mathscr{M}_{rcvg\ coil,spins} = \frac{\Phi_{rcvg\ coil}}{I_{spins}}$
is equal to the magnetic flux of the current loop induced by the current flowing through
the receiving coil, $\mathscr{M}_{spins, rcvg\ coil} = \frac{\Phi_{spins}}{I_{rcvg\ coil}}$.
In short, the magnetic flux of one coil is induced by the field produced by
the other coil's current. 

Furthermore, by expanding $\Phi_{spins}$ we can show
\begin{equation}
\mathscr{M}_{spins, rcvg\ coil} = \frac{A_{spins}B_{at\;spins}}{I_{coil}}
\label{M_spinscoil}
\end{equation}
where $B_{at\ spins}$ is the field at the sample.
Substituting $\mathscr{M}_{rcvg\ coil,spins} = \frac{\Phi_{rcvg\ coil}}{I_{spins}}$ into
\cref{Faraday1} we get:
\begin{equation}
    \xi_{NMR} = \omega \mathscr{M}_{rcvg\ coil,spins}I_{spins}
    \label{Faraday_2}
\end{equation}

Due to the law of mutual inductances (\cref{mutual_L}), we replace
$\mathscr{M}_{rcvg\ coil, spins}$ with $\mathscr{M}_{spins, rcvg\ coil}$.
Combining this with \cref{M_spinscoil} and \cref{spin_mag},
\cref{Faraday_2} simplifies to:
\begin{equation}
    \begin{aligned}
    \xi_{NMR} = \omega\mathscr{M}_{spins, rcvg\ coil}I_{spins} \\
    = \omega\left(\frac{B_{at\ spins}}{I}\right)\left(\frac{m_{spins}}{A_{spins}}\right) \\
    = \omega\left(\frac{B_{at\ spins}}{I}\right)m_{spins}
\end{aligned}
    \label{recip_deriv}
\end{equation}
Thus, we show that the field generated at the spins is directly related to the
produced emf (the principle of reciprocity).

Alternatively, we can start with the integral form of Faraday's law:
\begin{equation}
    \xi(\textbf{r,t}) \approx -\frac{1}{4\pi}\iiint_{V}\frac{(\frac{\partial \textbf{B}(\textbf{r}',t)}{\partial t})\times (\textbf{r}-\textbf{r}')}{\left| \textbf{r}-\textbf{r}' \right| ^{3}}d^{3}\textbf{r}',
    \label{eq:FaradayInt}
\end{equation}
where $B(\textbf{r}',t)$ is the magnetic field at position $\textbf{r}'$
as a function of time. It is not hard to recognize that this equation largely 
resembles Biot-Savart's integral form:
\begin{equation}
    \textbf{B}(\textbf{r}) = \frac{\mu_{0}}{4\pi}\int{\textbf{J}(\textbf{r}')\times \frac{\textbf{r}-\textbf{r}'}{\left| \textbf{r} - \textbf{r}'\right|^{3}}}d^{3}\textbf{r}',
    \label{eq:BiotSavInt}
\end{equation}
where $\textbf{J}(\textbf{r}')$ is the current density produced
by a current flowing through the wire, and
$\textbf{r}-\textbf{r}'$ is the vector between the wire/current
and the location of field.    
If we assume that Biot-Savart's law can be transformed to 
a time dependent field (e.g. $\textbf{B}(\textbf{r'},t) =
\textbf{B}_{1}(\textbf{r})cos(\omega t)$) and plug 
\cref{eq:BiotSavInt} into \cref{eq:FaradayInt} we find a
general proof of reciprocity relating the current induced by 
a magnetic field to the induced emf.
\subsection{Nutation Curve (Measurement of $\Lambda$)}\label{sec:t90_Ptx}
\figNutation
The text notes (\cref{sec:predictSignal}), that
because the output pulse from the amplifier has a finite rise time
and an imperfect shape
in the authors' lab
(\cref{fig:NutationPulseShape}),
it is necessary to calibrate 
the integral of $\sqrt{P_{tx}(t)}$,
rather than simply relying
on an assumption of a rectangular pulse shape
coupled with a single measurement of
$P_{tx}$ (as is typical).
For calibration, the oscilloscope captures pulses of varying pulse 
lengths output from the rf amplifier (attenuated with a calibrated attenuator).
Each captured pulse is converted to analytic signal (\cref{eq:RealToAnalytic}) and the
average power calculated via \cref{eq:analyticPavg}. The final $t_{pulse}\sqrt{P_{tx}}$ 
is calculated via:
\begin{equation}
    t_{pulse}\sqrt{P_{tx}} \equiv \frac{1}{\sqrt{50}}\int
    \frac{1}{\sqrt{2}}\left| s_a(t) \right|dt.
    \label{eq:integratedTPProduct}
\end{equation}
where $t_{pulse}$ is the length of the output pulse and
50 is the resistance in $\Omega$.  
To assist in this calculation a lorentzian filter with a width 
appropriate for the measured Q value is applied prior to 
integration.
The $t_{pulse}\sqrt{P_{tx}}$ vs output pulse length is fit to a polynomial so that
the desired $t_{90}\sqrt{P_{tx}}$ is obtained by extrapolating the proper pulse 
length from the calibration curve.

Armed with this calibration table, a simple nutation experiment
can then employ linearly spaced $t_{pulse}\sqrt{P_{tx}}$ values.
\Cref{fig:NutationDCCT} shows the raw data of the nutation
experiment as a domain colored coherence transfer pathway plot
that illustrates a clear inversion from red to blue (indicating
the $t_{180}\sqrt{P_{tx,rms}}$) at around
$34\;\text{μs}\sqrt{\text{W}}$.
This data is integrated along the time axis to yield
\cref{fig:NutationOneD} verifying the optimal (90° tip)
$t_{90}\sqrt{P_{tx}}$ at $17\;\mathrm{\mu s\sqrt{W}}$.

\subsection{Probe Circuit Diagrams}

In addition to building the 'single-sided' probe containing
the traditional tank circuit used
for most NMR probes, we also constructed a second balanced probe 
with the aim of concentrating 
the current within the circuit near the coil.
\cite{Mispelter2006NMRProBio} explains that this
design has the advantage of mitigating the
antenna effect by almost equalizing the impedance
to ground on either end of the coil. Both
circuits are shown in \cref{fig:ProbeCircuits}.
All parts used to build the circuits were purchased
from DigiKey and minicircuits.
The balanced probe was also built
to accommodate shim coils between the magnets 
shown in \cref{fig:hardwareSetup} and therefore 
the tuning box containing the circuit is both
slimmer and the frame itself is thinner compared
to the single-sided probe. Therefore, while the
balanced probe bears advantages in circuitry, the
single-sided probe bears the advantage of a thicker
encasement and therefore better shielding.

\begin{figure}
\hypertarget{fig:ProbeCircuits}{%
\centering
\includegraphics[width=\linewidth]{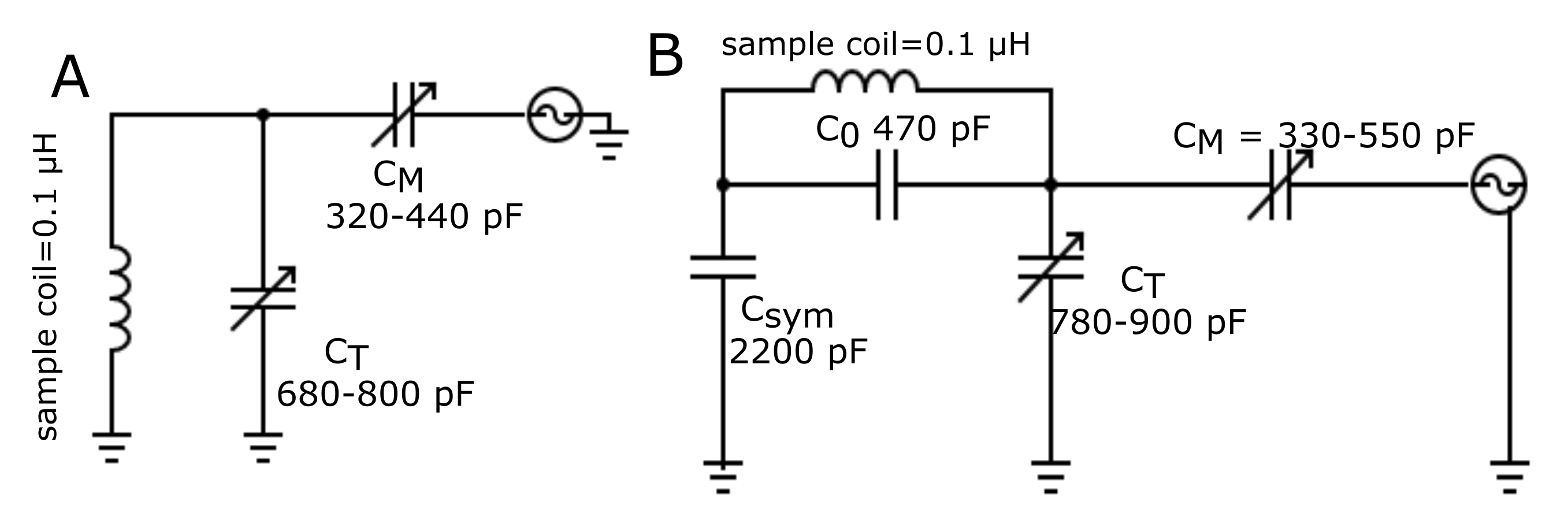}
\caption{Comparison of the two probe circuits.
A) The ``single sided'' tank circuit
consists of a tuning capacitance ($C_T$) and
matching capacitance ($C_M$).  Both include
120~pF variable capacitors in order to
span a range of frequencies.
The double hairpin coils for both probes have a volume of
8.55~$\mu\)L and an inductance of
\(\sim 0.1\;\mu\text{H}$ measured with a miniVNA.
B) The circuitry used for the balanced probe
includes two additional capacitors.
}\label{fig:ProbeCircuits}
} \end{figure} 

\subsection{Acquisition Scripts}\label{sec:acqScripts}
When utilizing the oscilloscope, the instrument
itself takes raw RS232 commands that utilize
\texttt{write}. All other commands, properties
and attributes were made in-house using 
\cite{InstNotes}. 
In the stage of calibrating the test signal,
the oscilloscope captures a single capture of the waveform 
which is converted to 
the analytic and a rough frequency filter is applied to display
the actual voltage amplitude using the following script:
{\footnotesize%
\par\noindent
\rule{\linewidth}{0.5pt}
\par\noindent
\vspace*{-5ex}
\par\noindent
\rule{\linewidth}{0.5pt}\begin{Verbatim}[commandchars=\\\{\},codes={\catcode`\$=3\catcode`\^=7\catcode`\_=8\relax}]
\PY{k+kn}{from} \PY{n+nn}{Instruments} \PY{k+kn}{import} \PY{n}{GDS\PYZus{}scope}
\PY{k+kn}{from} \PY{n+nn}{pyspecdata} \PY{k+kn}{import} \PY{n}{figlist\PYZus{}var}
\PY{k+kn}{from} \PY{n+nn}{matplotlib}\PY{n+nn}{.}\PY{n+nn}{pyplot} \PY{k+kn}{import} \PY{n}{text}\PY{p}{,} \PY{n}{gca}

\PY{n}{expected\PYZus{}Vamp} \PY{o}{=} \PY{l+m+mf}{15e\PYZhy{}3}
\PY{n}{filter\PYZus{}width\PYZus{}Hz} \PY{o}{=} \PY{l+m+mf}{10e6}  \PY{c+c1}{\PYZsh{} 10 MHz filter about the average}
\PY{c+c1}{\PYZsh{}                      frequency}
\PY{k}{with} \PY{n}{figlist\PYZus{}var}\PY{p}{(}\PY{p}{)} \PY{k}{as} \PY{n}{fl}\PY{p}{:}
    \PY{k}{with} \PY{n}{GDS\PYZus{}scope}\PY{p}{(}\PY{p}{)} \PY{k}{as} \PY{n}{g}\PY{p}{:}
        \PY{n}{g}\PY{o}{.}\PY{n}{reset}\PY{p}{(}\PY{p}{)}
        \PY{c+c1}{\PYZsh{} \PYZob{}\PYZob{}\PYZob{} display settings \PYZhy{} use channel 2}
        \PY{n}{g}\PY{o}{.}\PY{n}{CH1}\PY{o}{.}\PY{n}{disp} \PY{o}{=} \PY{k+kc}{False}
        \PY{n}{g}\PY{o}{.}\PY{n}{CH2}\PY{o}{.}\PY{n}{disp} \PY{o}{=} \PY{k+kc}{True}
        \PY{n}{g}\PY{o}{.}\PY{n}{CH3}\PY{o}{.}\PY{n}{disp} \PY{o}{=} \PY{k+kc}{False}
        \PY{c+c1}{\PYZsh{} \PYZcb{}\PYZcb{}\PYZcb{}}
        \PY{c+c1}{\PYZsh{} \PYZob{}\PYZob{}\PYZob{} voltage scale and acquisition settings}
        \PY{n}{g}\PY{o}{.}\PY{n}{CH2}\PY{o}{.}\PY{n}{voltscal} \PY{o}{=} \PY{p}{(}
            \PY{n}{expected\PYZus{}Vamp} \PY{o}{*} \PY{l+m+mf}{1.1} \PY{o}{/} \PY{l+m+mi}{4}
        \PY{p}{)}  \PY{c+c1}{\PYZsh{} set to a little more than $\frac{V_{amp}}{4}$}
        \PY{n}{g}\PY{o}{.}\PY{n}{timscal}\PY{p}{(}\PY{l+m+mf}{10e\PYZhy{}9}\PY{p}{,} \PY{n}{pos}\PY{o}{=}\PY{l+m+mi}{0}\PY{p}{)}
        \PY{n}{g}\PY{o}{.}\PY{n}{write}\PY{p}{(}\PY{l+s+s2}{\PYZdq{}}\PY{l+s+s2}{:TIM:MOD WIND}\PY{l+s+s2}{\PYZdq{}}\PY{p}{)}
        \PY{n}{g}\PY{o}{.}\PY{n}{write}\PY{p}{(}
            \PY{l+s+s2}{\PYZdq{}}\PY{l+s+s2}{:CHAN2:IMP 5.0E+1}\PY{l+s+s2}{\PYZdq{}}
        \PY{p}{)}  \PY{c+c1}{\PYZsh{} set impedance to 50 ohms}
        \PY{n}{g}\PY{o}{.}\PY{n}{write}\PY{p}{(}
            \PY{l+s+s2}{\PYZdq{}}\PY{l+s+s2}{:TRIG:SOUR CH2}\PY{l+s+s2}{\PYZdq{}}
        \PY{p}{)}  \PY{c+c1}{\PYZsh{} set the source of the trigger to channel 2}
        \PY{n}{g}\PY{o}{.}\PY{n}{write}\PY{p}{(}
            \PY{l+s+s2}{\PYZdq{}}\PY{l+s+s2}{:TRIG:MOD AUTO}\PY{l+s+s2}{\PYZdq{}}
        \PY{p}{)}  \PY{c+c1}{\PYZsh{} set trigger to auto}
        \PY{n}{g}\PY{o}{.}\PY{n}{write}\PY{p}{(}\PY{l+s+s2}{\PYZdq{}}\PY{l+s+s2}{:ACQ:MOD HIR}\PY{l+s+s2}{\PYZdq{}}\PY{p}{)}  \PY{c+c1}{\PYZsh{} high vertical res.}
        \PY{c+c1}{\PYZsh{} \PYZcb{}\PYZcb{}\PYZcb{}}
        \PY{c+c1}{\PYZsh{} \PYZob{}\PYZob{}\PYZob{} set horizontal cursors on oscilloscope display}
        \PY{n}{g}\PY{o}{.}\PY{n}{write}\PY{p}{(}
            \PY{l+s+s2}{\PYZdq{}}\PY{l+s+s2}{:CURS:MOD HV}\PY{l+s+s2}{\PYZdq{}}
        \PY{p}{)}  \PY{c+c1}{\PYZsh{} set horizontal cursors}
        \PY{n}{g}\PY{o}{.}\PY{n}{write}\PY{p}{(}
            \PY{l+s+s2}{\PYZdq{}}\PY{l+s+s2}{:CURS:SOUR CH2}\PY{l+s+s2}{\PYZdq{}}
        \PY{p}{)}  \PY{c+c1}{\PYZsh{} cursors pertain to channel 2}
        \PY{c+c1}{\PYZsh{} \PYZcb{}\PYZcb{}\PYZcb{}}
        \PY{c+c1}{\PYZsh{} \PYZob{}\PYZob{}\PYZob{} use expected amplitude to set initial}
        \PY{c+c1}{\PYZsh{}     position of cursors}
        \PY{n}{g}\PY{o}{.}\PY{n}{write}\PY{p}{(}
            \PY{l+s+s2}{\PYZdq{}}\PY{l+s+s2}{:CURS:V1P }\PY{l+s+s2}{\PYZdq{}}
            \PY{o}{+} \PY{p}{(}\PY{l+s+s2}{\PYZdq{}}\PY{l+s+si}{\PYZpc{}0.2e}\PY{l+s+s2}{\PYZdq{}} \PY{o}{\PYZpc{}} \PY{n}{expected\PYZus{}Vamp}\PY{p}{)}\PY{o}{.}\PY{n}{replace}\PY{p}{(}
                \PY{l+s+s2}{\PYZdq{}}\PY{l+s+s2}{e}\PY{l+s+s2}{\PYZdq{}}\PY{p}{,} \PY{l+s+s2}{\PYZdq{}}\PY{l+s+s2}{E}\PY{l+s+s2}{\PYZdq{}}
            \PY{p}{)}
        \PY{p}{)}
        \PY{n}{g}\PY{o}{.}\PY{n}{write}\PY{p}{(}
            \PY{l+s+s2}{\PYZdq{}}\PY{l+s+s2}{:CURS:V2P }\PY{l+s+s2}{\PYZdq{}}
            \PY{o}{+} \PY{p}{(}\PY{l+s+s2}{\PYZdq{}}\PY{l+s+si}{\PYZpc{}0.2e}\PY{l+s+s2}{\PYZdq{}} \PY{o}{\PYZpc{}} \PY{o}{\PYZhy{}}\PY{n}{expected\PYZus{}Vamp}\PY{p}{)}\PY{o}{.}\PY{n}{replace}\PY{p}{(}
                \PY{l+s+s2}{\PYZdq{}}\PY{l+s+s2}{e}\PY{l+s+s2}{\PYZdq{}}\PY{p}{,} \PY{l+s+s2}{\PYZdq{}}\PY{l+s+s2}{E}\PY{l+s+s2}{\PYZdq{}}
            \PY{p}{)}
        \PY{p}{)}
        \PY{c+c1}{\PYZsh{} \PYZcb{}\PYZcb{}\PYZcb{}}
        \PY{c+c1}{\PYZsh{} \PYZob{}\PYZob{}\PYZob{} grab waveform from oscilloscope}
        \PY{n}{g}\PY{o}{.}\PY{n}{write}\PY{p}{(}\PY{l+s+s2}{\PYZdq{}}\PY{l+s+s2}{:SING}\PY{l+s+s2}{\PYZdq{}}\PY{p}{)}  \PY{c+c1}{\PYZsh{} capture single acquisition}
        \PY{n}{data} \PY{o}{=} \PY{n}{g}\PY{o}{.}\PY{n}{waveform}\PY{p}{(}\PY{n}{ch}\PY{o}{=}\PY{l+m+mi}{2}\PY{p}{)}
        \PY{k}{assert} \PY{n}{data}\PY{o}{.}\PY{n}{get\PYZus{}units}\PY{p}{(}\PY{l+s+s2}{\PYZdq{}}\PY{l+s+s2}{t}\PY{l+s+s2}{\PYZdq{}}\PY{p}{)} \PY{o}{==} \PY{l+s+s2}{\PYZdq{}}\PY{l+s+s2}{s}\PY{l+s+s2}{\PYZdq{}}
        \PY{c+c1}{\PYZsh{} \PYZcb{}\PYZcb{}\PYZcb{}}
    \PY{n}{fl}\PY{o}{.}\PY{n}{next}\PY{p}{(}\PY{l+s+s2}{\PYZdq{}}\PY{l+s+s2}{Signal captured on GDS oscilloscope}\PY{l+s+s2}{\PYZdq{}}\PY{p}{)}
    \PY{n}{fl}\PY{o}{.}\PY{n}{plot}\PY{p}{(}\PY{n}{data}\PY{p}{,} \PY{n}{label}\PY{o}{=}\PY{l+s+s2}{\PYZdq{}}\PY{l+s+s2}{raw signal}\PY{l+s+s2}{\PYZdq{}}\PY{p}{)}
    \PY{c+c1}{\PYZsh{} \PYZob{}\PYZob{}\PYZob{} convert to analytic signal}
    \PY{n}{data}\PY{o}{.}\PY{n}{ft}\PY{p}{(}\PY{l+s+s2}{\PYZdq{}}\PY{l+s+s2}{t}\PY{l+s+s2}{\PYZdq{}}\PY{p}{,} \PY{n}{shift}\PY{o}{=}\PY{k+kc}{True}\PY{p}{)}
    \PY{n}{data} \PY{o}{=} \PY{n}{data}\PY{p}{[}\PY{l+s+s2}{\PYZdq{}}\PY{l+s+s2}{t}\PY{l+s+s2}{\PYZdq{}}\PY{p}{:}\PY{p}{(}\PY{l+m+mi}{0}\PY{p}{,} \PY{k+kc}{None}\PY{p}{)}\PY{p}{]}
    \PY{n}{data} \PY{o}{*}\PY{o}{=} \PY{l+m+mi}{2}
    \PY{n}{data}\PY{p}{[}\PY{l+s+s2}{\PYZdq{}}\PY{l+s+s2}{t}\PY{l+s+s2}{\PYZdq{}}\PY{p}{,} \PY{l+m+mi}{0}\PY{p}{]} \PY{o}{*}\PY{o}{=} \PY{l+m+mf}{0.5}
    \PY{n}{data}\PY{o}{.}\PY{n}{ift}\PY{p}{(}\PY{l+s+s2}{\PYZdq{}}\PY{l+s+s2}{t}\PY{l+s+s2}{\PYZdq{}}\PY{p}{)}
    \PY{c+c1}{\PYZsh{} \PYZcb{}\PYZcb{}\PYZcb{}}
    \PY{n}{fl}\PY{o}{.}\PY{n}{plot}\PY{p}{(}\PY{n+nb}{abs}\PY{p}{(}\PY{n}{data}\PY{p}{)}\PY{p}{,} \PY{n}{label}\PY{o}{=}\PY{l+s+s2}{\PYZdq{}}\PY{l+s+s2}{analytic signal}\PY{l+s+s2}{\PYZdq{}}\PY{p}{)}
    \PY{n}{frq} \PY{o}{=} \PY{p}{(}
        \PY{n}{data}\PY{o}{.}\PY{n}{C}\PY{o}{.}\PY{n}{phdiff}\PY{p}{(}\PY{l+s+s2}{\PYZdq{}}\PY{l+s+s2}{t}\PY{l+s+s2}{\PYZdq{}}\PY{p}{,} \PY{n}{return\PYZus{}error}\PY{o}{=}\PY{k+kc}{False}\PY{p}{)}
        \PY{o}{.}\PY{n}{mean}\PY{p}{(}\PY{l+s+s2}{\PYZdq{}}\PY{l+s+s2}{t}\PY{l+s+s2}{\PYZdq{}}\PY{p}{)}
        \PY{o}{.}\PY{n}{item}\PY{p}{(}\PY{p}{)}
    \PY{p}{)} \PY{c+c1}{\PYZsh{}calculate average frequency of signal}
    \PY{c+c1}{\PYZsh{} \PYZob{}\PYZob{}\PYZob{} now, filter the signal about the average}
    \PY{n}{data}\PY{o}{.}\PY{n}{ft}\PY{p}{(}\PY{l+s+s2}{\PYZdq{}}\PY{l+s+s2}{t}\PY{l+s+s2}{\PYZdq{}}\PY{p}{)}
    \PY{n}{data}\PY{p}{[}\PY{l+s+s2}{\PYZdq{}}\PY{l+s+s2}{t}\PY{l+s+s2}{\PYZdq{}} \PY{p}{:} \PY{p}{(}\PY{l+m+mi}{0}\PY{p}{,} \PY{n}{frq} \PY{o}{\PYZhy{}} \PY{n}{filter\PYZus{}width\PYZus{}Hz} \PY{o}{/} \PY{l+m+mi}{2}\PY{p}{)}\PY{p}{]} \PY{o}{=} \PY{l+m+mi}{0}
    \PY{n}{data}\PY{p}{[}\PY{l+s+s2}{\PYZdq{}}\PY{l+s+s2}{t}\PY{l+s+s2}{\PYZdq{}} \PY{p}{:} \PY{p}{(}\PY{n}{frq} \PY{o}{+} \PY{n}{filter\PYZus{}width\PYZus{}Hz} \PY{o}{/} \PY{l+m+mi}{2}\PY{p}{,} \PY{k+kc}{None}\PY{p}{)}\PY{p}{]} \PY{o}{=} \PY{l+m+mi}{0}
    \PY{n}{data}\PY{o}{.}\PY{n}{ift}\PY{p}{(}\PY{l+s+s2}{\PYZdq{}}\PY{l+s+s2}{t}\PY{l+s+s2}{\PYZdq{}}\PY{p}{)}
    \PY{c+c1}{\PYZsh{} \PYZcb{}\PYZcb{}\PYZcb{}}
    \PY{n}{fl}\PY{o}{.}\PY{n}{plot}\PY{p}{(}\PY{n}{data}\PY{p}{,} \PY{n}{label}\PY{o}{=}\PY{l+s+s2}{\PYZdq{}}\PY{l+s+s2}{filtered analytic signal}\PY{l+s+s2}{\PYZdq{}}\PY{p}{)}
    \PY{n}{fl}\PY{o}{.}\PY{n}{plot}\PY{p}{(}
        \PY{n+nb}{abs}\PY{p}{(}\PY{n}{data}\PY{p}{)}\PY{p}{,}
        \PY{n}{label}\PY{o}{=}\PY{l+s+s2}{\PYZdq{}}\PY{l+s+s2}{filtered analytic signal}\PY{l+s+s2}{\PYZdq{}}\PY{p}{,}
    \PY{p}{)}
    \PY{n}{Vamp} \PY{o}{=} \PY{p}{(}
        \PY{n+nb}{abs}\PY{p}{(}\PY{n}{data}\PY{p}{[}\PY{l+s+s2}{\PYZdq{}}\PY{l+s+s2}{t}\PY{l+s+s2}{\PYZdq{}}\PY{p}{:}\PY{p}{(}\PY{l+m+mf}{1e\PYZhy{}6}\PY{p}{,} \PY{l+m+mf}{4e\PYZhy{}6}\PY{p}{)}\PY{p}{]}\PY{p}{)}
        \PY{o}{.}\PY{n}{mean}\PY{p}{(}\PY{l+s+s2}{\PYZdq{}}\PY{l+s+s2}{t}\PY{l+s+s2}{\PYZdq{}}\PY{p}{)}
        \PY{o}{.}\PY{n}{real}\PY{o}{.}\PY{n}{item}\PY{p}{(}\PY{p}{)}
    \PY{p}{)}
    \PY{n}{text}\PY{p}{(}
        \PY{l+m+mf}{0.5}\PY{p}{,}
        \PY{l+m+mf}{0.05}\PY{p}{,}
        \PY{n}{s}\PY{o}{=}\PY{l+s+s2}{\PYZdq{}}\PY{l+s+s2}{\PYZdl{}V\PYZus{}}\PY{l+s+si}{\PYZob{}amp\PYZcb{}}\PY{l+s+s2}{ = }\PY{l+s+si}{\PYZpc{}0.6f}\PY{l+s+s2}{\PYZdl{} mV}\PY{l+s+s2}{\PYZdq{}} \PY{o}{\PYZpc{}} \PY{p}{(}\PY{n}{Vamp} \PY{o}{/} \PY{l+m+mf}{1e\PYZhy{}3}\PY{p}{)}\PY{p}{,}
        \PY{n}{transform}\PY{o}{=}\PY{n}{gca}\PY{p}{(}\PY{p}{)}\PY{o}{.}\PY{n}{transAxes}\PY{p}{,}
    \PY{p}{)}
\end{Verbatim}
\par\noindent
\rule{\linewidth}{0.5pt}
\label{single_GDS_acq}
}
\subsubsection{Capturing Noise Density}
In order to adequately capture the noise density on
the oscilloscope, the user must first manually adjust
the desired settings (voltscale, sampling rate, and trigger)
on the oscilloscope interface. The actual acquisition of 
the repeated captures is then acquired using the following
script:
{\footnotesize%
\par\noindent
\rule{\linewidth}{0.5pt}
\par\noindent
\vspace*{-5ex}
\par\noindent
\rule{\linewidth}{0.5pt}\begin{Verbatim}[commandchars=\\\{\},codes={\catcode`\$=3\catcode`\^=7\catcode`\_=8\relax}]
\PY{k+kn}{from} \PY{n+nn}{datetime} \PY{k+kn}{import} \PY{n}{datetime}
\PY{k+kn}{from} \PY{n+nn}{Instruments} \PY{k+kn}{import} \PY{n}{GDS\PYZus{}scope}
\PY{k+kn}{import} \PY{n+nn}{sys}
\PY{k+kn}{from} \PY{n+nn}{pyspecdata} \PY{k+kn}{import} \PY{n}{getDATADIR}
\PY{k+kn}{from} \PY{n+nn}{numpy} \PY{k+kn}{import} \PY{n}{float64}
\PY{k+kn}{import} \PY{n+nn}{time}


\PY{c+c1}{\PYZsh{} \PYZob{}\PYZob{}\PYZob{} Function for acquiring data}
\PY{k}{def} \PY{n+nf}{collect}\PY{p}{(}\PY{n}{file\PYZus{}string}\PY{p}{,} \PY{n}{N\PYZus{}capture}\PY{p}{)}\PY{p}{:}
    \PY{k}{with} \PY{n}{GDS\PYZus{}scope}\PY{p}{(}\PY{p}{)} \PY{k}{as} \PY{n}{g}\PY{p}{:}
        \PY{k}{for} \PY{n}{x} \PY{o+ow}{in} \PY{n+nb}{range}\PY{p}{(}\PY{n}{N\PYZus{}capture}\PY{p}{)}\PY{p}{:}
            \PY{n}{data} \PY{o}{=} \PY{n}{g}\PY{o}{.}\PY{n}{waveform}\PY{p}{(}
                \PY{n}{ch}\PY{o}{=}\PY{l+m+mi}{2}
            \PY{p}{)}  \PY{c+c1}{\PYZsh{} take the waveform from channel 2}
            \PY{k}{if} \PY{n}{x} \PY{o}{==} \PY{l+m+mi}{0}\PY{p}{:}
                \PY{n}{s} \PY{o}{=} \PY{p}{(}
                    \PY{n}{data}\PY{o}{.}\PY{n}{shape} \PY{o}{+} \PY{p}{(}\PY{l+s+s2}{\PYZdq{}}\PY{l+s+s2}{capture}\PY{l+s+s2}{\PYZdq{}}\PY{p}{,} \PY{n}{N\PYZus{}capture}\PY{p}{)}
                \PY{p}{)}\PY{o}{.}\PY{n}{alloc}\PY{p}{(}
                    \PY{n}{dtype}\PY{o}{=}\PY{n}{float64}
                \PY{p}{)}  \PY{c+c1}{\PYZsh{}  allocate an array that\PYZsq{}s shaped}
                \PY{c+c1}{\PYZsh{}    like a single capture,}
                \PY{c+c1}{\PYZsh{}    but with an additional \PYZdq{}capture\PYZdq{}}
                \PY{c+c1}{\PYZsh{}    dimension}
                \PY{n}{s}\PY{p}{[}\PY{l+s+s2}{\PYZdq{}}\PY{l+s+s2}{t}\PY{l+s+s2}{\PYZdq{}}\PY{p}{]} \PY{o}{=} \PY{n}{data}\PY{o}{.}\PY{n}{getaxis}\PY{p}{(}\PY{l+s+s2}{\PYZdq{}}\PY{l+s+s2}{t}\PY{l+s+s2}{\PYZdq{}}\PY{p}{)}
                \PY{n}{s}\PY{o}{.}\PY{n}{set\PYZus{}units}\PY{p}{(}\PY{l+s+s2}{\PYZdq{}}\PY{l+s+s2}{t}\PY{l+s+s2}{\PYZdq{}}\PY{p}{,} \PY{n}{data}\PY{o}{.}\PY{n}{get\PYZus{}units}\PY{p}{(}\PY{l+s+s2}{\PYZdq{}}\PY{l+s+s2}{t}\PY{l+s+s2}{\PYZdq{}}\PY{p}{)}\PY{p}{)}
            \PY{n}{s}\PY{p}{[}
                \PY{l+s+s2}{\PYZdq{}}\PY{l+s+s2}{capture}\PY{l+s+s2}{\PYZdq{}}\PY{p}{,} \PY{n}{x}
            \PY{p}{]} \PY{o}{=} \PY{n}{data}  \PY{c+c1}{\PYZsh{} store data for the capture in}
            \PY{c+c1}{\PYZsh{}           the appropriate index}
            \PY{n}{time}\PY{o}{.}\PY{n}{sleep}\PY{p}{(}\PY{l+m+mi}{1}\PY{p}{)}
    \PY{n}{s}\PY{o}{.}\PY{n}{setaxis}\PY{p}{(}
        \PY{l+s+s2}{\PYZdq{}}\PY{l+s+s2}{capture}\PY{l+s+s2}{\PYZdq{}}\PY{p}{,} \PY{l+s+s2}{\PYZdq{}}\PY{l+s+s2}{\PYZsh{}}\PY{l+s+s2}{\PYZdq{}}
    \PY{p}{)}  \PY{c+c1}{\PYZsh{} just set to a series of integers}
    \PY{n}{date} \PY{o}{=} \PY{n}{datetime}\PY{o}{.}\PY{n}{now}\PY{p}{(}\PY{p}{)}\PY{o}{.}\PY{n}{strftime}\PY{p}{(}\PY{l+s+s2}{\PYZdq{}}\PY{l+s+s2}{\PYZpc{}}\PY{l+s+s2}{y}\PY{l+s+s2}{\PYZpc{}}\PY{l+s+s2}{m}\PY{l+s+si}{\PYZpc{}d}\PY{l+s+s2}{\PYZdq{}}\PY{p}{)}
    \PY{n}{s}\PY{o}{.}\PY{n}{name}\PY{p}{(}
        \PY{l+s+s2}{\PYZdq{}}\PY{l+s+s2}{accumulated\PYZus{}}\PY{l+s+s2}{\PYZdq{}} \PY{o}{+} \PY{n}{date}
    \PY{p}{)}  \PY{c+c1}{\PYZsh{} nodename for h5 file}
    \PY{n}{s}\PY{o}{.}\PY{n}{hdf5\PYZus{}write}\PY{p}{(}
        \PY{n}{date} \PY{o}{+} \PY{l+s+s2}{\PYZdq{}}\PY{l+s+s2}{\PYZus{}}\PY{l+s+s2}{\PYZdq{}} \PY{o}{+} \PY{n}{file\PYZus{}string} \PY{o}{+} \PY{l+s+s2}{\PYZdq{}}\PY{l+s+s2}{.h5}\PY{l+s+s2}{\PYZdq{}}\PY{p}{,}
        \PY{n}{directory}\PY{o}{=}\PY{n}{getDATADIR}\PY{p}{(}  \PY{c+c1}{\PYZsh{} store in a directory}
            \PY{c+c1}{\PYZsh{} known to pyspecdata so}
            \PY{c+c1}{\PYZsh{} that storage location}
            \PY{c+c1}{\PYZsh{} doesn\PYZsq{}t depend on file}
            \PY{c+c1}{\PYZsh{} structure of computer it}
            \PY{c+c1}{\PYZsh{} is run on}
            \PY{n}{exp\PYZus{}type}\PY{o}{=}\PY{l+s+s2}{\PYZdq{}}\PY{l+s+s2}{ODNP\PYZus{}NMR\PYZus{}comp/noise\PYZus{}tests}\PY{l+s+s2}{\PYZdq{}}
        \PY{p}{)}\PY{p}{,}
    \PY{p}{)}
    \PY{k}{return}


\PY{c+c1}{\PYZsh{} \PYZcb{}\PYZcb{}\PYZcb{}}
\PY{k}{def} \PY{n+nf}{raise\PYZus{}arg\PYZus{}error}\PY{p}{(}\PY{p}{)}\PY{p}{:}
    \PY{k}{raise} \PY{n+ne}{ValueError}\PY{p}{(}
\PY{+w}{        }\PY{l+s+sd}{\PYZdq{}\PYZdq{}\PYZdq{}call like this:}

\PY{l+s+sd}{        python collect\PYZus{}GDS.py file\PYZus{}string N\PYZus{}capture}
\PY{l+s+sd}{        (where file\PYZus{}string is added to the filename}
\PY{l+s+sd}{        along with the date, and N\PYZus{}capture is the}
\PY{l+s+sd}{        number of captures you want to acquire)}
\PY{l+s+sd}{        \PYZdq{}\PYZdq{}\PYZdq{}}
    \PY{p}{)}


\PY{k}{if} \PY{n+nv+vm}{\PYZus{}\PYZus{}name\PYZus{}\PYZus{}} \PY{o}{==} \PY{l+s+s2}{\PYZdq{}}\PY{l+s+s2}{\PYZus{}\PYZus{}main\PYZus{}\PYZus{}}\PY{l+s+s2}{\PYZdq{}}\PY{p}{:}
    \PY{k}{if} \PY{n+nb}{len}\PY{p}{(}\PY{n}{sys}\PY{o}{.}\PY{n}{argv}\PY{p}{)} \PY{o}{\PYZlt{}} \PY{l+m+mi}{3}\PY{p}{:}
        \PY{n}{raise\PYZus{}arg\PYZus{}error}\PY{p}{(}\PY{p}{)}
    \PY{n}{file\PYZus{}string}\PY{p}{,} \PY{n}{N\PYZus{}capture} \PY{o}{=} \PY{n}{sys}\PY{o}{.}\PY{n}{argv}\PY{p}{[}
        \PY{l+m+mi}{1}\PY{p}{:}\PY{l+m+mi}{3}
    \PY{p}{]}  \PY{c+c1}{\PYZsh{} the filename will be the date followed by the}
    \PY{c+c1}{\PYZsh{}    first argument in the terminal}
    \PY{k}{try}\PY{p}{:}
        \PY{n}{N\PYZus{}capture} \PY{o}{=} \PY{n+nb}{int}\PY{p}{(}\PY{n}{N\PYZus{}capture}\PY{p}{)}
    \PY{k}{except}\PY{p}{:}
        \PY{n}{raise\PYZus{}arg\PYZus{}error}\PY{p}{(}\PY{p}{)}
    \PY{n}{collect}\PY{p}{(}
        \PY{n}{file\PYZus{}string}\PY{p}{,} \PY{n+nb}{int}\PY{p}{(}\PY{n}{N\PYZus{}capture}\PY{p}{)}
    \PY{p}{)}  \PY{c+c1}{\PYZsh{} acquire the data}
\end{Verbatim}
\par\noindent
\rule{\linewidth}{0.5pt}
\label{GDS_acq}
}

Unlike the oscilloscope the SpinCore 
has no need for manual set up and simply 
requires the user to input the
number of averages, spectral width
and carrier into the following script before
running:
{\footnotesize%
\par\noindent
\rule{\linewidth}{0.5pt}
\par\noindent
\vspace*{-5ex}
\par\noindent
\rule{\linewidth}{0.5pt}\begin{Verbatim}[commandchars=\\\{\},codes={\catcode`\$=3\catcode`\^=7\catcode`\_=8\relax}]
\PY{k+kn}{import} \PY{n+nn}{pyspecdata} \PY{k}{as} \PY{n+nn}{ps}
\PY{k+kn}{import} \PY{n+nn}{numpy} \PY{k}{as} \PY{n+nn}{np}
\PY{k+kn}{from} \PY{n+nn}{numpy} \PY{k+kn}{import} \PY{n}{r\PYZus{}}
\PY{k+kn}{import} \PY{n+nn}{SpinCore\PYZus{}pp} \PY{k}{as} \PY{n+nn}{sc}
\PY{k+kn}{from} \PY{n+nn}{datetime} \PY{k+kn}{import} \PY{n}{datetime}

\PY{c+c1}{\PYZsh{} \PYZob{}\PYZob{}\PYZob{} set filename, spectral width and carrier}
\PY{n}{date} \PY{o}{=} \PY{n}{datetime}\PY{o}{.}\PY{n}{now}\PY{p}{(}\PY{p}{)}\PY{o}{.}\PY{n}{strftime}\PY{p}{(}\PY{l+s+s2}{\PYZdq{}}\PY{l+s+s2}{\PYZpc{}}\PY{l+s+s2}{y}\PY{l+s+s2}{\PYZpc{}}\PY{l+s+s2}{m}\PY{l+s+si}{\PYZpc{}d}\PY{l+s+s2}{\PYZdq{}}\PY{p}{)}
\PY{n}{description} \PY{o}{=} \PY{l+s+s2}{\PYZdq{}}\PY{l+s+s2}{SpinCore\PYZus{}noise}\PY{l+s+s2}{\PYZdq{}}
\PY{n}{SW\PYZus{}kHz} \PY{o}{=} \PY{l+m+mi}{75000}
\PY{n}{output\PYZus{}name} \PY{o}{=} \PY{p}{(}\PY{n}{date} \PY{o}{+} \PY{l+s+s2}{\PYZdq{}}\PY{l+s+s2}{\PYZus{}}\PY{l+s+s2}{\PYZdq{}} \PY{o}{+} \PY{n}{description} \PY{o}{+} \PY{l+s+s2}{\PYZdq{}}\PY{l+s+s2}{\PYZus{}}\PY{l+s+s2}{\PYZdq{}} 
        \PY{o}{+} \PY{n+nb}{str}\PY{p}{(}\PY{n}{SW\PYZus{}kHz}\PY{p}{)} \PY{o}{+} \PY{l+s+s2}{\PYZdq{}}\PY{l+s+s2}{kHz}\PY{l+s+s2}{\PYZdq{}}\PY{p}{)}
\PY{n}{carrierFreq\PYZus{}MHz} \PY{o}{=} \PY{l+m+mi}{20}
\PY{c+c1}{\PYZsh{} \PYZcb{}\PYZcb{}\PYZcb{}}
\PY{c+c1}{\PYZsh{} \PYZob{}\PYZob{}\PYZob{} SpinCore settings}
\PY{n}{adcOffset} \PY{o}{=} \PY{l+m+mi}{38}
\PY{n}{tx\PYZus{}phases} \PY{o}{=} \PY{n}{r\PYZus{}}\PY{p}{[}\PY{l+m+mf}{0.0}\PY{p}{,} \PY{l+m+mf}{90.0}\PY{p}{,} \PY{l+m+mf}{180.0}\PY{p}{,} \PY{l+m+mf}{270.0}\PY{p}{]}
\PY{n}{nScans} \PY{o}{=} \PY{l+m+mi}{100}
\PY{n}{nPoints} \PY{o}{=} \PY{l+m+mi}{1024} \PY{o}{*} \PY{l+m+mi}{2}
\PY{c+c1}{\PYZsh{} \PYZcb{}\PYZcb{}\PYZcb{}}
\PY{c+c1}{\PYZsh{} \PYZob{}\PYZob{}\PYZob{} Acquire data}
\PY{k}{for} \PY{n}{x} \PY{o+ow}{in} \PY{n+nb}{range}\PY{p}{(}\PY{n}{nScans}\PY{p}{)}\PY{p}{:}
    \PY{c+c1}{\PYZsh{} \PYZob{}\PYZob{}\PYZob{} configure SpinCore}
    \PY{n}{sc}\PY{o}{.}\PY{n}{configureTX}\PY{p}{(}
        \PY{n}{adcOffset}\PY{p}{,}
        \PY{n}{carrierFreq\PYZus{}MHz}\PY{p}{,}
        \PY{n}{tx\PYZus{}phases}\PY{p}{,}
        \PY{l+m+mf}{1.0}\PY{p}{,}
        \PY{n}{nPoints}\PY{p}{,}
    \PY{p}{)}
    \PY{n}{acq\PYZus{}time\PYZus{}ms} \PY{o}{=} \PY{n}{sc}\PY{o}{.}\PY{n}{configureRX}\PY{p}{(}
        \PY{n}{SW\PYZus{}kHz}\PY{p}{,}
        \PY{n}{nPoints}\PY{p}{,}
        \PY{l+m+mi}{1}\PY{p}{,}
        \PY{l+m+mi}{1}\PY{p}{,}  \PY{c+c1}{\PYZsh{} assume nEchoes = 1}
        \PY{l+m+mi}{1}\PY{p}{,}  \PY{c+c1}{\PYZsh{} assume nPhaseSteps = 1}
    \PY{p}{)}
    \PY{n}{sc}\PY{o}{.}\PY{n}{init\PYZus{}ppg}\PY{p}{(}\PY{p}{)}
    \PY{c+c1}{\PYZsh{} \PYZcb{}\PYZcb{}\PYZcb{}}
    \PY{c+c1}{\PYZsh{} \PYZob{}\PYZob{}\PYZob{} ppg to generate the SpinCore data}
    \PY{n}{sc}\PY{o}{.}\PY{n}{load}\PY{p}{(}
        \PY{p}{[}
            \PY{p}{(}\PY{l+s+s2}{\PYZdq{}}\PY{l+s+s2}{marker}\PY{l+s+s2}{\PYZdq{}}\PY{p}{,} \PY{l+s+s2}{\PYZdq{}}\PY{l+s+s2}{start}\PY{l+s+s2}{\PYZdq{}}\PY{p}{,} \PY{l+m+mi}{1}\PY{p}{)}\PY{p}{,}
            \PY{p}{(}\PY{l+s+s2}{\PYZdq{}}\PY{l+s+s2}{phase\PYZus{}reset}\PY{l+s+s2}{\PYZdq{}}\PY{p}{,} \PY{l+m+mi}{1}\PY{p}{)}\PY{p}{,}
            \PY{p}{(}\PY{l+s+s2}{\PYZdq{}}\PY{l+s+s2}{delay}\PY{l+s+s2}{\PYZdq{}}\PY{p}{,} \PY{l+m+mf}{0.5e3}\PY{p}{)}\PY{p}{,}  \PY{c+c1}{\PYZsh{} pick short delay (ms)}
            \PY{p}{(}\PY{l+s+s2}{\PYZdq{}}\PY{l+s+s2}{acquire}\PY{l+s+s2}{\PYZdq{}}\PY{p}{,} \PY{n}{acq\PYZus{}time\PYZus{}ms}\PY{p}{)}\PY{p}{,}
            \PY{p}{(}\PY{l+s+s2}{\PYZdq{}}\PY{l+s+s2}{delay}\PY{l+s+s2}{\PYZdq{}}\PY{p}{,} \PY{l+m+mf}{1e4}\PY{p}{)}\PY{p}{,}  \PY{c+c1}{\PYZsh{} short rep\PYZus{}delay\PYZus{}us}
            \PY{p}{(}\PY{l+s+s2}{\PYZdq{}}\PY{l+s+s2}{jumpto}\PY{l+s+s2}{\PYZdq{}}\PY{p}{,} \PY{l+s+s2}{\PYZdq{}}\PY{l+s+s2}{start}\PY{l+s+s2}{\PYZdq{}}\PY{p}{)}\PY{p}{,}
        \PY{p}{]}
    \PY{p}{)}
    \PY{c+c1}{\PYZsh{} \PYZcb{}\PYZcb{}\PYZcb{}}
    \PY{n}{sc}\PY{o}{.}\PY{n}{stop\PYZus{}ppg}\PY{p}{(}\PY{p}{)}
    \PY{n}{sc}\PY{o}{.}\PY{n}{runBoard}\PY{p}{(}\PY{p}{)}
    \PY{n}{raw\PYZus{}data} \PY{o}{=} \PY{p}{(}  \PY{c+c1}{\PYZsh{} grab data for the singl}
    \PY{c+c1}{\PYZsh{}               capture as a complex value        }
        \PY{n}{sc}\PY{o}{.}\PY{n}{getData}\PY{p}{(}
            \PY{p}{(}\PY{l+m+mi}{2} \PY{o}{*} \PY{n}{nPoints} \PY{o}{*} \PY{l+m+mi}{1} \PY{o}{*} \PY{l+m+mi}{1}\PY{p}{)}\PY{p}{,} 
            \PY{n}{nPoints}\PY{p}{,} 
            \PY{l+m+mi}{1}\PY{p}{,} 
            \PY{l+m+mi}{1}
            \PY{p}{)}
        \PY{o}{.}\PY{n}{astype}\PY{p}{(}\PY{n+nb}{float}\PY{p}{)}
        \PY{o}{.}\PY{n}{view}\PY{p}{(}\PY{n+nb}{complex}\PY{p}{)}
    \PY{p}{)}  \PY{c+c1}{\PYZsh{} assume nEchoes and nPhaseSteps = 1}
    \PY{c+c1}{\PYZsh{} \PYZcb{}\PYZcb{}\PYZcb{}}
    \PY{c+c1}{\PYZsh{} \PYZob{}\PYZob{}\PYZob{} If this is the first scan, then allocate an}
    \PY{c+c1}{\PYZsh{}     array to drop the data into, and assign the }
    \PY{c+c1}{\PYZsh{}     axis coordinates, etc.}
    \PY{k}{if} \PY{n}{x} \PY{o}{==} \PY{l+m+mi}{0}\PY{p}{:}
        \PY{n}{time\PYZus{}axis} \PY{o}{=} \PY{n}{np}\PY{o}{.}\PY{n}{linspace}\PY{p}{(}\PY{l+m+mf}{0.0}\PY{p}{,} 
                \PY{n}{acq\PYZus{}time\PYZus{}ms} \PY{o}{*} \PY{l+m+mf}{1e\PYZhy{}3}\PY{p}{,} 
                \PY{n}{raw\PYZus{}data}\PY{o}{.}\PY{n}{size}\PY{p}{)}
        \PY{n}{data} \PY{o}{=} \PY{p}{(}
            \PY{n}{ps}\PY{o}{.}\PY{n}{ndshape}\PY{p}{(}
                \PY{p}{[}\PY{n}{raw\PYZus{}data}\PY{o}{.}\PY{n}{size}\PY{p}{,} \PY{n}{nScans}\PY{p}{]}\PY{p}{,}
                \PY{p}{[}\PY{l+s+s2}{\PYZdq{}}\PY{l+s+s2}{t}\PY{l+s+s2}{\PYZdq{}}\PY{p}{,} \PY{l+s+s2}{\PYZdq{}}\PY{l+s+s2}{nScans}\PY{l+s+s2}{\PYZdq{}}\PY{p}{]}\PY{p}{,}
            \PY{p}{)}
            \PY{o}{.}\PY{n}{alloc}\PY{p}{(}\PY{n}{dtype}\PY{o}{=}\PY{n}{np}\PY{o}{.}\PY{n}{complex128}\PY{p}{)}
            \PY{o}{.}\PY{n}{setaxis}\PY{p}{(}\PY{l+s+s2}{\PYZdq{}}\PY{l+s+s2}{t}\PY{l+s+s2}{\PYZdq{}}\PY{p}{,} \PY{n}{time\PYZus{}axis}\PY{p}{)}
            \PY{o}{.}\PY{n}{set\PYZus{}units}\PY{p}{(}\PY{l+s+s2}{\PYZdq{}}\PY{l+s+s2}{t}\PY{l+s+s2}{\PYZdq{}}\PY{p}{,} \PY{l+s+s2}{\PYZdq{}}\PY{l+s+s2}{s}\PY{l+s+s2}{\PYZdq{}}\PY{p}{)}
            \PY{o}{.}\PY{n}{setaxis}\PY{p}{(}\PY{l+s+s2}{\PYZdq{}}\PY{l+s+s2}{nScans}\PY{l+s+s2}{\PYZdq{}}\PY{p}{,} \PY{n}{r\PYZus{}}\PY{p}{[}\PY{l+m+mi}{0}\PY{p}{:}\PY{n}{nScans}\PY{p}{]}\PY{p}{)}
            \PY{o}{.}\PY{n}{name}\PY{p}{(}\PY{l+s+s2}{\PYZdq{}}\PY{l+s+s2}{signal}\PY{l+s+s2}{\PYZdq{}}\PY{p}{)}
        \PY{p}{)}
    \PY{c+c1}{\PYZsh{} \PYZcb{}\PYZcb{}\PYZcb{}}
    \PY{c+c1}{\PYZsh{} drop the data into appropriate index}
    \PY{n}{data}\PY{p}{[}\PY{l+s+s2}{\PYZdq{}}\PY{l+s+s2}{nScans}\PY{l+s+s2}{\PYZdq{}}\PY{p}{,} \PY{n}{x}\PY{p}{]} \PY{o}{=} \PY{n}{raw\PYZus{}data}      
    \PY{n}{sc}\PY{o}{.}\PY{n}{stopBoard}\PY{p}{(}\PY{p}{)}
\PY{c+c1}{\PYZsh{} \PYZcb{}\PYZcb{}\PYZcb{}}
\PY{n}{data}\PY{o}{.}\PY{n}{hdf5\PYZus{}write}\PY{p}{(}
    \PY{n}{output\PYZus{}name} \PY{o}{+} \PY{l+s+s2}{\PYZdq{}}\PY{l+s+s2}{.h5}\PY{l+s+s2}{\PYZdq{}}\PY{p}{,}
    \PY{n}{directory}\PY{o}{=}\PY{n}{ps}\PY{o}{.}\PY{n}{getDATADIR}\PY{p}{(}\PY{n}{exp\PYZus{}type}\PY{o}{=}\PY{l+s+s2}{\PYZdq{}}\PY{l+s+s2}{ODNP\PYZus{}NMR\PYZus{}comp/noise\PYZus{}tests}\PY{l+s+s2}{\PYZdq{}}\PY{p}{)}\PY{p}{,}
\PY{p}{)}  \PY{c+c1}{\PYZsh{} save data as h5 file}
\end{Verbatim}
\par\noindent
\rule{\linewidth}{0.5pt}
    \label{SC_acq}
}

\subsubsection{Capturing Receiver Response}\label{sec:CodeRecResp}
The receiver acquired test signal of constant voltage output by
the AFG over a set range of frequencies (defined by the user) to
obtain the digital filter datasets. Again \cite{InstNotes} interacts with
the AFG controlling the USB connection and initialization.
Similar to the GDS, an in-house class with methods and
attributes was made specifically to interact with the
AFG. At each output frequency, the SpinCore acquires a 
user defined number of scans at a set carrier frequency
and spectral width (again defined by the user).
Each frequency dataset is saved as a node within 
an H5 file where the nodename includes the output
frequency of the AFG in kHz.
{\footnotesize%
\par\noindent
\rule{\linewidth}{0.5pt}
\par\noindent
\vspace*{-5ex}
\par\noindent
\rule{\linewidth}{0.5pt}\begin{Verbatim}[commandchars=\\\{\},codes={\catcode`\$=3\catcode`\^=7\catcode`\_=8\relax}]
\PY{k+kn}{from} \PY{n+nn}{Instruments} \PY{k+kn}{import} \PY{n}{AFG}
\PY{k+kn}{from} \PY{n+nn}{pyspecdata} \PY{k+kn}{import} \PY{n}{r\PYZus{}}\PY{p}{,} \PY{n}{ndshape}
\PY{k+kn}{import} \PY{n+nn}{time}
\PY{k+kn}{from} \PY{n+nn}{numpy} \PY{k+kn}{import} \PY{n}{linspace}\PY{p}{,} \PY{n}{complex128}
\PY{k+kn}{import} \PY{n+nn}{SpinCore\PYZus{}pp} \PY{k}{as} \PY{n+nn}{sc}
\PY{k+kn}{from} \PY{n+nn}{datetime} \PY{k+kn}{import} \PY{n}{datetime}

\PY{c+c1}{\PYZsh{} \PYZob{}\PYZob{}\PYZob{} set filename, spectral width and carrier}
\PY{n}{date} \PY{o}{=} \PY{n}{datetime}\PY{o}{.}\PY{n}{now}\PY{p}{(}\PY{p}{)}\PY{o}{.}\PY{n}{strftime}\PY{p}{(}\PY{l+s+s2}{\PYZdq{}}\PY{l+s+s2}{\PYZpc{}}\PY{l+s+s2}{y}\PY{l+s+s2}{\PYZpc{}}\PY{l+s+s2}{m}\PY{l+s+si}{\PYZpc{}d}\PY{l+s+s2}{\PYZdq{}}\PY{p}{)}
\PY{n}{description} \PY{o}{=} \PY{l+s+s2}{\PYZdq{}}\PY{l+s+s2}{3p9kHz\PYZus{}filter}\PY{l+s+s2}{\PYZdq{}}
\PY{n}{SW\PYZus{}kHz} \PY{o}{=} \PY{l+m+mf}{3.9}
\PY{n}{output\PYZus{}name} \PY{o}{=} \PY{n}{date} \PY{o}{+} \PY{l+s+s2}{\PYZdq{}}\PY{l+s+s2}{\PYZus{}}\PY{l+s+s2}{\PYZdq{}} \PY{o}{+} \PY{n}{description} \PY{o}{+} \PY{l+s+s2}{\PYZdq{}}\PY{l+s+s2}{.h5}\PY{l+s+s2}{\PYZdq{}}
\PY{n}{carrierFreq\PYZus{}MHz} \PY{o}{=} \PY{l+m+mf}{14.9}
\PY{c+c1}{\PYZsh{} \PYZcb{}\PYZcb{}\PYZcb{}}
\PY{c+c1}{\PYZsh{} \PYZob{}\PYZob{}\PYZob{} AFG settings}
\PY{c+c1}{\PYZsh{} Make a list of the desired output frequencies}
\PY{n}{freq\PYZus{}list} \PY{o}{=} \PY{n}{linspace}\PY{p}{(}\PY{l+m+mf}{14.8766e6}\PY{p}{,} \PY{l+m+mf}{14.9234e6}\PY{p}{,} \PY{l+m+mi}{300}\PY{p}{)}  \PY{c+c1}{\PYZsh{} Hz}
\PY{n}{amplitude} \PY{o}{=} \PY{l+m+mf}{0.01}  \PY{c+c1}{\PYZsh{} desired Vpp}
\PY{c+c1}{\PYZsh{} \PYZcb{}\PYZcb{}\PYZcb{}}
\PY{c+c1}{\PYZsh{} \PYZob{}\PYZob{}\PYZob{} Spincore settings}
\PY{n}{adcOffset} \PY{o}{=} \PY{l+m+mi}{42}
\PY{n}{tx\PYZus{}phases} \PY{o}{=} \PY{n}{r\PYZus{}}\PY{p}{[}\PY{l+m+mf}{0.0}\PY{p}{,} \PY{l+m+mf}{90.0}\PY{p}{,} \PY{l+m+mf}{180.0}\PY{p}{,} \PY{l+m+mf}{270.0}\PY{p}{]}
\PY{n}{nScans} \PY{o}{=} \PY{l+m+mi}{25}
\PY{n}{nPoints} \PY{o}{=} \PY{l+m+mi}{1024} \PY{o}{*} \PY{l+m+mi}{2}
\PY{c+c1}{\PYZsh{} \PYZcb{}\PYZcb{}\PYZcb{}}
\PY{k}{with} \PY{n}{AFG}\PY{p}{(}\PY{p}{)} \PY{k}{as} \PY{n}{a}\PY{p}{:}  \PY{c+c1}{\PYZsh{} context block that automatically}
    \PY{c+c1}{\PYZsh{}              handles routines to initiate}
    \PY{c+c1}{\PYZsh{}              communication with AFG (arbitrary}
    \PY{c+c1}{\PYZsh{}              function generator), perform checks, and}
    \PY{c+c1}{\PYZsh{}              to close the (USB serial) connection at}
    \PY{c+c1}{\PYZsh{}              the end of the block}
    \PY{n}{a}\PY{o}{.}\PY{n}{reset}\PY{p}{(}\PY{p}{)}
    \PY{k}{for} \PY{n}{j}\PY{p}{,} \PY{n}{frq} \PY{o+ow}{in} \PY{n+nb}{enumerate}\PY{p}{(}\PY{n}{freq\PYZus{}list}\PY{p}{)}\PY{p}{:}
        \PY{n}{a}\PY{p}{[}\PY{l+m+mi}{0}\PY{p}{]}\PY{o}{.}\PY{n}{output} \PY{o}{=} \PY{k+kc}{True}  \PY{c+c1}{\PYZsh{} turn on first channel}
        \PY{n}{a}\PY{o}{.}\PY{n}{sin}\PY{p}{(}\PY{n}{ch}\PY{o}{=}\PY{l+m+mi}{1}\PY{p}{,} \PY{n}{V}\PY{o}{=}\PY{n}{amplitude}\PY{p}{,} \PY{n}{f}\PY{o}{=}\PY{n}{frq}\PY{p}{)}  \PY{c+c1}{\PYZsh{} set a sine wave}
        \PY{c+c1}{\PYZsh{}                                 output with the}
        \PY{c+c1}{\PYZsh{}                                 desired amplitude}
        \PY{c+c1}{\PYZsh{}                                 and frequency}
        \PY{n}{time}\PY{o}{.}\PY{n}{sleep}\PY{p}{(}\PY{l+m+mi}{2}\PY{p}{)}
        \PY{k}{for} \PY{n}{x} \PY{o+ow}{in} \PY{n+nb}{range}\PY{p}{(}\PY{n}{nScans}\PY{p}{)}\PY{p}{:}
            \PY{c+c1}{\PYZsh{} \PYZob{}\PYZob{}\PYZob{} configure SpinCore receiver}
            \PY{n}{sc}\PY{o}{.}\PY{n}{configureTX}\PY{p}{(}
                \PY{n}{adcOffset}\PY{p}{,}
                \PY{n}{carrierFreq\PYZus{}MHz}\PY{p}{,}
                \PY{n}{tx\PYZus{}phases}\PY{p}{,}
                \PY{l+m+mf}{1.0}\PY{p}{,}
                \PY{n}{nPoints}\PY{p}{,}
            \PY{p}{)}
            \PY{n}{acq\PYZus{}time\PYZus{}ms} \PY{o}{=} \PY{n}{sc}\PY{o}{.}\PY{n}{configureRX}\PY{p}{(}
                \PY{n}{SW\PYZus{}kHz}\PY{p}{,}
                \PY{n}{nPoints}\PY{p}{,}
                \PY{l+m+mi}{1}\PY{p}{,}
                \PY{l+m+mi}{1}\PY{p}{,}  \PY{c+c1}{\PYZsh{} assume nEchoes = 1}
                \PY{l+m+mi}{1}\PY{p}{,}  \PY{c+c1}{\PYZsh{} assume nPhaseSteps = 1}
            \PY{p}{)}
            \PY{n}{sc}\PY{o}{.}\PY{n}{init\PYZus{}ppg}\PY{p}{(}\PY{p}{)}
            \PY{c+c1}{\PYZsh{} \PYZcb{}\PYZcb{}\PYZcb{}}
            \PY{c+c1}{\PYZsh{} \PYZob{}\PYZob{}\PYZob{} ppg to generate the SpinCore Data}
            \PY{n}{sc}\PY{o}{.}\PY{n}{load}\PY{p}{(}
                \PY{p}{[}
                    \PY{p}{(}\PY{l+s+s2}{\PYZdq{}}\PY{l+s+s2}{marker}\PY{l+s+s2}{\PYZdq{}}\PY{p}{,} \PY{l+s+s2}{\PYZdq{}}\PY{l+s+s2}{start}\PY{l+s+s2}{\PYZdq{}}\PY{p}{,} \PY{l+m+mi}{1}\PY{p}{)}\PY{p}{,}
                    \PY{p}{(}\PY{l+s+s2}{\PYZdq{}}\PY{l+s+s2}{phase\PYZus{}reset}\PY{l+s+s2}{\PYZdq{}}\PY{p}{,} \PY{l+m+mi}{1}\PY{p}{)}\PY{p}{,}
                    \PY{p}{(}\PY{l+s+s2}{\PYZdq{}}\PY{l+s+s2}{delay}\PY{l+s+s2}{\PYZdq{}}\PY{p}{,} \PY{l+m+mf}{0.5e3}\PY{p}{)}\PY{p}{,} \PY{c+c1}{\PYZsh{} pick short delay (ms)}
                    \PY{p}{(}\PY{l+s+s2}{\PYZdq{}}\PY{l+s+s2}{delay}\PY{l+s+s2}{\PYZdq{}}\PY{p}{,} \PY{l+m+mf}{10.0}\PY{p}{)}\PY{p}{,}
                    \PY{p}{(}\PY{l+s+s2}{\PYZdq{}}\PY{l+s+s2}{acquire}\PY{l+s+s2}{\PYZdq{}}\PY{p}{,} \PY{n}{acq\PYZus{}time\PYZus{}ms}\PY{p}{)}\PY{p}{,}
                    \PY{p}{(}\PY{l+s+s2}{\PYZdq{}}\PY{l+s+s2}{delay}\PY{l+s+s2}{\PYZdq{}}\PY{p}{,} \PY{l+m+mf}{1e4}\PY{p}{)}\PY{p}{,} \PY{c+c1}{\PYZsh{} short rep\PYZus{}delay\PYZus{}us}
                    \PY{p}{(}\PY{l+s+s2}{\PYZdq{}}\PY{l+s+s2}{jumpto}\PY{l+s+s2}{\PYZdq{}}\PY{p}{,} \PY{l+s+s2}{\PYZdq{}}\PY{l+s+s2}{start}\PY{l+s+s2}{\PYZdq{}}\PY{p}{)}\PY{p}{,}
                \PY{p}{]}
            \PY{p}{)}
            \PY{c+c1}{\PYZsh{} \PYZcb{}\PYZcb{}\PYZcb{}}
            \PY{n}{sc}\PY{o}{.}\PY{n}{stop\PYZus{}ppg}\PY{p}{(}\PY{p}{)}
            \PY{n}{sc}\PY{o}{.}\PY{n}{runBoard}\PY{p}{(}\PY{p}{)}
            \PY{n}{raw\PYZus{}data} \PY{o}{=} \PY{p}{(}  \PY{c+c1}{\PYZsh{} grab data for the single}
            \PY{c+c1}{\PYZsh{}              capture as a complex value}
                \PY{n}{sc}\PY{o}{.}\PY{n}{getData}\PY{p}{(}
                    \PY{p}{(}\PY{l+m+mi}{2} \PY{o}{*} \PY{n}{nPoints} \PY{o}{*} \PY{l+m+mi}{1} \PY{o}{*} \PY{l+m+mi}{1}\PY{p}{)}\PY{p}{,} 
                    \PY{n}{nPoints}\PY{p}{,} 
                    \PY{l+m+mi}{1}\PY{p}{,} 
                    \PY{l+m+mi}{1}
                    \PY{p}{)}
                \PY{o}{.}\PY{n}{astype}\PY{p}{(}\PY{n+nb}{float}\PY{p}{)}
                \PY{o}{.}\PY{n}{view}\PY{p}{(}\PY{n+nb}{complex}\PY{p}{)}
            \PY{p}{)}  \PY{c+c1}{\PYZsh{} assume nEchoes and nPhaseSteps = 1}
            \PY{c+c1}{\PYZsh{} \PYZob{}\PYZob{}\PYZob{} If this is the first scan, then allocate an}
            \PY{c+c1}{\PYZsh{}     array to drop the data into, and assign the}
            \PY{c+c1}{\PYZsh{}     axis coordinates, etc.}
            \PY{k}{if} \PY{n}{x} \PY{o}{==} \PY{l+m+mi}{0}\PY{p}{:}
                \PY{n}{time\PYZus{}axis} \PY{o}{=} \PY{n}{linspace}\PY{p}{(}\PY{l+m+mf}{0.0}\PY{p}{,} 
                        \PY{n}{acq\PYZus{}time\PYZus{}ms} \PY{o}{*} \PY{l+m+mf}{1e\PYZhy{}3}\PY{p}{,} 
                        \PY{n}{raw\PYZus{}data}\PY{o}{.}\PY{n}{size}\PY{p}{)}
                \PY{n}{data} \PY{o}{=} \PY{p}{(}
                    \PY{n}{ndshape}\PY{p}{(}
                        \PY{p}{[}\PY{n}{raw\PYZus{}data}\PY{o}{.}\PY{n}{size}\PY{p}{,} \PY{n}{nScans}\PY{p}{]}\PY{p}{,}
                        \PY{p}{[}\PY{l+s+s2}{\PYZdq{}}\PY{l+s+s2}{t}\PY{l+s+s2}{\PYZdq{}}\PY{p}{,} \PY{l+s+s2}{\PYZdq{}}\PY{l+s+s2}{nScans}\PY{l+s+s2}{\PYZdq{}}\PY{p}{]}\PY{p}{,}
                    \PY{p}{)}
                    \PY{o}{.}\PY{n}{alloc}\PY{p}{(}\PY{n}{dtype}\PY{o}{=}\PY{n}{complex128}\PY{p}{)}
                    \PY{o}{.}\PY{n}{setaxis}\PY{p}{(}\PY{l+s+s2}{\PYZdq{}}\PY{l+s+s2}{t}\PY{l+s+s2}{\PYZdq{}}\PY{p}{,} \PY{n}{time\PYZus{}axis}\PY{p}{)}
                    \PY{o}{.}\PY{n}{set\PYZus{}units}\PY{p}{(}\PY{l+s+s2}{\PYZdq{}}\PY{l+s+s2}{t}\PY{l+s+s2}{\PYZdq{}}\PY{p}{,} \PY{l+s+s2}{\PYZdq{}}\PY{l+s+s2}{s}\PY{l+s+s2}{\PYZdq{}}\PY{p}{)}
                    \PY{o}{.}\PY{n}{setaxis}\PY{p}{(}\PY{l+s+s2}{\PYZdq{}}\PY{l+s+s2}{nScans}\PY{l+s+s2}{\PYZdq{}}\PY{p}{,} \PY{n}{r\PYZus{}}\PY{p}{[}\PY{l+m+mi}{0}\PY{p}{:}\PY{n}{nScans}\PY{p}{]}\PY{p}{)}
                    \PY{o}{.}\PY{n}{name}\PY{p}{(}\PY{l+s+s2}{\PYZdq{}}\PY{l+s+s2}{signal }\PY{l+s+si}{\PYZpc{}f}\PY{l+s+s2}{ kHz}\PY{l+s+s2}{\PYZdq{}} \PY{o}{\PYZpc{}} \PY{n}{frq} \PY{o}{/} \PY{l+m+mf}{1e3}\PY{p}{)}
                \PY{p}{)}
            \PY{c+c1}{\PYZsh{} \PYZcb{}\PYZcb{}\PYZcb{}}
            \PY{c+c1}{\PYZsh{} drop the data into appropriate index}
            \PY{n}{data}\PY{p}{[}\PY{l+s+s2}{\PYZdq{}}\PY{l+s+s2}{nScans}\PY{l+s+s2}{\PYZdq{}}\PY{p}{,} \PY{n}{x}\PY{p}{]} \PY{o}{=} \PY{n}{raw\PYZus{}data}  
            \PY{n}{sc}\PY{o}{.}\PY{n}{stopBoard}\PY{p}{(}\PY{p}{)}
        \PY{n}{data}\PY{o}{.}\PY{n}{set\PYZus{}prop}\PY{p}{(}
            \PY{l+s+s2}{\PYZdq{}}\PY{l+s+s2}{afg\PYZus{}frq}\PY{l+s+s2}{\PYZdq{}}\PY{p}{,} \PY{n}{frq} \PY{o}{/} \PY{l+m+mf}{1e3}
        \PY{p}{)}  \PY{c+c1}{\PYZsh{} store the AFG frequency in units of kHz}
        \PY{n}{data}\PY{o}{.}\PY{n}{name}\PY{p}{(}\PY{l+s+s2}{\PYZdq{}}\PY{l+s+s2}{afg\PYZus{}}\PY{l+s+si}{\PYZpc{}d}\PY{l+s+s2}{\PYZdq{}} \PY{o}{\PYZpc{}} \PY{n}{frq}\PY{p}{)}  \PY{c+c1}{\PYZsh{} the nddata name}
        \PY{c+c1}{\PYZsh{}                           determines the node name}
        \PY{c+c1}{\PYZsh{}                           of the data in the HDF5}
        \PY{c+c1}{\PYZsh{}                           file, below}
        \PY{n}{nodename} \PY{o}{=} \PY{n}{data}\PY{o}{.}\PY{n}{name}\PY{p}{(}\PY{p}{)}
        \PY{n}{data}\PY{o}{.}\PY{n}{hdf5\PYZus{}write}\PY{p}{(}
            \PY{n}{output}\PY{p}{,}
            \PY{n}{directory}\PY{o}{=}\PY{l+s+s2}{\PYZdq{}}\PY{l+s+s2}{ODNP\PYZus{}NMR\PYZus{}comp/noise\PYZus{}tests}\PY{l+s+s2}{\PYZdq{}}\PY{p}{,}
        \PY{p}{)}
\end{Verbatim}
\par\noindent
\rule{\linewidth}{0.5pt}
\label{afg_receiver_response}
}
\subsubsection{Processing PSDs}
The following script calculates the PSD for datasets acquired
on the oscilloscope. The final result (unconvolved and convolved) 
is plotted on a semilog plot.
Note it is assumed the acquired data has already been
converted to analytic:
{\footnotesize%
\par\noindent
\rule{\linewidth}{0.5pt}
\par\noindent
\vspace*{-5ex}
\par\noindent
\rule{\linewidth}{0.5pt}\begin{Verbatim}[commandchars=\\\{\},codes={\catcode`\$=3\catcode`\^=7\catcode`\_=8\relax}]
\PY{k+kn}{from} \PY{n+nn}{numpy} \PY{k+kn}{import} \PY{n}{r\PYZus{}}
\PY{k+kn}{from} \PY{n+nn}{pyspecdata} \PY{k+kn}{import} \PY{p}{(}
    \PY{n}{figlist\PYZus{}var}\PY{p}{,}
    \PY{n}{nddata\PYZus{}hdf5}\PY{p}{,}
    \PY{n}{getDATADIR}\PY{p}{,}
\PY{p}{)}
\PY{k+kn}{from} \PY{n+nn}{pylab} \PY{k+kn}{import} \PY{n}{diff}\PY{p}{,} \PY{n}{sqrt}\PY{p}{,} \PY{n}{ylim}\PY{p}{,} \PY{n}{ylabel}

\PY{n}{width} \PY{o}{=} \PY{p}{(}
    \PY{l+m+mf}{0.1e6}  \PY{c+c1}{\PYZsh{} width used for Gaussian convolution filter}
\PY{p}{)}
\PY{n}{filename} \PY{o}{=} \PY{l+s+s2}{\PYZdq{}}\PY{l+s+s2}{230914\PYZus{}RX\PYZus{}GDS\PYZus{}4.h5}\PY{l+s+s2}{\PYZdq{}}
\PY{n}{nodename} \PY{o}{=} \PY{l+s+s2}{\PYZdq{}}\PY{l+s+s2}{accumulated\PYZus{}230914}\PY{l+s+s2}{\PYZdq{}}
\PY{k}{with} \PY{n}{figlist\PYZus{}var}\PY{p}{(}\PY{p}{)} \PY{k}{as} \PY{n}{fl}\PY{p}{:}
    \PY{c+c1}{\PYZsh{} load data according to the filename and nodename}
    \PY{n}{s} \PY{o}{=} \PY{n}{nddata\PYZus{}hdf5}\PY{p}{(}
        \PY{n}{filename} \PY{o}{+} \PY{n}{nodename}\PY{p}{,}
        \PY{n}{directory}\PY{o}{=}\PY{n}{getDATADIR}\PY{p}{(}
            \PY{n}{exp\PYZus{}type}\PY{o}{=}\PY{l+s+s2}{\PYZdq{}}\PY{l+s+s2}{ODNP\PYZus{}NMR\PYZus{}comp/noise\PYZus{}tests}\PY{l+s+s2}{\PYZdq{}}
        \PY{p}{)}\PY{p}{,}
    \PY{p}{)}
    \PY{n}{s}\PY{o}{.}\PY{n}{set\PYZus{}units}\PY{p}{(}\PY{l+s+s2}{\PYZdq{}}\PY{l+s+s2}{t}\PY{l+s+s2}{\PYZdq{}}\PY{p}{,} \PY{l+s+s2}{\PYZdq{}}\PY{l+s+s2}{s}\PY{l+s+s2}{\PYZdq{}}\PY{p}{)}
    \PY{n}{s} \PY{o}{=} \PY{n}{s}\PY{p}{[}
        \PY{l+s+s2}{\PYZdq{}}\PY{l+s+s2}{ch}\PY{l+s+s2}{\PYZdq{}}\PY{p}{,} \PY{l+m+mi}{0}
    \PY{p}{]}  \PY{c+c1}{\PYZsh{} select the first channel, note multiple}
    \PY{c+c1}{\PYZsh{}    channels may be acquired}
    \PY{n}{acq\PYZus{}time} \PY{o}{=} \PY{n}{diff}\PY{p}{(}\PY{n}{s}\PY{o}{.}\PY{n}{getaxis}\PY{p}{(}\PY{l+s+s2}{\PYZdq{}}\PY{l+s+s2}{t}\PY{l+s+s2}{\PYZdq{}}\PY{p}{)}\PY{p}{[}\PY{n}{r\PYZus{}}\PY{p}{[}\PY{l+m+mi}{0}\PY{p}{,} \PY{o}{\PYZhy{}}\PY{l+m+mi}{1}\PY{p}{]}\PY{p}{]}\PY{p}{)}\PY{p}{[}
        \PY{l+m+mi}{0}
    \PY{p}{]}  \PY{c+c1}{\PYZsh{} calculate acquisition time}
    \PY{n}{s}\PY{o}{.}\PY{n}{ft}\PY{p}{(}\PY{l+s+s2}{\PYZdq{}}\PY{l+s+s2}{t}\PY{l+s+s2}{\PYZdq{}}\PY{p}{)} \PY{c+c1}{\PYZsh{} $\frac{V_{p}\sqrt{s}}{\sqrt{Hz}}$}
    \PY{n}{s} \PY{o}{/}\PY{o}{=} \PY{n}{sqrt}\PY{p}{(}\PY{l+m+mi}{2}\PY{p}{)} \PY{c+c1}{\PYZsh{} instantaneous $\frac{V_{p}\sqrt{s}}{\sqrt{Hz}} \rightarrow \frac{V_{rms}\sqrt{s}}{\sqrt{Hz}}$}
    \PY{c+c1}{\PYZsh{} \PYZob{}\PYZob{}\PYZob{} equation 21}
    \PY{n}{s} \PY{o}{=} \PY{p}{(}
        \PY{n+nb}{abs}\PY{p}{(}\PY{n}{s}\PY{p}{)} \PY{o}{*}\PY{o}{*} \PY{l+m+mi}{2}
    \PY{p}{)}  \PY{c+c1}{\PYZsh{} take mod squared to convert to energy $\frac{V_{rms}^{2} \cdot s}{Hz}$}
    \PY{n}{s}\PY{o}{.}\PY{n}{mean}\PY{p}{(}\PY{l+s+s2}{\PYZdq{}}\PY{l+s+s2}{capture}\PY{l+s+s2}{\PYZdq{}}\PY{p}{)}  \PY{c+c1}{\PYZsh{} average over all captures}
    \PY{n}{s} \PY{o}{/}\PY{o}{=} \PY{n}{acq\PYZus{}time}  \PY{c+c1}{\PYZsh{} convert to Power $\frac{V_{rms}^2}{Hz} = W$}
    \PY{n}{s} \PY{o}{/}\PY{o}{=} \PY{l+m+mi}{50}  \PY{c+c1}{\PYZsh{} W/Hz}
    \PY{c+c1}{\PYZsh{} \PYZcb{}\PYZcb{}\PYZcb{}}
    \PY{n}{s}\PY{o}{.}\PY{n}{set\PYZus{}units}\PY{p}{(}\PY{l+s+s2}{\PYZdq{}}\PY{l+s+s2}{t}\PY{l+s+s2}{\PYZdq{}}\PY{p}{,} \PY{l+s+s2}{\PYZdq{}}\PY{l+s+s2}{Hz}\PY{l+s+s2}{\PYZdq{}}\PY{p}{)}
    \PY{c+c1}{\PYZsh{} Plot unconvolved PSD on a semilog plot}
    \PY{n}{fl}\PY{o}{.}\PY{n}{next}\PY{p}{(}\PY{l+s+s2}{\PYZdq{}}\PY{l+s+s2}{\PYZdq{}}\PY{p}{,} \PY{n}{figsize}\PY{o}{=}\PY{p}{(}\PY{l+m+mi}{6}\PY{p}{,} \PY{l+m+mi}{4}\PY{p}{)}\PY{p}{)}
    \PY{n}{fl}\PY{o}{.}\PY{n}{plot}\PY{p}{(}
        \PY{n}{s}\PY{p}{[}\PY{l+s+s2}{\PYZdq{}}\PY{l+s+s2}{t}\PY{l+s+s2}{\PYZdq{}}\PY{p}{:}\PY{p}{(}\PY{l+m+mi}{0}\PY{p}{,} \PY{l+m+mf}{49e6}\PY{p}{)}\PY{p}{]}\PY{p}{,}
        \PY{n}{color}\PY{o}{=}\PY{l+s+s2}{\PYZdq{}}\PY{l+s+s2}{blue}\PY{l+s+s2}{\PYZdq{}}\PY{p}{,}
        \PY{n}{label}\PY{o}{=}\PY{k+kc}{None}\PY{p}{,}
        \PY{n}{alpha}\PY{o}{=}\PY{l+m+mf}{0.1}\PY{p}{,}
        \PY{n}{plottype}\PY{o}{=}\PY{l+s+s2}{\PYZdq{}}\PY{l+s+s2}{semilogy}\PY{l+s+s2}{\PYZdq{}}\PY{p}{,}
    \PY{p}{)}
    \PY{c+c1}{\PYZsh{} convolve using the width specified above}
    \PY{c+c1}{\PYZsh{} the convolve function of pySpecData automatically}
    \PY{c+c1}{\PYZsh{} applies the division by $\sigma\sqrt{2\pi}$}
    \PY{n}{s}\PY{o}{.}\PY{n}{convolve}\PY{p}{(}\PY{l+s+s2}{\PYZdq{}}\PY{l+s+s2}{t}\PY{l+s+s2}{\PYZdq{}}\PY{p}{,} \PY{n}{width}\PY{p}{,} \PY{n}{enforce\PYZus{}causality}\PY{o}{=}\PY{k+kc}{False}\PY{p}{)}
    \PY{c+c1}{\PYZsh{} plot the convolved PSD on the semilog plot with the}
    \PY{c+c1}{\PYZsh{} unconvolved}
    \PY{n}{fl}\PY{o}{.}\PY{n}{plot}\PY{p}{(}
        \PY{n}{s}\PY{p}{[}\PY{l+s+s2}{\PYZdq{}}\PY{l+s+s2}{t}\PY{l+s+s2}{\PYZdq{}}\PY{p}{:}\PY{p}{(}\PY{l+m+mi}{0}\PY{p}{,} \PY{l+m+mf}{49e6}\PY{p}{)}\PY{p}{]}\PY{p}{,}
        \PY{n}{color}\PY{o}{=}\PY{l+s+s2}{\PYZdq{}}\PY{l+s+s2}{blue}\PY{l+s+s2}{\PYZdq{}}\PY{p}{,}
        \PY{n}{label}\PY{o}{=}\PY{k+kc}{None}\PY{p}{,}
        \PY{n}{alpha}\PY{o}{=}\PY{l+m+mf}{0.3}\PY{p}{,}
        \PY{n}{plottype}\PY{o}{=}\PY{l+s+s2}{\PYZdq{}}\PY{l+s+s2}{semilogy}\PY{l+s+s2}{\PYZdq{}}\PY{p}{,}
    \PY{p}{)}
    \PY{n}{ylim}\PY{p}{(}\PY{l+m+mf}{1e\PYZhy{}17}\PY{p}{,} \PY{l+m+mf}{1.8e\PYZhy{}12}\PY{p}{)}  \PY{c+c1}{\PYZsh{} set y limits}
    \PY{n}{ylabel}\PY{p}{(}\PY{l+s+s2}{\PYZdq{}}\PY{l+s+s2}{\PYZdl{}}\PY{l+s+s2}{\PYZob{}}\PY{l+s+s2}{S(}\PY{l+s+se}{\PYZbs{}\PYZbs{}}\PY{l+s+s2}{nu)\PYZcb{}\PYZcb{}\PYZdl{} / (W/Hz)}\PY{l+s+s2}{\PYZdq{}}\PY{p}{)}
\end{Verbatim}
\par\noindent
\rule{\linewidth}{0.5pt}
}

Initially, the raw data acquired on the SpinCore is fed to a pre-processing
function.  
The function performs a Fourier transform of the time domain and the phase
cycling dimensions (when applicable) and finally applies the appropriate sinc
digital filter (\cref{eq:transceiverTimeDomain}) before returning the data:
{\footnotesize%
\par\noindent
\rule{\linewidth}{0.5pt}
\par\noindent
\vspace*{-5ex}
\par\noindent
\rule{\linewidth}{0.5pt}\begin{Verbatim}[commandchars=\\\{\},codes={\catcode`\$=3\catcode`\^=7\catcode`\_=8\relax}]
\PY{k+kn}{import} \PY{n+nn}{numpy} \PY{k}{as} \PY{n+nn}{np}
\PY{k}{def} \PY{n+nf}{proc\PYZus{}spincore\PYZus{}generalproc\PYZus{}v1}\PY{p}{(}\PY{n}{s}\PY{p}{,} \PY{n}{direct}\PY{o}{=}\PY{l+s+s2}{\PYZdq{}}\PY{l+s+s2}{t2}\PY{l+s+s2}{\PYZdq{}}\PY{p}{,}
        \PY{n}{include\PYZus{}tau\PYZus{}sub}\PY{o}{=}\PY{k+kc}{False}\PY{p}{)}\PY{p}{:}
\PY{+w}{    }\PY{l+s+sd}{\PYZdq{}\PYZdq{}\PYZdq{} Preprocessing function that takes the raw data}
\PY{l+s+sd}{    and fourier transforms both the time domain into}
\PY{l+s+sd}{    the frequency domain and the phase cycling domain}
\PY{l+s+sd}{    into the coherence transfer pathway domain. The}
\PY{l+s+sd}{    receiver response is also applied prior to}
\PY{l+s+sd}{    returning the data.}

\PY{l+s+sd}{    Parameters }
\PY{l+s+sd}{    ========== }
\PY{l+s+sd}{    s: nddata }
\PY{l+s+sd}{        Raw dataset that was acquired on the SpinCore.}
\PY{l+s+sd}{    direct: str }
\PY{l+s+sd}{        Direct axis of the data.  }
\PY{l+s+sd}{    include\PYZus{}tau\PYZus{}sub: boolean }
\PY{l+s+sd}{        Option to subtract the echo time in order to}
\PY{l+s+sd}{        center the echo about zero.  }
\PY{l+s+sd}{    Returns }
\PY{l+s+sd}{    ======= }
\PY{l+s+sd}{    s: nddata }
\PY{l+s+sd}{        The original data with the receiver response}
\PY{l+s+sd}{        applied returned in the frequency and coherence}
\PY{l+s+sd}{        transfer pathway domain }
\PY{l+s+sd}{    \PYZdq{}\PYZdq{}\PYZdq{}}

    \PY{n}{s}\PY{o}{.}\PY{n}{run}\PY{p}{(}\PY{n}{np}\PY{o}{.}\PY{n}{conj}\PY{p}{)}  \PY{c+c1}{\PYZsh{} SC flips data in a weird way,}
    \PY{c+c1}{\PYZsh{}                 this corrects for that}
    \PY{k}{if} \PY{n}{include\PYZus{}tau\PYZus{}sub}\PY{p}{:} \PY{c+c1}{\PYZsh{} option to subtract echo time}
    \PY{c+c1}{\PYZsh{}                     to center about zero}
        \PY{k}{if} \PY{l+s+s2}{\PYZdq{}}\PY{l+s+s2}{tau\PYZus{}us}\PY{l+s+s2}{\PYZdq{}} \PY{o+ow}{in} \PY{n}{s}\PY{o}{.}\PY{n}{get\PYZus{}prop}\PY{p}{(}\PY{l+s+s2}{\PYZdq{}}\PY{l+s+s2}{acq\PYZus{}params}\PY{l+s+s2}{\PYZdq{}}\PY{p}{)}\PY{o}{.}\PY{n}{keys}\PY{p}{(}\PY{p}{)}\PY{p}{:}
            \PY{n}{s}\PY{p}{[}\PY{n}{direct}\PY{p}{]} \PY{o}{\PYZhy{}}\PY{o}{=} \PY{p}{(}\PY{n}{s}\PY{o}{.}\PY{n}{get\PYZus{}prop}\PY{p}{(}\PY{l+s+s2}{\PYZdq{}}\PY{l+s+s2}{acq\PYZus{}params}\PY{l+s+s2}{\PYZdq{}}\PY{p}{)}\PY{p}{[}\PY{l+s+s2}{\PYZdq{}}\PY{l+s+s2}{tau\PYZus{}us}\PY{l+s+s2}{\PYZdq{}}\PY{p}{]} 
                    \PY{o}{*} \PY{l+m+mf}{1e\PYZhy{}6}\PY{p}{)}
    \PY{n}{s}\PY{o}{.}\PY{n}{ft}\PY{p}{(}\PY{n}{direct}\PY{p}{,} \PY{n}{shift}\PY{o}{=}\PY{k+kc}{True}\PY{p}{)}
    \PY{k}{for} \PY{n}{j} \PY{o+ow}{in} \PY{p}{[}\PY{n}{k} \PY{k}{for} \PY{n}{k} \PY{o+ow}{in} \PY{n}{s}\PY{o}{.}\PY{n}{dimlabels} \PY{k}{if} \PY{n}{k}\PY{o}{.}\PY{n}{startswith}\PY{p}{(}\PY{l+s+s2}{\PYZdq{}}\PY{l+s+s2}{ph}\PY{l+s+s2}{\PYZdq{}}\PY{p}{)}\PY{p}{]}\PY{p}{:}
        \PY{n}{dph} \PY{o}{=} \PY{n}{s}\PY{p}{[}\PY{n}{j}\PY{p}{]}\PY{p}{[}\PY{l+m+mi}{1}\PY{p}{]} \PY{o}{\PYZhy{}} \PY{n}{s}\PY{p}{[}\PY{n}{j}\PY{p}{]}\PY{p}{[}\PY{l+m+mi}{0}\PY{p}{]}
        \PY{n}{Dph} \PY{o}{=} \PY{n}{s}\PY{p}{[}\PY{n}{j}\PY{p}{]}\PY{p}{[}\PY{o}{\PYZhy{}}\PY{l+m+mi}{1}\PY{p}{]} \PY{o}{+} \PY{n}{dph} \PY{o}{\PYZhy{}} \PY{n}{s}\PY{p}{[}\PY{n}{j}\PY{p}{]}\PY{p}{[}\PY{l+m+mi}{0}\PY{p}{]}
        \PY{k}{if} \PY{n}{Dph} \PY{o}{==} \PY{l+m+mi}{1}\PY{p}{:}
            \PY{n}{s}\PY{p}{[}\PY{n}{j}\PY{p}{]} \PY{o}{=} \PY{p}{(}\PY{o}{\PYZhy{}}\PY{n}{s}\PY{p}{[}\PY{n}{j}\PY{p}{]} \PY{o}{+} \PY{l+m+mi}{1}\PY{p}{)} \PY{o}{\PYZpc{}} \PY{l+m+mi}{1} \PY{c+c1}{\PYZsh{} when we take the complex}
            \PY{c+c1}{\PYZsh{}                        conjugate, that changes }
            \PY{c+c1}{\PYZsh{}                        the phase cycle, as well,}
            \PY{c+c1}{\PYZsh{}                        so we have to re\PYZhy{}label the}
            \PY{c+c1}{\PYZsh{}                        axis coordinates for the }
            \PY{c+c1}{\PYZsh{}                        phase cycle to the negative}
            \PY{c+c1}{\PYZsh{}                        of what they were before.}
            \PY{c+c1}{\PYZsh{}                        We also apply phase wrapping }
            \PY{c+c1}{\PYZsh{}                        to get positive numbers.}
        \PY{k}{elif} \PY{n}{Dph} \PY{o}{==} \PY{l+m+mi}{4}\PY{p}{:} \PY{c+c1}{\PYZsh{} uses units of quarter cycle}
            \PY{n}{s}\PY{p}{[}\PY{n}{j}\PY{p}{]} \PY{o}{=} \PY{p}{(}\PY{o}{\PYZhy{}}\PY{n}{s}\PY{p}{[}\PY{n}{j}\PY{p}{]} \PY{o}{+} \PY{l+m+mi}{4}\PY{p}{)} \PY{o}{\PYZpc{}} \PY{l+m+mi}{4}
        \PY{k}{else}\PY{p}{:}
            \PY{k}{raise} \PY{n+ne}{ValueError}\PY{p}{(}
                \PY{l+s+s2}{\PYZdq{}}\PY{l+s+s2}{the phase cycling dimension }\PY{l+s+s2}{\PYZdq{}}
                \PY{o}{+} \PY{n}{j}
                \PY{o}{+} \PY{l+s+s2}{\PYZdq{}}\PY{l+s+s2}{ appears not to go all the way around}\PY{l+s+s2}{\PYZdq{}}
                \PY{o}{+} \PY{l+s+s2}{\PYZdq{}}\PY{l+s+s2}{ the circle!}\PY{l+s+s2}{\PYZdq{}}
            \PY{p}{)}
        \PY{n}{s}\PY{o}{.}\PY{n}{sort}\PY{p}{(}\PY{n}{j}\PY{p}{)}    
        \PY{n}{s}\PY{o}{.}\PY{n}{ft}\PY{p}{(}\PY{p}{[}\PY{n}{j}\PY{p}{]}\PY{p}{)}  \PY{c+c1}{\PYZsh{} FT the phase cycling dimensions}
        \PY{c+c1}{\PYZsh{}            into the coherence transfer}
        \PY{c+c1}{\PYZsh{}            pathways}
    \PY{c+c1}{\PYZsh{} \PYZob{}\PYZob{}\PYZob{} always put the phase cycling dimensions on the outside}
    \PY{n}{neworder} \PY{o}{=} \PY{p}{[}\PY{n}{j} \PY{k}{for} \PY{n}{j} \PY{o+ow}{in} \PY{n}{s}\PY{o}{.}\PY{n}{dimlabels} \PY{k}{if} \PY{n}{j}\PY{o}{.}\PY{n}{startswith}\PY{p}{(}\PY{l+s+s2}{\PYZdq{}}\PY{l+s+s2}{ph}\PY{l+s+s2}{\PYZdq{}}\PY{p}{)}\PY{p}{]}
    \PY{c+c1}{\PYZsh{} \PYZcb{}\PYZcb{}\PYZcb{}}
    \PY{c+c1}{\PYZsh{} \PYZob{}\PYZob{}\PYZob{} reorder the rest based on size}
    \PY{n}{nonphdims} \PY{o}{=} \PY{p}{[}\PY{n}{j} \PY{k}{for} \PY{n}{j} \PY{o+ow}{in} \PY{n}{s}\PY{o}{.}\PY{n}{dimlabels} \PY{k}{if} \PY{o+ow}{not}
            \PY{n}{j}\PY{o}{.}\PY{n}{startswith}\PY{p}{(}\PY{l+s+s2}{\PYZdq{}}\PY{l+s+s2}{ph}\PY{l+s+s2}{\PYZdq{}}\PY{p}{)}\PY{p}{]}
    \PY{k}{if} \PY{n+nb}{len}\PY{p}{(}\PY{n}{nonphdims}\PY{p}{)} \PY{o}{\PYZgt{}} \PY{l+m+mi}{1}\PY{p}{:}
        \PY{n}{sizeidx} \PY{o}{=} \PY{n}{np}\PY{o}{.}\PY{n}{argsort}\PY{p}{(}\PY{p}{[}\PY{n}{s}\PY{o}{.}\PY{n}{shape}\PY{p}{[}\PY{n}{j}\PY{p}{]} \PY{k}{for} \PY{n}{j} \PY{o+ow}{in} \PY{n}{nonphdims}\PY{p}{]}\PY{p}{)}
        \PY{n}{neworder} \PY{o}{+}\PY{o}{=} \PY{p}{[}\PY{n}{nonphdims}\PY{p}{[}\PY{n}{j}\PY{p}{]} \PY{k}{for} \PY{n}{j} \PY{o+ow}{in} \PY{n}{sizeidx}\PY{p}{]}
    \PY{c+c1}{\PYZsh{} \PYZcb{}\PYZcb{}\PYZcb{}}
    \PY{n}{s}\PY{o}{.}\PY{n}{reorder}\PY{p}{(}\PY{n}{neworder}\PY{p}{)}
    \PY{c+c1}{\PYZsh{} \PYZob{}\PYZob{}\PYZob{} apply the receiver response}
    \PY{n}{s}\PY{o}{.}\PY{n}{set\PYZus{}prop}\PY{p}{(}
        \PY{l+s+s2}{\PYZdq{}}\PY{l+s+s2}{dig\PYZus{}filter}\PY{l+s+s2}{\PYZdq{}}\PY{p}{,}
        \PY{n}{s}\PY{o}{.}\PY{n}{fromaxis}\PY{p}{(}\PY{n}{direct}\PY{p}{)}\PY{o}{.}\PY{n}{run}\PY{p}{(}
            \PY{k}{lambda} \PY{n}{x}\PY{p}{:} \PY{n}{np}\PY{o}{.}\PY{n}{sinc}\PY{p}{(}\PY{n}{x} \PY{o}{/}
                \PY{p}{(}\PY{n}{s}\PY{o}{.}\PY{n}{get\PYZus{}prop}\PY{p}{(}\PY{l+s+s2}{\PYZdq{}}\PY{l+s+s2}{acq\PYZus{}params}\PY{l+s+s2}{\PYZdq{}}\PY{p}{)}\PY{p}{[}\PY{l+s+s2}{\PYZdq{}}\PY{l+s+s2}{SW\PYZus{}kHz}\PY{l+s+s2}{\PYZdq{}}\PY{p}{]} \PY{o}{*}
                    \PY{l+m+mf}{1e3}\PY{p}{)}\PY{p}{)}
        \PY{p}{)}\PY{p}{,}
    \PY{p}{)}
    \PY{n}{s} \PY{o}{/}\PY{o}{=} \PY{n}{s}\PY{o}{.}\PY{n}{get\PYZus{}prop}\PY{p}{(}\PY{l+s+s2}{\PYZdq{}}\PY{l+s+s2}{dig\PYZus{}filter}\PY{l+s+s2}{\PYZdq{}}\PY{p}{)} \PY{c+c1}{\PYZsh{} set the filter}
    \PY{c+c1}{\PYZsh{}                               response as a}
    \PY{c+c1}{\PYZsh{}                               property so that we}
    \PY{c+c1}{\PYZsh{}                               can access at a}
    \PY{c+c1}{\PYZsh{}                               later time if}
    \PY{c+c1}{\PYZsh{}                               needed}
    \PY{c+c1}{\PYZsh{} \PYZcb{}\PYZcb{}\PYZcb{}}
    \PY{n}{s}\PY{o}{.}\PY{n}{squeeze}\PY{p}{(}\PY{p}{)}
    \PY{k}{return} \PY{n}{s}
\end{Verbatim}
\par\noindent
\rule{\linewidth}{0.5pt}
}

Following the pre-processing, the data
is converted to voltage using the acquired 
calibration factors prior to the calculation of
the PSD (\cref{eq:convPSD}). 
The final PSD is then plotted (unconvolved 
and convolved) on a semilog plot using the following
script:
{\footnotesize%
\par\noindent
\rule{\linewidth}{0.5pt}
\par\noindent
\vspace*{-5ex}
\par\noindent
\rule{\linewidth}{0.5pt}\begin{Verbatim}[commandchars=\\\{\},codes={\catcode`\$=3\catcode`\^=7\catcode`\_=8\relax}]
\PY{k+kn}{from} \PY{n+nn}{numpy} \PY{k+kn}{import} \PY{n}{r\PYZus{}}
\PY{k+kn}{from} \PY{n+nn}{pyspecdata} \PY{k+kn}{import} \PY{n}{figlist\PYZus{}var}\PY{p}{,} \PY{n}{find\PYZus{}file}
\PY{k+kn}{from} \PY{n+nn}{pyspecProcScripts} \PY{k+kn}{import} \PY{n}{lookup\PYZus{}table}
\PY{k+kn}{from} \PY{n+nn}{pylab} \PY{k+kn}{import} \PY{n}{diff}\PY{p}{,} \PY{n}{ylabel}\PY{p}{,} \PY{n}{sqrt}

\PY{n}{width} \PY{o}{=} \PY{p}{(}
    \PY{l+m+mf}{4e3}  \PY{c+c1}{\PYZsh{} width used for Gaussian convolution filter}
\PY{p}{)}
\PY{n}{dg\PYZus{}per\PYZus{}V} \PY{o}{=} \PY{l+m+mf}{583e6}  \PY{c+c1}{\PYZsh{} calibration coefficient to convert}
\PY{c+c1}{\PYZsh{}                   the intrinsic SC units to V. Note this value}
\PY{c+c1}{\PYZsh{}                   will change with different SW}
\PY{n}{filename} \PY{o}{=} \PY{l+s+s2}{\PYZdq{}}\PY{l+s+s2}{230822\PYZus{}BNC\PYZus{}RX\PYZus{}magon\PYZus{}200kHz.h5}\PY{l+s+s2}{\PYZdq{}}
\PY{k}{with} \PY{n}{figlist\PYZus{}var}\PY{p}{(}\PY{p}{)} \PY{k}{as} \PY{n}{fl}\PY{p}{:}
    \PY{c+c1}{\PYZsh{} load data and apply preprocessing}
    \PY{n}{s} \PY{o}{=} \PY{n}{find\PYZus{}file}\PY{p}{(}
        \PY{n}{filename}\PY{p}{,}
        \PY{n}{exp\PYZus{}type}\PY{o}{=}\PY{l+s+s2}{\PYZdq{}}\PY{l+s+s2}{ODNP\PYZus{}NMR\PYZus{}comp/Echoes}\PY{l+s+s2}{\PYZdq{}}\PY{p}{,}
        \PY{n}{expno}\PY{o}{=}\PY{l+s+s2}{\PYZdq{}}\PY{l+s+s2}{signal}\PY{l+s+s2}{\PYZdq{}}\PY{p}{,}
        \PY{n}{postproc} \PY{o}{=} \PY{l+s+s2}{\PYZdq{}}\PY{l+s+s2}{spincore\PYZus{}general}\PY{l+s+s2}{\PYZdq{}}\PY{p}{,}
        \PY{n}{lookup} \PY{o}{=} \PY{n}{lookup\PYZus{}table}
    \PY{p}{)} \PY{c+c1}{\PYZsh{} digital filter is applied in preprocessing}
    \PY{n}{s}\PY{o}{.}\PY{n}{rename}\PY{p}{(}
        \PY{l+s+s2}{\PYZdq{}}\PY{l+s+s2}{nScans}\PY{l+s+s2}{\PYZdq{}}\PY{p}{,} \PY{l+s+s2}{\PYZdq{}}\PY{l+s+s2}{capture}\PY{l+s+s2}{\PYZdq{}}
    \PY{p}{)}  \PY{c+c1}{\PYZsh{} to be more consistent with the oscilloscope data,}
    \PY{c+c1}{\PYZsh{}    rename the scans axis}
    \PY{n}{s} \PY{o}{/}\PY{o}{=} \PY{n}{dg\PYZus{}per\PYZus{}V}  \PY{c+c1}{\PYZsh{} convert the intrinsic units of the SC}
    \PY{c+c1}{\PYZsh{}                into actual $V_p$}
    \PY{n}{s}\PY{o}{.}\PY{n}{set\PYZus{}units}\PY{p}{(}\PY{l+s+s2}{\PYZdq{}}\PY{l+s+s2}{t}\PY{l+s+s2}{\PYZdq{}}\PY{p}{,} \PY{l+s+s2}{\PYZdq{}}\PY{l+s+s2}{s}\PY{l+s+s2}{\PYZdq{}}\PY{p}{)}  
    \PY{c+c1}{\PYZsh{} calculate acquisition time}
    \PY{n}{acq\PYZus{}time} \PY{o}{=} \PY{n}{diff}\PY{p}{(}\PY{n}{s}\PY{o}{.}\PY{n}{getaxis}\PY{p}{(}\PY{l+s+s2}{\PYZdq{}}\PY{l+s+s2}{t}\PY{l+s+s2}{\PYZdq{}}\PY{p}{)}\PY{p}{[}\PY{n}{r\PYZus{}}\PY{p}{[}\PY{l+m+mi}{0}\PY{p}{,} \PY{o}{\PYZhy{}}\PY{l+m+mi}{1}\PY{p}{]}\PY{p}{]}\PY{p}{)}\PY{p}{[}\PY{l+m+mi}{0}\PY{p}{]}
    \PY{n}{s}\PY{o}{.}\PY{n}{ft}\PY{p}{(}
        \PY{l+s+s2}{\PYZdq{}}\PY{l+s+s2}{t}\PY{l+s+s2}{\PYZdq{}}\PY{p}{,} \PY{n}{shift}\PY{o}{=}\PY{k+kc}{True}
    \PY{p}{)}  \PY{c+c1}{\PYZsh{} $\frac{V_{p}\sqrt{s}}{\sqrt{Hz}}$}
    \PY{n}{s} \PY{o}{/}\PY{o}{=} \PY{n}{sqrt}\PY{p}{(}\PY{l+m+mi}{2}\PY{p}{)}  \PY{c+c1}{\PYZsh{} instantaneous $\frac{V_{p}\sqrt{s}}{\sqrt{Hz}} \rightarrow \frac{V_{rms}\sqrt{s}}{\sqrt{Hz}}$}
    \PY{c+c1}{\PYZsh{} \PYZob{}\PYZob{}\PYZob{} equation 21}
    \PY{n}{s} \PY{o}{=} \PY{p}{(}
        \PY{n+nb}{abs}\PY{p}{(}\PY{n}{s}\PY{p}{)} \PY{o}{*}\PY{o}{*} \PY{l+m+mi}{2}
    \PY{p}{)}  \PY{c+c1}{\PYZsh{} take mod squared to convert to energy $\frac{V_{rms}^{2} \cdot s}{Hz}$}
    \PY{n}{s}\PY{o}{.}\PY{n}{mean}\PY{p}{(}\PY{l+s+s2}{\PYZdq{}}\PY{l+s+s2}{capture}\PY{l+s+s2}{\PYZdq{}}\PY{p}{)}  \PY{c+c1}{\PYZsh{} average over all captures}
    \PY{n}{s} \PY{o}{/}\PY{o}{=} \PY{n}{acq\PYZus{}time}  \PY{c+c1}{\PYZsh{} convert to Power $\frac{V_{rms}^2}{Hz} = W$}
    \PY{n}{s} \PY{o}{/}\PY{o}{=} \PY{l+m+mi}{50}  \PY{c+c1}{\PYZsh{} W/Hz}
    \PY{c+c1}{\PYZsh{} \PYZcb{}\PYZcb{}\PYZcb{}}
    \PY{c+c1}{\PYZsh{} plot the unconvolved PSD on a semilog plot}
    \PY{n}{fl}\PY{o}{.}\PY{n}{next}\PY{p}{(}\PY{l+s+s2}{\PYZdq{}}\PY{l+s+s2}{\PYZdq{}}\PY{p}{,} \PY{n}{figsize}\PY{o}{=}\PY{p}{(}\PY{l+m+mi}{6}\PY{p}{,} \PY{l+m+mi}{4}\PY{p}{)}\PY{p}{)}
    \PY{n}{fl}\PY{o}{.}\PY{n}{plot}\PY{p}{(}
        \PY{n}{s}\PY{p}{,} \PY{n}{color}\PY{o}{=}\PY{l+s+s2}{\PYZdq{}}\PY{l+s+s2}{blue}\PY{l+s+s2}{\PYZdq{}}\PY{p}{,} \PY{n}{alpha}\PY{o}{=}\PY{l+m+mf}{0.1}\PY{p}{,} \PY{n}{plottype}\PY{o}{=}\PY{l+s+s2}{\PYZdq{}}\PY{l+s+s2}{semilogy}\PY{l+s+s2}{\PYZdq{}}
    \PY{p}{)}
    \PY{c+c1}{\PYZsh{} convolve using the width specified above}
    \PY{c+c1}{\PYZsh{} the convolve function of pySpecData automatically}
    \PY{c+c1}{\PYZsh{} applies the division by $\sigma\sqrt{2\pi}$}
    \PY{n}{s}\PY{o}{.}\PY{n}{convolve}\PY{p}{(}\PY{l+s+s2}{\PYZdq{}}\PY{l+s+s2}{t}\PY{l+s+s2}{\PYZdq{}}\PY{p}{,} \PY{n}{width}\PY{p}{,} \PY{n}{enforce\PYZus{}causality}\PY{o}{=}\PY{k+kc}{False}\PY{p}{)}
    \PY{c+c1}{\PYZsh{} plot the convolved PSD on the semilog plot with the}
    \PY{c+c1}{\PYZsh{} unconvolved}
    \PY{n}{fl}\PY{o}{.}\PY{n}{plot}\PY{p}{(}
        \PY{n}{s}\PY{p}{,} \PY{n}{color}\PY{o}{=}\PY{l+s+s2}{\PYZdq{}}\PY{l+s+s2}{blue}\PY{l+s+s2}{\PYZdq{}}\PY{p}{,} \PY{n}{alpha}\PY{o}{=}\PY{l+m+mf}{0.5}\PY{p}{,} \PY{n}{plottype}\PY{o}{=}\PY{l+s+s2}{\PYZdq{}}\PY{l+s+s2}{semilogy}\PY{l+s+s2}{\PYZdq{}}
    \PY{p}{)}
    \PY{n}{ylabel}\PY{p}{(}\PY{l+s+s2}{\PYZdq{}}\PY{l+s+s2}{\PYZdl{}}\PY{l+s+s2}{\PYZob{}}\PY{l+s+s2}{S(}\PY{l+s+se}{\PYZbs{}\PYZbs{}}\PY{l+s+s2}{nu)\PYZcb{}\PYZcb{}\PYZdl{} / (W/Hz)}\PY{l+s+s2}{\PYZdq{}}\PY{p}{)}
\end{Verbatim}
\par\noindent
\rule{\linewidth}{0.5pt}
}

\putbib
\end{bibunit}
\else
\bibliography{references}
\fi
\end{document}